\pdfoutput=1 

\documentclass[journal]{IEEEtran}
\usepackage{graphicx} 
\usepackage{caption} 
\usepackage{amsmath}
\usepackage{amsmath,amssymb,amsfonts}
\usepackage[switch, mathlines]{lineno}
\usepackage[subrefformat=parens,labelformat=parens]{subfig}
\usepackage{tikz}
\usepackage{cite}
\usetikzlibrary{shapes.geometric, arrows}
\usepackage{array}
\usepackage{textcomp}
\newcolumntype{C}[1]{>{\centering\arraybackslash}m{#1}}
\usepackage{relsize}
\usepackage{stackengine}
\usepackage{comment}
\usepackage[font={footnotesize}]{caption}
\usepackage{threeparttable}
\usepackage[nottoc]{tocbibind}
\usepackage{multirow}

\ifCLASSINFOpdf
\else
\fi
\begin{document}

%

\title{Computational Scatter Correction for High-\\ Resolution Flat-Panel CT Based on a Fast \\ Monte Carlo Photon Transport Model }

%
%
%

\author{Ammar~Alsaffar,
        Steffen~Kie\ss, Kaicong~Sun, 
        and~Sven~Simon
}

\maketitle

\begin{abstract}


In computed tomography (CT) reconstruction, scattering causes server quality degradation of the reconstructed CT images by introducing streaks and cupping artifacts which reduce the detectability of low contrast objects. Monte Carlo (MC) simulation is considered as the most accurate approach for scatter estimation. However, the existing MC estimators are computationally expensive especially for the considered high-resolution flat-panel CT. In this paper, we propose a fast and accurate photon transport model which describes the physics within the 1 keV to 1 MeV range using multiple controllable key parameters. Based on this model, scatter computation for a single projection can be completed within a range of few seconds under well-defined model parameters. Smoothing and interpolation are performed on the estimated scatter to accelerate the scatter calculation without compromising accuracy too much compared to measured near scatter-free projection images. Combining the scatter estimation with the filtered backprojection (FBP), scatter correction is performed effectively in an iterative manner. In order to evaluate the proposed MC model, we have conducted extensive experiments on the simulated data and real-world high-resolution flat-panel CT. Comparing to the state-of-the-art MC simulators, our photon transport model achieved a 202$\times$ speed-up on a four GPU system comparing to the multi-threaded state-of-the-art EGSnrc MC simulator. Besides, it is shown that for real-world high-resolution flat-panel CT, scatter correction with sufficient accuracy is accomplished within one to three iterations using a FBP and a forward projection computed with the proposed fast MC photon transport model. 

\end{abstract}


%
\IEEEpeerreviewmaketitle

\section{Introduction}
%
%
%
%
\IEEEPARstart{C}{omputed} tomography (CT) imaging is the state-of-the-art non-destructive testing technique. It provides inception about the inner of the scanned object and is widely used for industrial and medical applications. However, this technique suffers from severe quality degradation artifacts. Among these artifacts the scatter, which results from the change in the direction or the direction and the energy of the photon penetrating the object, degrades the quality of the CT reconstructed image by inserting cupping and streak artifacts. These artifacts reduce the contrast of this image \cite{Zbijewski,Endo}, and the contrast-to-noise \cite{Siewerdsen,Ding,Watson}. Accurate scatter computation could be done using the MC method due to the accurate modeling of the physics involved in the photon matter interactions, the shape and the composition of the object require no simplification and could be easily simulated, the ability of this method to model multiple scattered photons, and the real-world scanner configuration which could be easily be considered in the MC simulation \cite{Zbijewski}. Although it is accurate, the MC method requires a huge computation time.

An accurate and fast MC model accelerated over multiple-GPU is implemented in this paper, which simulates the fundamentals physics involved within the keV range including Compton scattering, Rayleigh scattering, and photoelectric absorption. This MC model has been extensively evaluated against other MC simulators and real-world CT scanner. On the other hand, an iterative scatter correction algorithm which requires a few iterations for scatter compensation is also implemented, the latter has been embedded with the implemented MC model for fast scatter estimation. The estimated scatter from this MC model is used to correct the scatter-corrupted raw projections from the real-world scanner. Moreover, smoothing and interpolation are used to accelerate the iterative scatter correction algorithm. Smoothing reduces the time needed for the MC simulation by denoising the estimated scatter acquired using a reduced number of photons, while interpolation is used to estimates a full set of projections from half the number, and to up-sampling the simulated projection from low-resolution to a high-resolution. The correction result from the iterative scatter correction algorithm is in fine agreement with the near scatter-free collimator based reconstructed data set.

  
\subsection{Contributions}
\subsubsection{Implementation of an Accurate and Parameterized MC Forward Projection Model with Postprocessing}
The proposed MC forward projection model incorporates basic physical properties such as Compton scattering, Rayleigh scattering, and photoelectric absorption, as well as parameterized post-processing such as filtering and resolution interpolation to reduce the computation time through slight approximations that lead to minor but tolerable deviations in the computation result. In addition, the basic parameters of the MC model, such as the step size of the ray tracing and the division number of photons, can be controlled to balance the trade-off between the speed-up and the accuracy of the scattering estimate.


\subsubsection{Multi-GPU Acceleration of the MC Simulation}
Since CT reconstruction algorithms are usually accelerated on multi-GPUs and thus these GPU hardware resources are also available for scattering correction, a multi-GPU accelerated MC simulation model was implemented such that thousands of photons are simulated simultaneously. In comparison with the state-of-the-art multi-threaded CPU MC simulator EGSnrc, our model achieves 202× speed-up on a four GPUs system.



\subsubsection{Fast Iterative Scatter Correction Algorithm}

MC simulation is considered as the gold standard for the scatter estimation\cite{Watson,Jarry,Jarryx,Hing,Zhang,Ruhrnschopf,Maier,Chan}, due to the accurate modeling of the physics involved in the photon transport. However, it is computationally expensive. This has led to a choice of methods other than MC simulation for scatter correction for the reconstruction of volume representations in computed tomography in the past. The huge computation time is particularly true for the high resolutions in non-medical computed tomography, i.e., computed tomography for materials, engineering, and natural sciences in general. In this work, however, we show that with careful choice of the parameters of the MC simulation, such as the step size of the ray tracing, the number of photons of the simulation, and by the use of smoothing and interpolation techniques which have been shown to lead to acceptable image quality, as well as by hardware accelerators (GPUs), such as those used for CT reconstruction, two things are achieved simultaneously. These two things are both acceptable quality of results and speed in terms of X-ray scatter calculation and correction and are achieved simultaneously, making MC simulation and correction of scattering suitable for integration into a real-world high-resolution flat-panel CT reconstruction. Based on this approach, we achieved a speed-up of MC simulation in a demonstrated use case from 45.4 h to 0.28 h for 3000 projections of a computed tomography scan with a flat-panel detector of $2k\times3k$ resolution. This computation time for MC simulation is even lower than the time required for the Filtered Back-Projection (FBP) being the fastest and the most widely used reconstruction technique which takes almost 0.46 h in our implementation for this case.

\subsubsection{Experimentally Proven Sufficient Accuracy for the Fast MC-based Scatter Correction}

Furthermore, it is shown that the proposed MC model and implementation for scatter correction provide a solution to the problem of scatter correction with an accuracy that is acceptable compared to experimental results. These experimental reference results were generated with CT scans using a collimator, which is commonly used to produce a nearly scatter-free reference result for individual 2D-like layers of a 3D object.

\subsection{Related Work}

The scatter correction of projections from the X-ray scanner results in an enhanced CT reconstructed image quality. Several methods have been introduced to compensate for scatter. According to \cite{Jin}, these methods are divided into two approaches. The first approach is the direct elimination of the scatter during the scan process by the use of anti-scatter grid \cite{Sorensonanti}, bowite filter \cite{Graham}, or optimizing the geometry by increasing the distance between the scattering object and detector, \cite{Neitzel,Sorenson,Persliden}. These methods are only able to reduce the scatter in the reconstructed image due to a better CT-scanner setup and do not lead to a complete scatter correction. The second approach is by computing the scatter and its subtraction in each projection using different computational models like an analytical or empirical approach, by neural networks \cite{Maier} or by MC simulation. 

MC simulation is an accurate approach to model the scatter due to the accurate modeling of the physics involved in the photon-matter interaction. Many MC simulators have been introduced for this purpose. Among these simulators are MCNP \cite{MCNP}, aRTist \cite{aRTist}, \cite{Bellon}, EGSnrc \cite{EGSnrc}, Penelope \cite{Penelope}, and others. The main drawback of the existing MC simulators is their required computational effort. To reduce the computation time, many acceleration methods are adopted. In \cite{Hing,Watson,Xu,ThingRR,Mainergra}, it is mentioned that the use of what is known as the variance reduction techniques, such as photon splitting, Russian Roulette, and forced detection could enhance the efficiency of the photon transport and reduce the execution time by several orders of magnitude. 

Apart from the aforementioned algorithmic acceleration techniques, GPU is employed to further accelerate the MC simulation from the hardware aspect. In \cite{Badal}, the authors have accelerated the MC simulation in a voxelized geometry using the Penelope MC simulation physics by the use of the CUDA programming model. A speed-up factor of 27 compared to the CPU version of this simulator is achieved. The authors in \cite{Lippuner} have implemented the simulation of the photon transport of the EGSnrc simulator on the GPU by the use of the CUDA programming model. Between 20 to 40$\times$ speed-up is achieved depending on the number of voxels used in the simulation. The Geant4 MC simulator is the base of the work in \cite{Bert} in which the authors took the advantage of the varieties of the physics available in this simulator and performed the acceleration on the GPU with a speed-up factor of 86. In contrast to the all methods mentioned above, which perform the MC simulation using a voxelized object, the method in \cite{Chi} uses quadratic functions to represent the bounding surfaces of a region. As a result, a $\sim$3$\times$ higher computation time required in comparison to the use of a voxelized geometry. A hybrid approach of using a GPU accelerated MC simulation combined with the use of the variance reduction technique is introduced in\cite{Sisniega}.

    Scatter correction based on MC computation is the focus of several works. In \cite{Hing}, the authors have used the EGSnrc simulator supported by the use of the variance reduction techniques to correct the scatter-corrupted projections iteratively for low resolution projections. The same approach is extended for a real phantom study in \cite{Watson} also for low resolution projections. Other works based on CPU MC simulators are found in \cite{Jarry}, \cite{Jarryx}, \cite{ThingR}, \cite{Bertram}. Although the number of projections and photons used in these works are low, their simulations required long computation time beyond the applicability to high resolution flat-panel CT and with compromised correction quality \cite{Xu}. Acceleration techniques are employed to accelerate the MC simulators and the scatter correction algorithms. The iterative scatter correction in \cite{Zbijewski} is based on the fast estimation of the scatter by using a low number of photons. The resultant noisy scatter estimation was then efficiently denoised by a three-dimensional fitting of Gaussian basis function. However, this approach works well for small objects only \cite{Mainergra}. The MC-based scatter correction algorithm proposed in \cite{Poludniowski} estimates the scatter on a small number of detector nodes combined with a reduced number of projections. Linear interpolation is then used between the nodes and projections to derive the complete scatter estimation of the scan. Such an approach could not track possible high spatial frequencies in the scatter distribution especially if the interpolation grid is too coarse \cite{Mainergra}. The fast scatter correction algorithm proposed in \cite{Xu}, which is based on a single-GPU MC scatter simulation and extended in \cite{Zhang} for multi-GPU, relies again on the usage of a very low number of photons and projections and requires the availability of a priori information such as the planning CT scan.
    
 In this work, we have implemented a fast and accurate MC forward projection model and an iterative scatter correction algorithm for high-resolution flat-panel CT. Unlike most of the previously mentioned works, the scatter correction algorithm introduced here does not rely on heavily smoothing or interpolating the simulated projections. By applying certain acceleration techniques, the proposed MC model could achieve a 162$\times$ speed-up in comparison to the standard case without acceleration using the same model. As a consequence, the time needed for the MC simulation for high-resolution flat-panel CT is less than the time needed for the FBP reconstruction. This accelerates the speed of the proposed iterative scatter correction algorithm a lot without compromising the accuracy of the scatter correction results as they match the near scatter-free results acquired using the collimator.

\section{ Material and Methods}
\subsection{The Proposed MC Photon Transport Model}

The MC photon transport model implemented is based on the physics of the photon-matter interaction. The major photon interaction processes such as the photoelectric, the Compton, and the Rayleigh scattering are integrated in this model. As in reality, the photon interaction with the object occurs with a bound non-stationary electron of an atom, the binding effect of the electron-atom is considered in both the Compton and the Rayleigh scattering by the use of the form factors tabulated in \cite{Hubbel}. The Doppler broadening in the Compton scattering, which is due to the non-static nature of the atom is taking into consideration \cite{MCNP}, \cite{Namito}. Including these effects is necessary to produce an accurate simulation that mimics the realistic interaction of a photon with the atom's electron\cite{MCNP}. The photon's scatter angle after the interaction is calculated by randomly sampling using the rejection and inversion method \cite{Plante}. This sampling is done using the Compton scattering PDF (extended from the Klein Nishina PDF by including the form factor) shown in (\ref{eqn:KleinNishina}) and using the Rayleigh scattering PDF (extended from the Thomson PDF by including the form factor) shown in (\ref{eqn:Thomson}) respectively \cite{MCNP}.


\begin{figure}[ht!]
\centering
\def\stackalignment{l}
\subfloat{\topinset{\bfseries \textcolor{black}{(a)}}{\label{fig:SamplingKleinNishina}\includegraphics[width=0.5\linewidth]{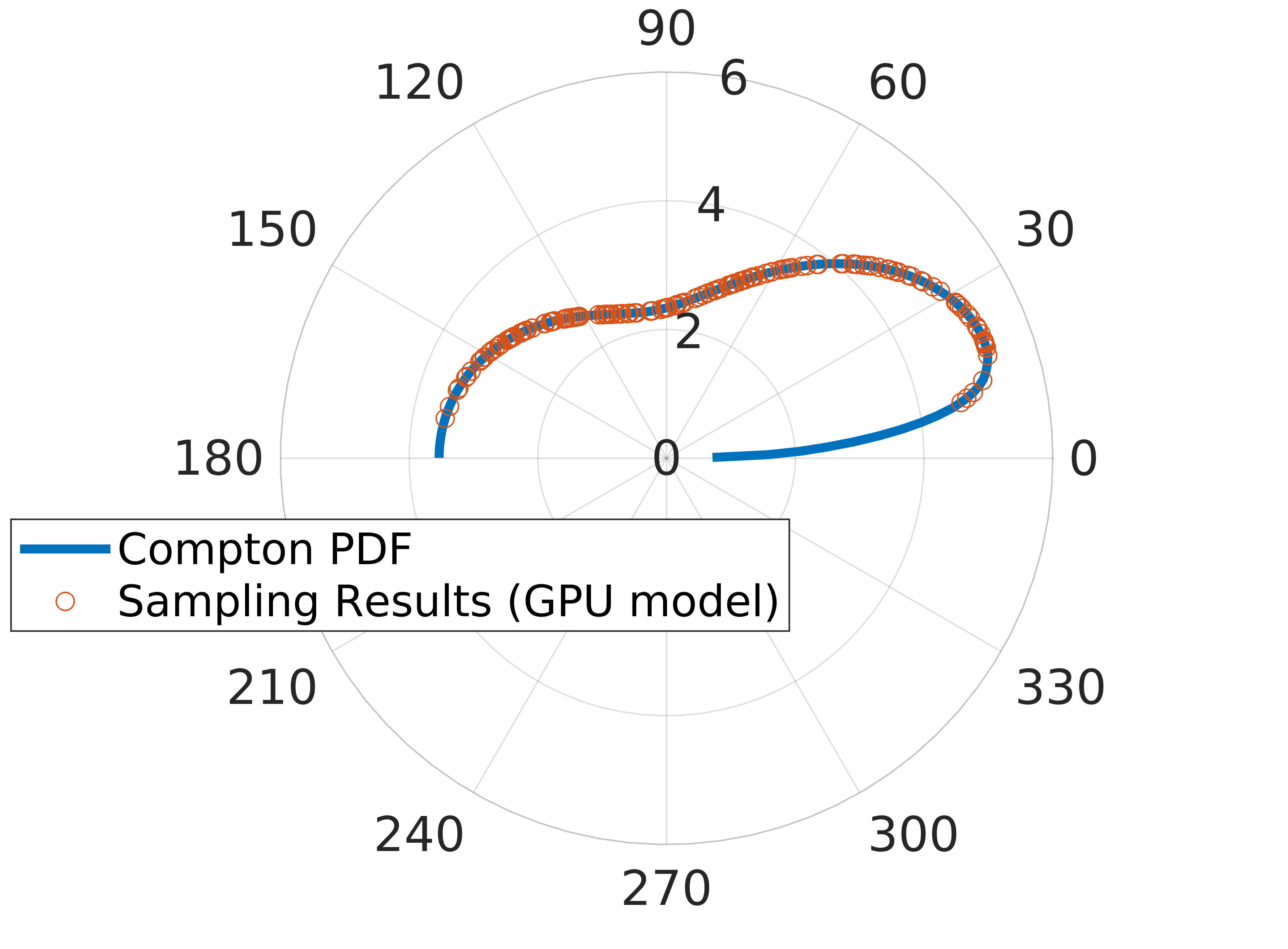}}{0.12in}{.27in}}
\subfloat{\topinset{\bfseries \textcolor{black}{(b)}}{\label{fig:SamplingThomson}\includegraphics[width=0.48\linewidth]{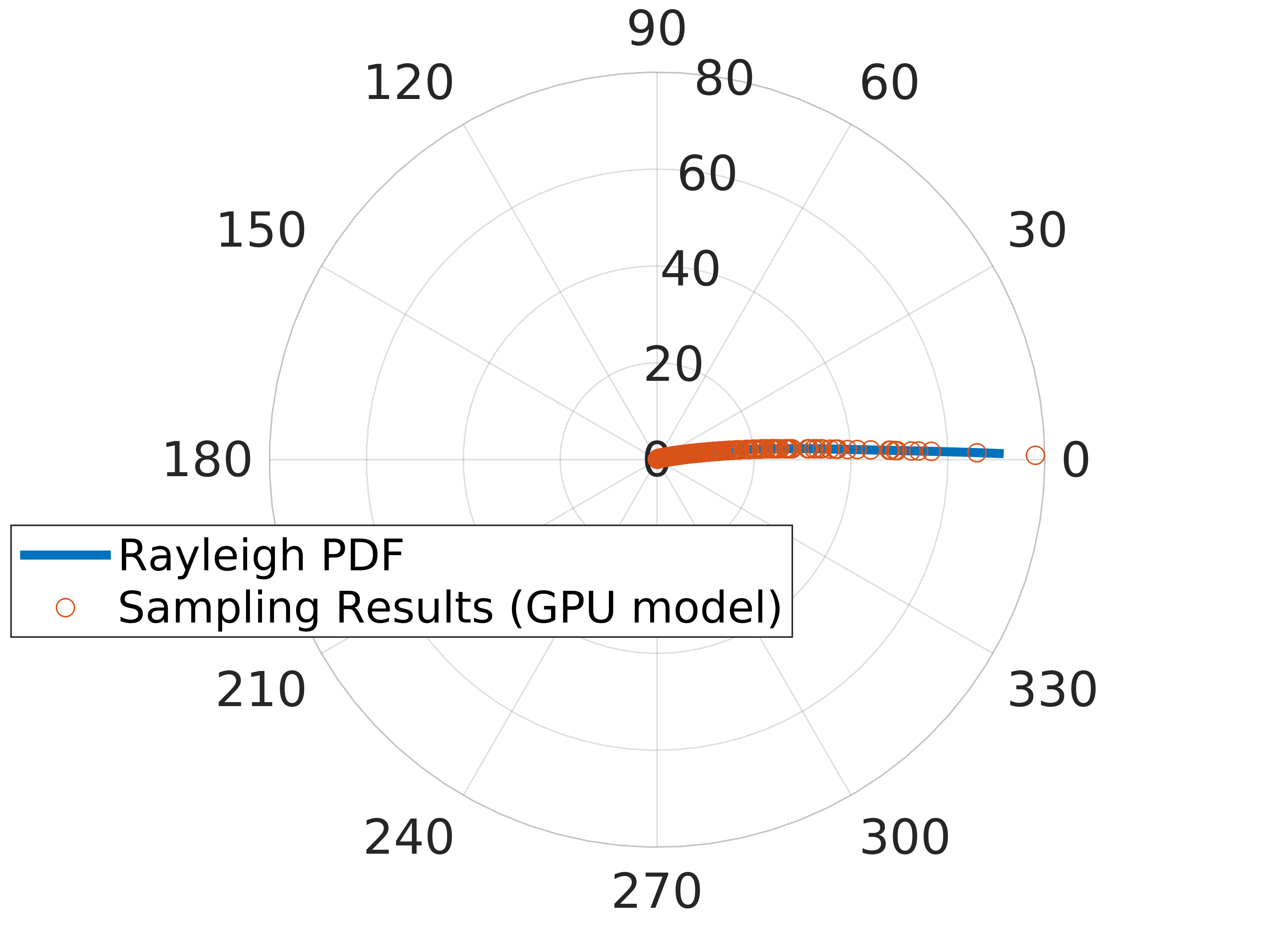}}{0.12in}{.27in}}\\[-0.1ex]

\caption{Sampling from the Compton and the Rayleigh scattering PDFs. (a) Compton scattering cross section and the sampling result from the proposed MC model, (b) Rayleigh scattering cross section and the sampling result from the proposed MC model. Both results are for aluminum at photon energies of 100 keV.}
\label{fig:Sampling of PDFs}
\end{figure}

\begin{align}
\label{eqn:KleinNishina}
\frac{d\sigma\textsubscript{Incoherent}}{d\Omega}(\theta)=\frac{r_0 ^2}{2}\Big(\frac{\alpha^\prime}{\alpha}\Big)^2\Big(\frac{\alpha^\prime}{\alpha}+\frac{\alpha}{\alpha^\prime}-sin(\theta)^2\Big)S(q,z),  \  
\end{align}

where $\Omega=2\pi (1-cos(\theta))$, $\theta$ is the scatter angle of the photon, $r_0$ is the electron radius $2.9817938\times10^{-13}$ cm, $\alpha$ and $\alpha^\prime$ are the incident and final photon energies in units of 0.511 MeV, $\alpha=E/(mc^2)$, where $m$ is the mass of the electron and $c$ is the speed of light, and $\alpha^\prime=\alpha/ [1+\alpha(1-cos(\theta))]$, 
$S(q,Z)$ is an appropriate scattering factor modifying the Klein-Nishina cross section taken from \cite{Hubbel} with $q$ being the inverse length and $Z$ being the atomic number of the material.


\begin{align}
\label{eqn:Thomson}
\frac{d\sigma\textsubscript{coherent}}{d\Omega}(\theta)=\frac{r_0 ^2}{2}\Big(1+cos(\theta)^2\Big)F(q,z)^2,  \  
\end{align}  

where $F(q,z)$ is a form factor modifying the energy-independent Thomson cross section taken from \cite{Hubbel}.
The sampling from the Compton scattering PDF is performed according to Kahn's method\cite{Kahn}, \cite{Raeside}. After the photon encounters an interaction, the new angle of the scattered photon, and the change in its energy after the interaction are determined. Fig.~\ref{fig:Sampling of PDFs}\subref{fig:SamplingKleinNishina} and Fig.~\ref{fig:Sampling of PDFs}\subref{fig:SamplingThomson} show the PDFs of the Compton and the Rayleigh scattering respectively and the results of the sampling from these PDFs taken from the proposed MC model. The continuous lines show the actual PDFs while the dots on the lines show the results of the sampling from the proposed MC model.  

The scoring of the scatter on the detector in this model is done by the use of the point detector method, i.e., a contribution to the detector from each photon's interaction point is calculated using the following equation \cite{MCNP}.

\begin{align}
\label{eqn:Th}
x=\frac{\eta(E) p(\lambda) W}{2\pi d^2}exp(-\mu(E) L), \ 
\end{align} 

where $\eta(E)$ is the energy response of the detector, $p(\lambda)$ represents the probability of scatter toward the detector given by (\ref{eqn:pmueforIncoherent}) and (\ref{eqn:pmueforcoherent}) for the Compton and the Rayleigh scatter respectively, $\lambda$ is the cosine of the angle between the photon path and the direction to the detector \cite{MCNP}, $W$ is the photon's weight, $d$ is the distance from the interaction point to the detector, and $exp(-\mu(E) L)$ accounts for the attenuation from the interaction point to the detector.

\begin{align}
\label{eqn:pmueforIncoherent}
p(\lambda)=\frac{\pi r_0 ^2}{\sigma\textsubscript{Incoh}(Z,\alpha)}\Big(\frac{\alpha^\prime}{\alpha}\Big)^2\Big(\frac{\alpha^\prime}{\alpha}+\frac{\alpha}{\alpha^\prime}-sin(\theta)^2\Big)S(q,z).  \  
\end{align}

\begin{align}
\label{eqn:pmueforcoherent}
p(\lambda)=\frac{\pi r_0 ^2}{\sigma\textsubscript{Coh}(Z,\alpha)}\Big(1+cos(\theta)^2\Big)F(q,z)^2.  \   
\end{align} 

$\pi r_0 ^2$ is a constant and $\sigma\textsubscript{Incoh}$, $\sigma\textsubscript{Coh}$, are the integrated incoherent and coherent cross section respectively taken from \cite{Hubbel}.


The scatter intensity projection on the detector is calculated by summing up the contributions from all the interactions points of all the simulated photons (\ref{eqn:Is}). The selection of the pixel where the scoring occurs is done randomly.

\begin{align}
\label{eqn:Is}
Is=\sum_{0}^{N}\sum_{0}^{M}\sum_{0}^{K}x, \ 
\end{align}

where $N$ is the number of energy bins of the polychromatic source, $M$ is the number of photons in the energy bin, and $K$ is the number of interactions. The scoring from the primary photons that reach the detector without encountering any interaction is given by (\ref{eqn:Ip}).

\begin{align}
\label{eqn:Ip}
I_p= \int_{0}^{E\textsubscript{max}} \, \frac{S(E) \eta(E)}{d^2}exp(-\mu(E) L)dE, \ 
\end{align} 

where $E\textsubscript{max}$ is the maximum energy of the polychromatic spectrum, and $S(E)$ is the spectrum.
The polychromatic nature of the source is considered by dividing the spectrum into discrete energy bins. Each photon from each energy bin is simulated individually. The model is originally designed to perform simulation on CAD surface model using triangular elements, it is then extended to work on a voxelized object to enable the use of this model in the correction of the scatter iteratively\cite{Watson}. 



\subsection{Multi-GPU Implementation of the Forward Projection Model}
The GPU implementation is achieved using the OpenCL C++ platform in which multiple thousands of photons are simulated simultaneously by exploiting the threads of the GPU. Regarding multi-GPU implementation, the number of projections is equally assigned to the available GPUs so that these projections can be simulated simultaneously.

Tables \ref{tab:time1}, \ref{tab:time2}, and \ref{tab:time3} in subsection \ref{Evaluation on Execution Time} show the execution time of the proposed MC model for both the GPU and the CPU versions in comparison to the EGSnrc and the aRTist simulators using four different test objects. 


\subsection{Applied Acceleration Schemes}
In addition to the GPU acceleration, other techniques are utilized to accelerate the scatter correction process, i.e., variance reduction techniques, the controllable step size of the ray tracing, interpolation, and smoothing. Particularly, the variance reduction and the adjustable step size are used to speed-up the proposed MC model while the smoothing and the interpolation are adopted to accelerate the iterative scatter correction algorithm.

\subsubsection{Variance Reduction Techniques}

For scatter signal, the variance per particle is much larger than the primary signal, as the contributions on the detector from some of the points where the photons encounter scatter tend to have larger values in comparison to others. Therefore, acceptable statistical measures are only achievable by using a high number of photons and thus longer computation time is required \cite{Mainergra}. For this reason, special techniques are employed in the MC simulators to increase the efficiency and reduce the simulation time. These techniques are known as the variance reduction techniques. Examples of these techniques are splitting and Russian Roulette. Splitting in which the photon from each interaction point is split into several numbers of what is known as a pseudo particle. This process reduces the variance and enhances the contrast in the CT reconstructed image, as any possible high contribution on the detector is distributed among many pixels. Russian Roulette is used to discard some of the pseudo particles that have a low contribution on the result overall to save the time of the expensive tracking process \cite{MCNP}. 

\subsubsection{Controllable Step Size of the Ray Tracing}

The ray-tracing through a voxelized object is the most expensive part of the MC simulations. To reduce the time of this process, the number of voxels in the object is down-sampled and the step size of the scoring through the voxelized object is made adjustable in the proposed MC model. A study of the effect of different step sizes on the accuracy of the scatter correction can be found in subsection \ref{optimization effect}.


\subsubsection{Dual-Interpolation Techniques}
The proposed MC model is used to calculate the primary and the scatter intensities. These intensities are then imported to the iterative scatter correction algorithm to correct raw projections from the scanner. As the target of this work is to correct high-resolution projections of (2304,3200) pixels, estimated projections from the proposed MC model should also be calculated using the same resolution. This requires the use of a high number of photons to produce adequate quality with the expense of a long computation time. To reduce the number of photons, scatter and primary projections are simulated using a resolution of (576,800) pixels. These projections are then up-sampled to the same resolution from the scanner using cubic interpolation. Moreover, as the scatter projections tend to change slowly between projections \cite{Ning}, half of these projections is calculated through the proposed MC model, the rest is calculated by using the interpolation technique. Whereas the full set of the primary projections are simulated using the proposed MC model. This approach has been tested in the iterative scatter correction algorithm, a comparison between the results of the scatter correction with and without the use of the interpolation technique can be found in subsection \ref{interpolation effect}.

\subsubsection{Smoothing the Noisy Data}
Smoothing operation enhances the estimation speed of the scatter by more than $50\%$ by reducing the number of photons required in the MC simulation \cite{Hing}. Many previous works adopt smoothing filters \cite{Zbijewski}, \cite{Colijn}. The EGSnrc simulator uses the Savitzky-Golay filter \cite{Hing}, \cite{Savitzky}. 

\begin{figure}[ht!]
\centering
\def\stackalignment{l}
\subfloat{\topinset{\bfseries \textcolor{white}{(a)}}{\includegraphics[width=0.5\linewidth]{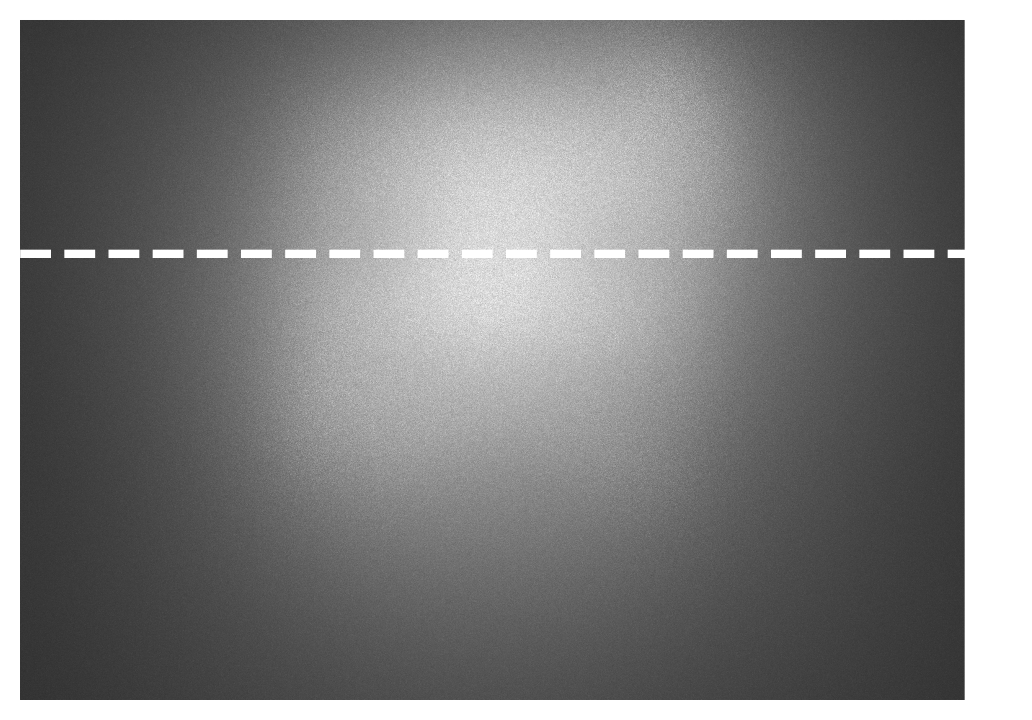}}{0.1in}{.1in}}
\subfloat{\topinset{\bfseries \textcolor{white}{(b)}}{\includegraphics[width=0.5\linewidth]{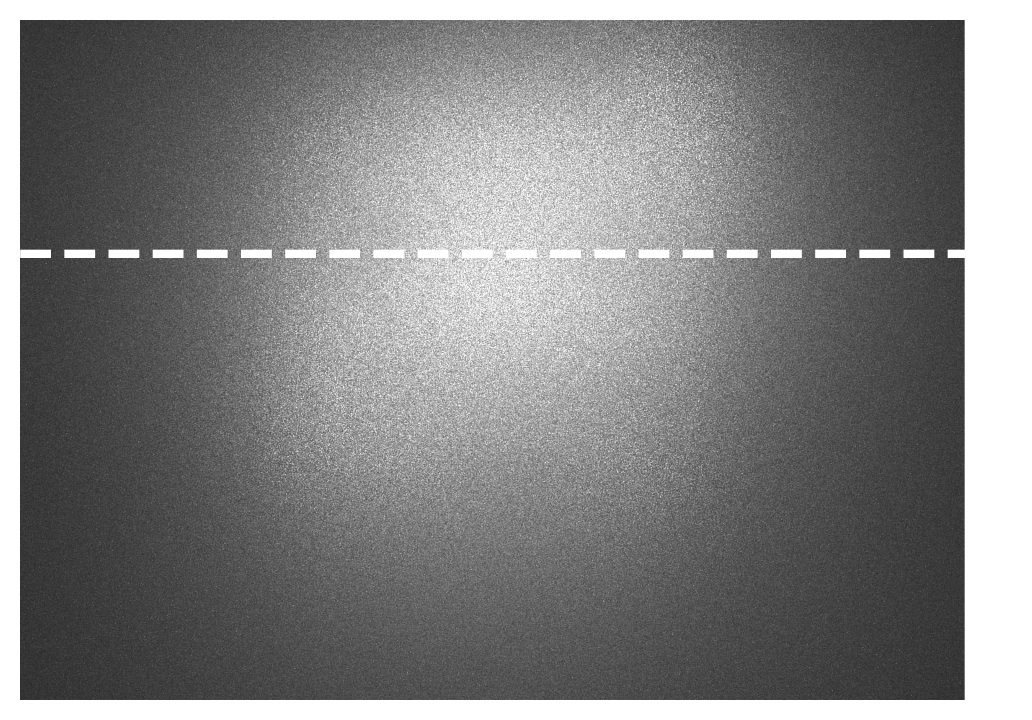}}{0.1in}{.1in}}\\[-0.1ex]
\subfloat{\topinset{\bfseries \textcolor{black}{(c)}}{\includegraphics[width=0.5\linewidth]{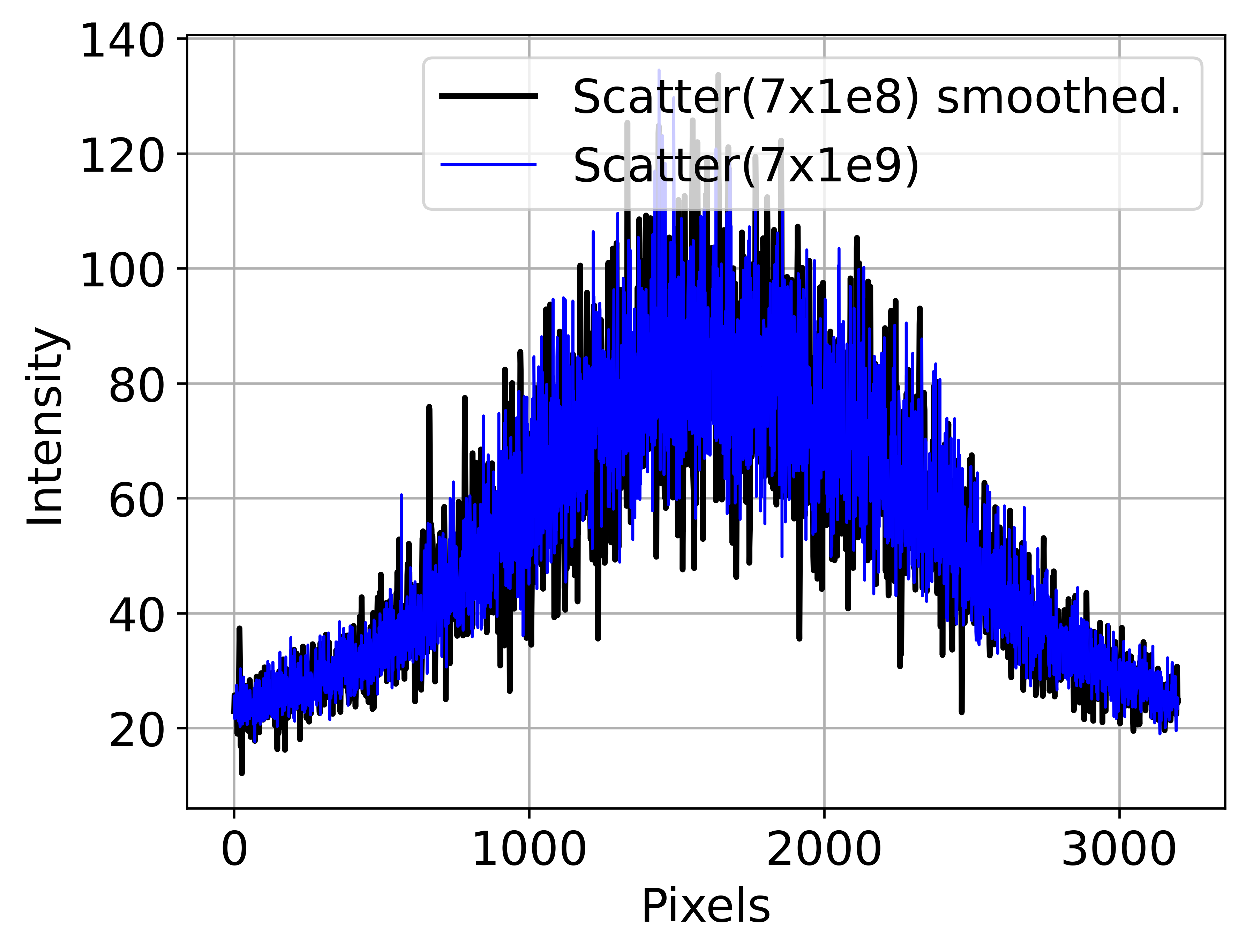}}{0.35in}{.3in}}
\subfloat{\topinset{\bfseries \textcolor{black}{(d)}}{\includegraphics[width=0.5\linewidth]{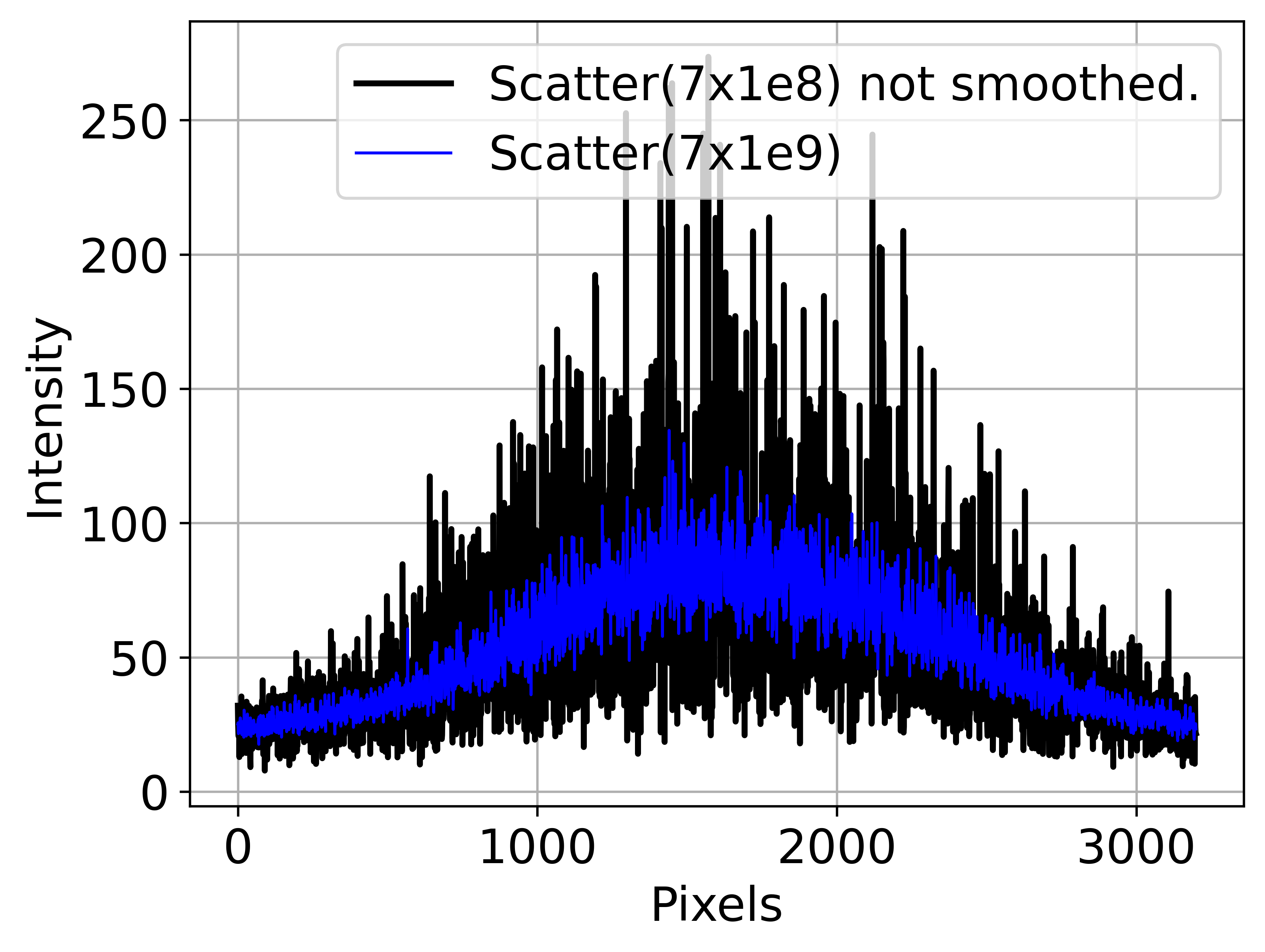}}{0.35in}{.3in}}

\caption{The effectiveness of the smoothing operation using the Savitzky-Golay filter. (a) Scatter projection from the proposed MC model using $7\times10^9$ photons, (b) scatter projection from the proposed MC model using $7\times10^8$ photons smoothed using this filter, (c) profile lines of (a) and (b), (d)  profile lines of (a) and the scatter result from the proposed MC model using $7\times10^8$ photons without smoothing (not shown here).}
\label{fig:Effect of Smoothing}
\end{figure}

The Savitzky-Golay filter is used to denoise the noisy scatter estimates in this work, as this type of filter is able to preserve the high frequency components of the smoothed image \cite{Chakraborty}. This filter applies the use of the least square estimates to fit a polynomial of a certain degree to a subset from the noisy data. This is done in a way that the data attributes such as the peaks and width are preserved. 


 

 
The effect of the use of this kind of filter on denoising scatter projection is shown in Fig. \ref{fig:Effect of Smoothing}. In this figure, the Poisson noise accompanying the estimated scatter projection using low number of photons is suppressed a lot due to the use of this filter. It is shown that the profile line from the denoised scatter projection fits well with the profile line of the scatter projection acquired with a 10$\times$ higher number of photons in comparison to the first case \cite{Hing}. The order of the filter used in this work is three, i.e., a cubic polynomial is being used in the smoothing operation.

\subsection{Simulation of the Polychromatic Behavior}
To achieve an accurate CT scan simulation, the polychromatic behavior of the source and the detector should be simulated accurately. Therefore, the simulations of both were done using MC simulation taking into account the internal construction and the components setup.

\subsubsection{Simulation of the Source}
\begin{figure}[ht!]
\centering
\includegraphics[width=0.7\linewidth]{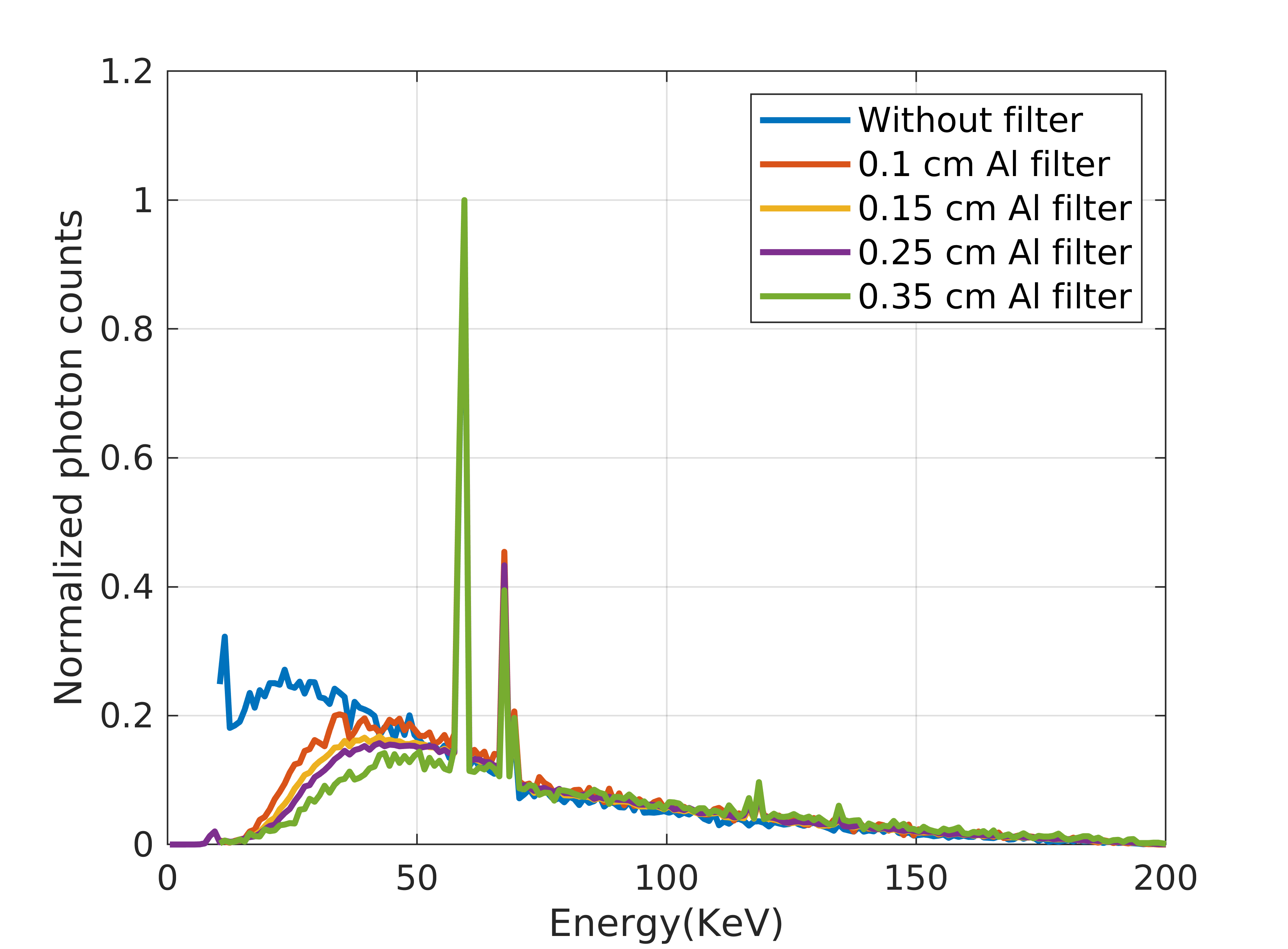}
\caption{Spectrum of 200 keV simulated using the Geant4 MC simulator for different aluminum filtration.}
\label{fig:Source}
\end{figure}

The simulation of the source is implemented using the Geant4 MC simulator. The parameters in the simulation are those of the source from the Nikon XT H 225 ST CT machine. This source simply generates an electrons beam, which is accelerated to hit a tungsten target. The photons generated by this process propagate through a 500 $\mu$m thick beryllium window and the installed filter(s). Fig. \ref{fig:Source} shows examples of the spectrum generated using simulations with different filtration.

\subsubsection{Simulation of the Detector} 
\begin{figure}[ht!]
\centering
\def\stackalignment{l}
\subfloat{\topinset{\bfseries \textcolor{black}{(a)}}{\label{fig:detectorefficiency}\includegraphics[width=0.5\linewidth]{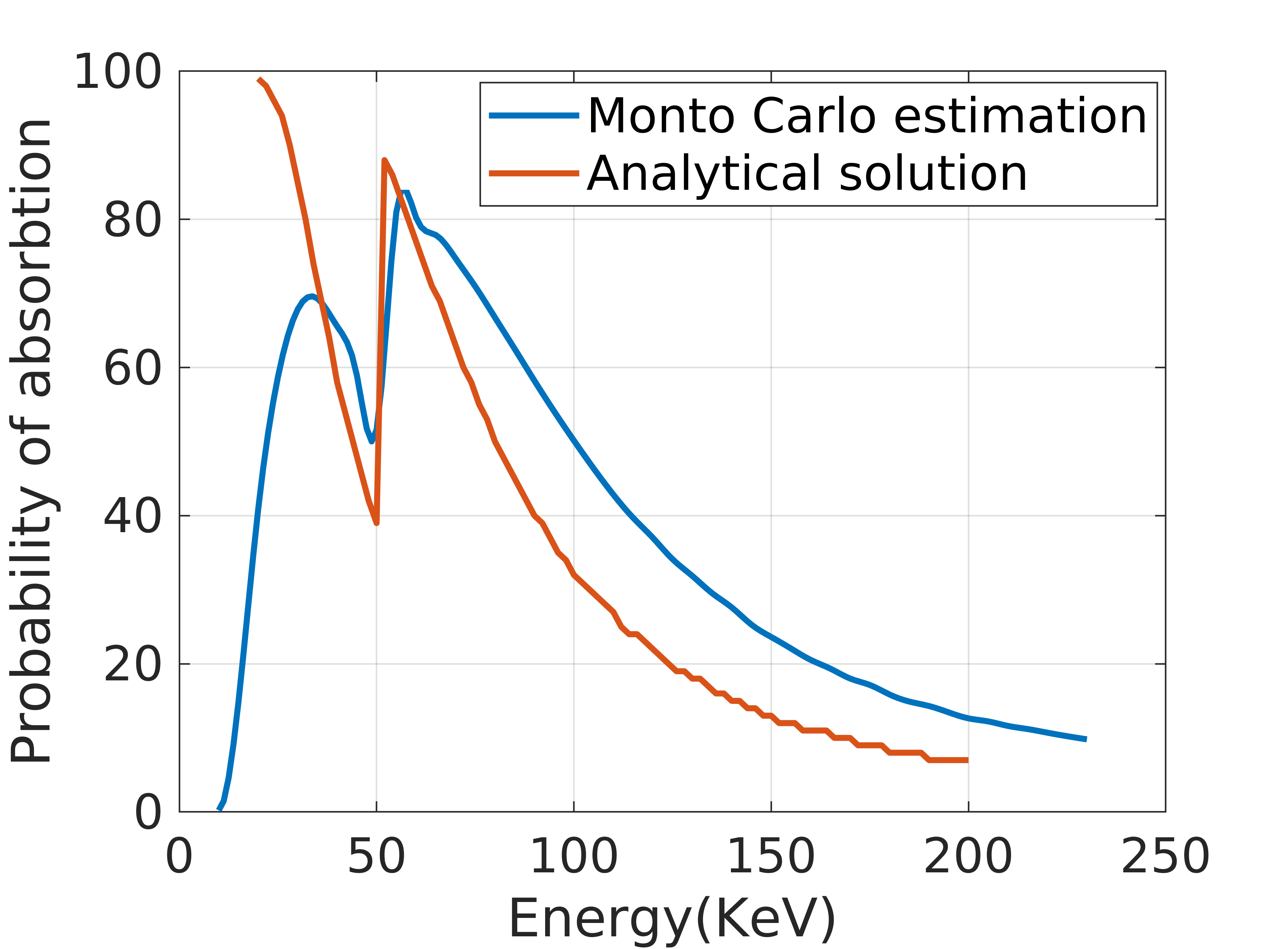}}{0.12in}{.27in}}
\subfloat{\topinset{\bfseries \textcolor{black}{(b)}}{\label{fig:depositenergy}\includegraphics[width=0.5\linewidth]{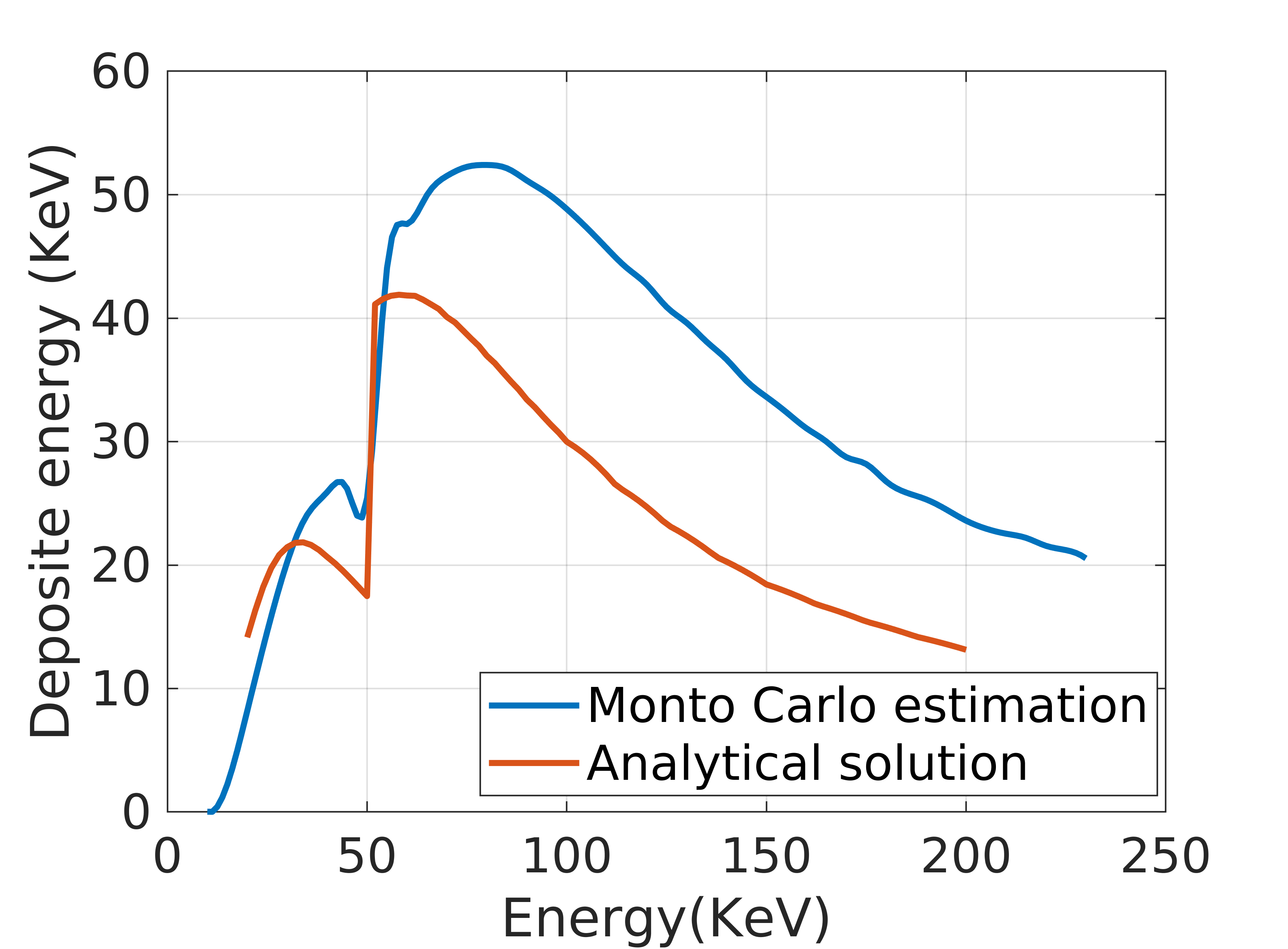}}{0.12in}{.27in}}\\[-0.1ex]

\caption{Simulation of the detector. (a) The DQE simulated analytically and using the Geant4 MC simulator, (b) the deposit energy of the photon on the detector simulated analytically and using the Geant4 MC simulator.}
\label{fig:detector response}
\end{figure}

The parameters of the detector in the simulation are those of the used PaxScan 4030E detector from Varian. This detector contains a 2.86 cm vacuum layer followed by a 0.25 cm carbon layer, a thin aluminum foil, a 0.0208 cm scintillator layer, and the silicon layer. The scintillator layer of this detector is made from Gadolinium Oxysulfide (cd2o2s).
To measure the efficiency of the X-ray detector, the detective quantum efficiency (DQE) (or the absorption probability of the photon) and the energy response of this detector have been simulated analytically and by the use of the Geant4 MC simulator for more accurate simulation. Fig.~\ref{fig:detector response}\subref{fig:detectorefficiency} shows the DQE of this detector. While Fig.~\ref{fig:detector response}\subref{fig:depositenergy} shows the detector response.





The proposed MC model maps the detected photon energy on the detector according to the energy response curve (Fig.~\ref{fig:detector response}\subref{fig:depositenergy}). 

\subsection{Iterative Scatter Correction}
The iterative scatter correction algorithm based on a fast FBP and the proposed fast photon transport model is shown in Fig. \ref{fig:flowchart}. Usually one to three iterations are required for the scatter correction in considered cases shown later.

\tikzstyle{decision} = [diamond, draw, fill=blue!3, 
    text width=4em, text badly centered, node distance=3cm, inner sep=0pt]
\tikzstyle{block} = [rectangle, draw, fill=blue!3, 
    text width=6.5em, text centered, rounded corners, minimum height=4em]
\tikzstyle{line} = [draw, -latex']
\tikzstyle{cloud} = [draw, ellipse,fill=red!3, node distance=3cm,
    minimum height=2em]
\begin{figure}[ht!]
\centering 
\begin{tikzpicture}[node distance = 2cm, auto]
    \node [block] (init) {Scanner Corrupted Projs.};
    \node [block, below of=init] (Convert) {Convert to $a$ };
    \node [block, below of=Convert] (Reconstruction) {Reconstruction FBP};
    \node [block, right of=Reconstruction, node distance=3cm] (Otsu) {OTSU Segmentation};
    \node [block, above of=Otsu] (VToD) {Convert Volume to Density};
    \node [block, above of=VToD] (GPU) {GPU MC Simulation ($Ip$ and $Is$)};
    \node [block, right of=GPU, node distance=3cm] (SmoothAndInterp) {Smoothing and Interpolation};
    \node [block, below of=SmoothAndInterp] (Correction) {$c=a-ln\Big(\frac{Ip}{Ip+Is}\Big)$ };
    \path [line,dashed] (init) -- (Convert);
    \path [line,dashed] (Convert) -- (Reconstruction);
    \path [line] (Reconstruction) -- (Otsu);
    \path [line] (Otsu) -- (VToD);
    \path [line] (VToD) -- (GPU);
    \path [line] (GPU) -- (SmoothAndInterp);
    \path [line] (SmoothAndInterp) -- (Correction);
    \path[line] (Convert.west)  -| (-1.35,-2cm) -| ([xshift=-2.61cm,yshift=-1.3cm]Reconstruction.east) -- ([xshift=-2.6cm,yshift=-3.3cm]Correction.east) -|  (Correction.south);
    
    \path[line] (Correction.south)+(-0.5,0.02)  |- ++(-4,-2.3)  -|(Reconstruction.south);
\end{tikzpicture}
\caption{Flowchart of the iterative scatter correction algorithm. The dotted line is a onetime execution.}
\label{fig:flowchart}
\end{figure}
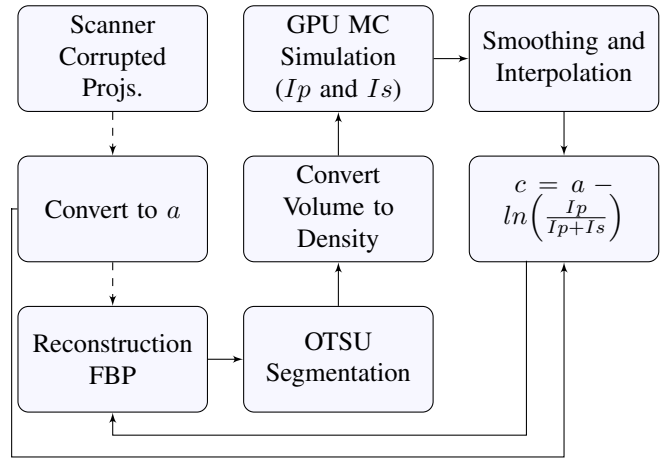

  The iterative scatter correction algorithm starts by importing the scatter-corrupted raw intensity projections from the scanner, these projections are of (2304,3200) pixels resolution. The intensity projections are then converted into linear attenuation projections to prepare them for the reconstruction algorithm. A reconstruction using the FBP method is then performed to get a volume using these projections. The reconstructed volume is down-sampled to (791,576,791) voxels to fit the GPU. The different materials in the down-sampled reconstructed volume are segmented using the Otsu segmentation method \cite{Otsu}. After performing the segmentation, the different materials within the volume are then converted into densities. A MC simulation is then applied using the proposed MC model to acquire the scatter and the primary intensity projections using the voxelized volume. These projections are then used in (\ref{eqn:Correction}) to perform the scatter correction in the iterative algorithm.

\begin{align}
\label{eqn:Correction}
c=a-ln\Big(\frac{Ip}{Ip+Is}\Big), \ 
\end{align} 
 
where $c$ is the corrected projection, $a$ is the scatter-corrupted linear attenuation projection from the scanner, and $Is$, $Ip$ are the scatter and the primary intensity projections taken from the proposed MC model respectively.

  
 As mentioned before, this work aims to correct high-resolution projection from the scanner. Two approaches can be followed in the iterative scatter correction algorithm. The first approach is to calculate the scatter and the primary projections using the proposed MC model with the same resolution of (2304,3200) from the scanner. These projections should be denoised by the smoothing filter before using them in the correction step. This approach is time consuming, as performing the MC simulation on a high-resolution detector requires the use of a high number of photons to get a low noise projection. The second approach is to calculate the scatter and the primary projections using lower resolution i.e., (576,800) with less number of photons. These projections are then up-sampled to the high-resolution case using cubic interpolation to prepare them for the correction step. 

  In CT, a high number of projections should be used to get a proper reconstruction quality using the FBP method. Simulating the scatter for all these projections is timely expensive. Since the scatter normally tends to change slowly between projections \cite{Ning}, only half of the projections is simulated and the other half is acquired using linear interpolation between every two projections. In the subsection \ref{interpolation effect}, it is shown that the use of the intra-and-inter-projections only shows a minor effect on the accuracy of the correction.


  
 The primary and the scatter intensities projections, prepared in the previous step, are used in the initial correction of the projections from the scanner using (\ref{eqn:Correction}). This initial correction lacks accuracy since the estimation of the scatter and the primary intensities in this step has been done using a scatter-corrupted volume.
  
 The correction process should continue iteratively by performing a FBP again on the partially corrected projections from the initial step to reconstruct a better volume. This partially corrected volume is used in the MC simulation to derive an improved estimation of the scatter and the primary projections. The latter are imported again after performing the smoothing and interpolation steps together with the original corrupted projections into (\ref{eqn:Correction}) to perform the correction again. The overall correction process should be repeated until an adequate correction is derived. Different parameters of the MC setting were tested for the sake of optimizing the time. The total number of the photons, the splitting number of the photons during the scoring of the scatter, and the scoring step size of the ray tracer through the volume. It is shown that certain settings of these parameters minimize the MC simulation time to the lowest extend while maintaining the accuracy of the scatter correction. This is one of the main differences for the iterative scatter correction algorithm given in Fig. \ref{fig:flowchart} compared to previous works \cite {Watson}, \cite{Hing}. Further differences are the low resolution, low number of photons, low number of projections and the CPU instead of GPU based scattering corrections. A study of the effect of these parameters on the accuracy of the scatter correction is presented in subsection \ref{optimization effect}.

\subsection{Quantitative Evaluation}
\label{Quantitative Evaluation}
 In order to quantitatively evaluate the effect of the interpolation technique and the optimization of the key parameters of the proposed MC model on the scatter correction quality, the mean square error (MSE), the normalized cross-correlation (NCC), and the contrast-to-noise ratio (CNR) are used. The CNR, which is calculated using (\ref{eqn:CNR}), is evaluated in a region-of-interest (ROI) of $14\times14$ pixels (shown in Fig.~\ref{fig:Effect of Interpolation}\subref{fig:a}) within the target images. The MSE is calculated by equation (\ref{eqn:MSE}). While the NCC is given by (\ref{eqn:NCC}).

\begin{align}
MSE=\mathop{\mathbb{E}}\Big[\parallel\chi_{1}-\chi_{2}\parallel_2^2\Big],\label{eqn:MSE}
\end{align}

Where $\chi_{1}$ and $\chi_{2}$ represent the first and the second images which are used in the evaluation.

\begin{align}
\label{eqn:NCC}
NCC=\frac{\mathop{\mathbb{E}}((\chi_{1}-\mu_{1})(\chi_{2}-\mu_{2})) }{\sigma_1\sigma_2},\ 
\end{align}

$\mu_1$ and $\mu_2$ are the mean values of image 1 and image 2 respectively, while $\sigma_1$ and $\sigma_2$ are the standard deviation of image 1 and image 2 respectively.


\begin{align}
\label{eqn:CNR}
CNR=\frac{\lvert\mu_{ROI}-\mu_{Background}\lvert }{\sigma_{Background}},\ 
\end{align}

where $\mu_{ROI}$ and $\mu_{Background}$  are the mean values of the ROI and the background respectively, while $\sigma_{Background}$ is the standard deviation of the background.

\section{Experiments and Results}
\label{Results}
In this section, the proposed MC model is evaluated on the real-world data set and compared with other simulators. In addition, results of the scatter correction algorithm and comparison of these results with and without the use of different acceleration and optimization techniques are also shown.

\subsection{Collimator Scatter-Suppressed Results}
As mentioned before, cone beam CT suffers much from the scatter effect due to the large X-ray field of view. Reducing this field of view results in much less scatter. To produce a scatter-suppressed result that could be served as ground truth in this work, the cone beam X-rays from the scanner has been converted into a fan beam through the use of a simple collimator. Two copper blocks of $2\,cm\times2\,cm\times4\,cm$ dimension have been used as a simple collimator. These blocks were placed in front of the X-ray source in which the long side was positioned perpendicularly with the exit window of the X-ray source. Thus the X-ray was passing only through the slit between the two blocks, and the rest of the photons outside this slit is heavily attenuated. As a result, the original cone beam is converted into a fan beam and the scanned slit with this fan beam is used as ground truth in this work.

\subsection{Verification of the Proposed MC Model}

Several methods were used to verify the proposed model. A comparison of the primary and the scatter projections from the proposed MC model with the same projections from the aRTist simulator has been conducted. The proposed MC model was used to correct the scatter in a scatter-corrupted projection from the real-scanner. The scatter-corrected projection has been compared with the near scatter-free result acquired experimentally using the collimator for the same scan. Finally, the proposed MC model was also used to correct scatter-corrupted projection from the EGSnrc simulator. The result of this correction has been compared with the scatter-free projection available from this simulator.

\subsubsection{Correction of Scatter-Corrupted Projection from Scanner}

\begin{figure}[ht!]
\centering
\def\stackalignment{l}
\subfloat{\topinset{\bfseries \textcolor{white}{(a)}}{\includegraphics[width=0.5\linewidth]{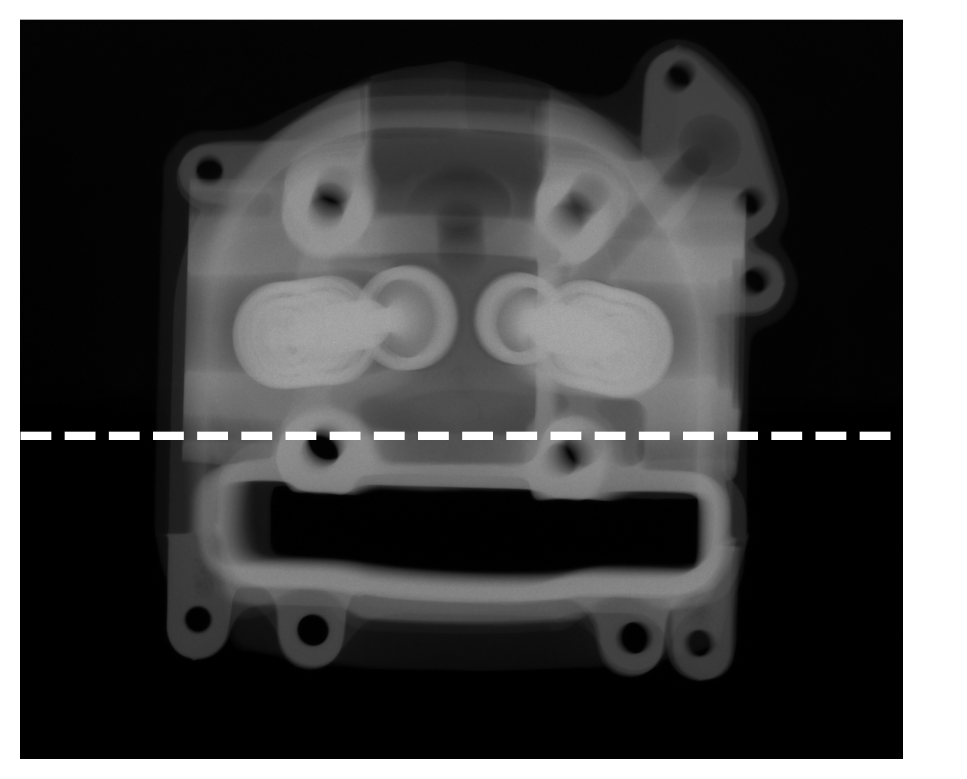}}{0.1in}{.1in}}
\subfloat{\topinset{\bfseries \textcolor{white}{(b)}}{\includegraphics[width=0.5\linewidth]{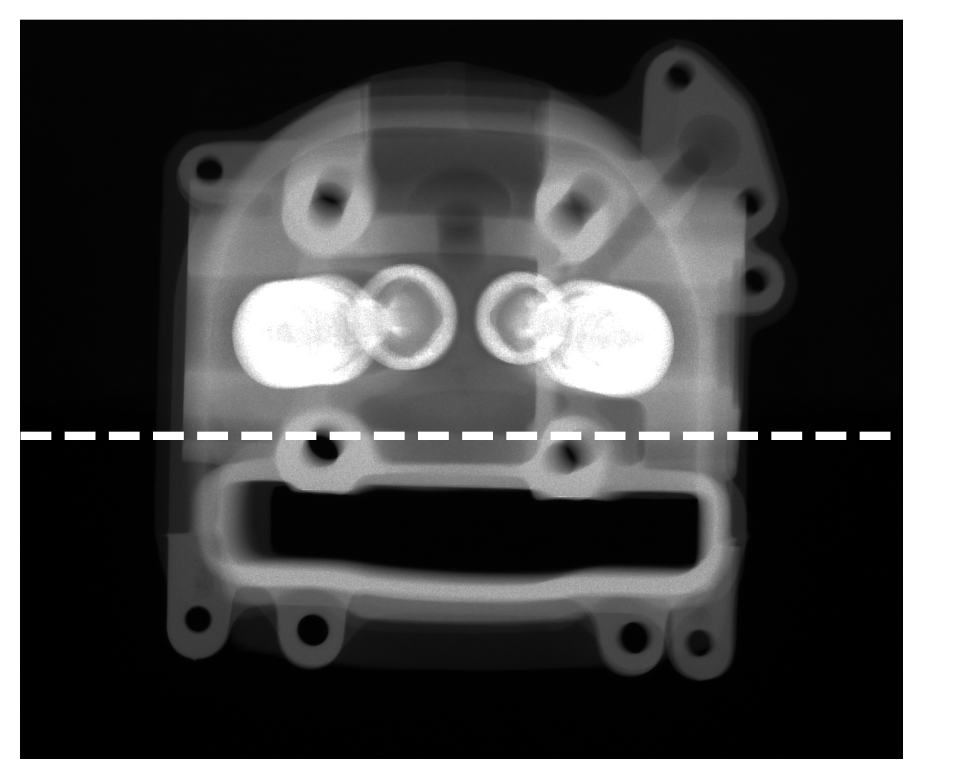}}{0.1in}{.1in}}\\[-0.15ex]
\subfloat{\topinset{\bfseries \textcolor{black}{(c)}}{\includegraphics[width=0.7\linewidth]{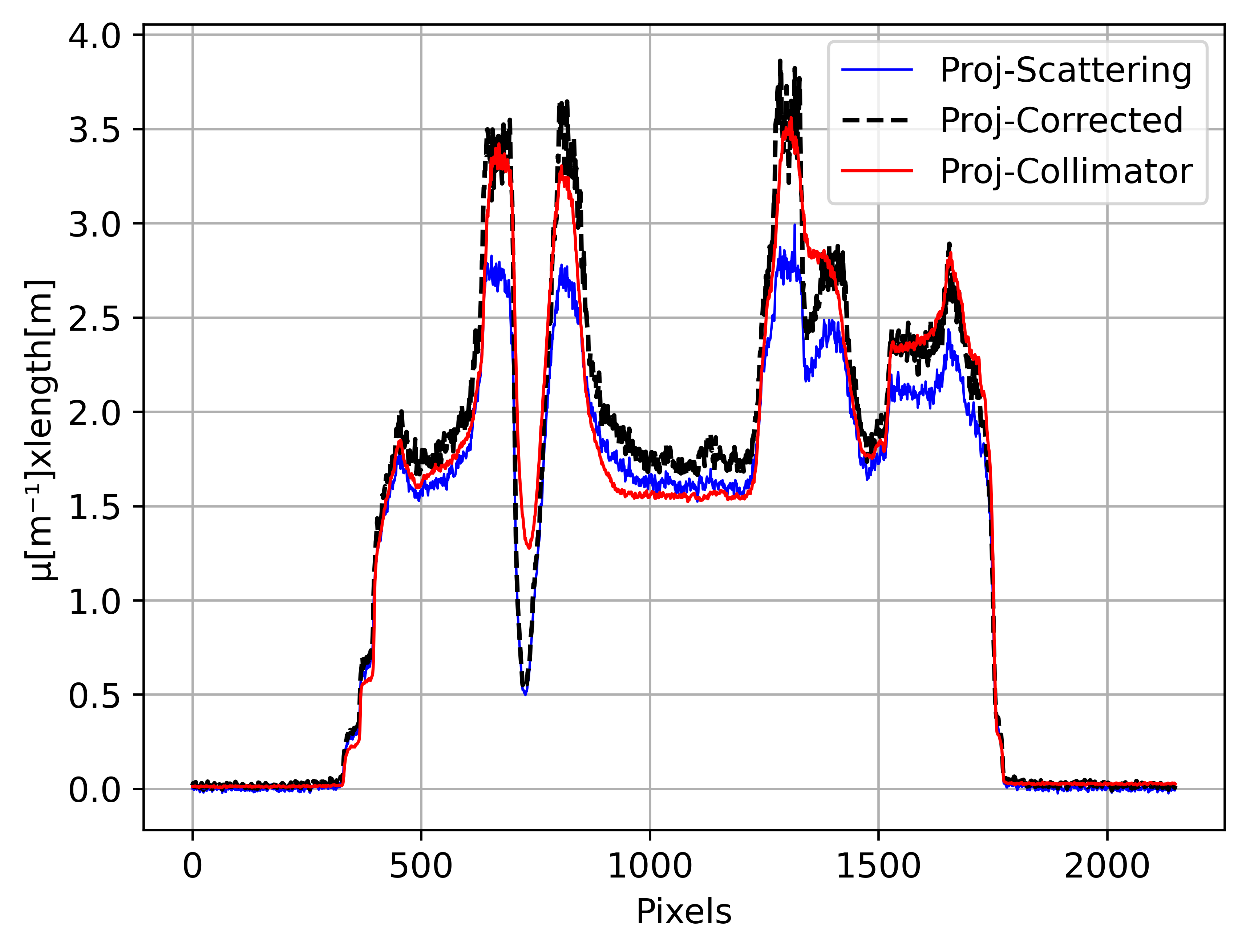}}{0.1in}{.35in}}\\[-1ex]

\caption{Correction of a scatter-corrupted projection from the real-world scanner using the proposed MC model for an aluminum motorcycle cylinder head. (a) Scatter-corrupted projection, (b) scatter-corrected projection using the proposed MC model, (c) profile lines of (a), (b) and from the near scatter-free projection acquired using a collimator. The profiles are marked by white dashed lines in (a) and (b).}
\label{fig:scatter correction of proj from scanner}
\end{figure}

Fig. \ref{fig:scatter correction of proj from scanner} shows the result of the scatter correction of a scatter-corrupted projection from the scanner. The correction was done using (\ref{eqn:Correction}) utilizing the scatter and the primary projections simulated by the proposed MC model. The result of the correction of this projection matches well with the near scatter-free result of the same scan acquired using the collimator.

\subsubsection{Comparison with MC Simulators}
\begin{figure}[ht]
\centering
\def\stackalignment{l}
\subfloat{\topinset{\bfseries \textcolor{white}{(a)}}{\includegraphics[width=0.5\linewidth]{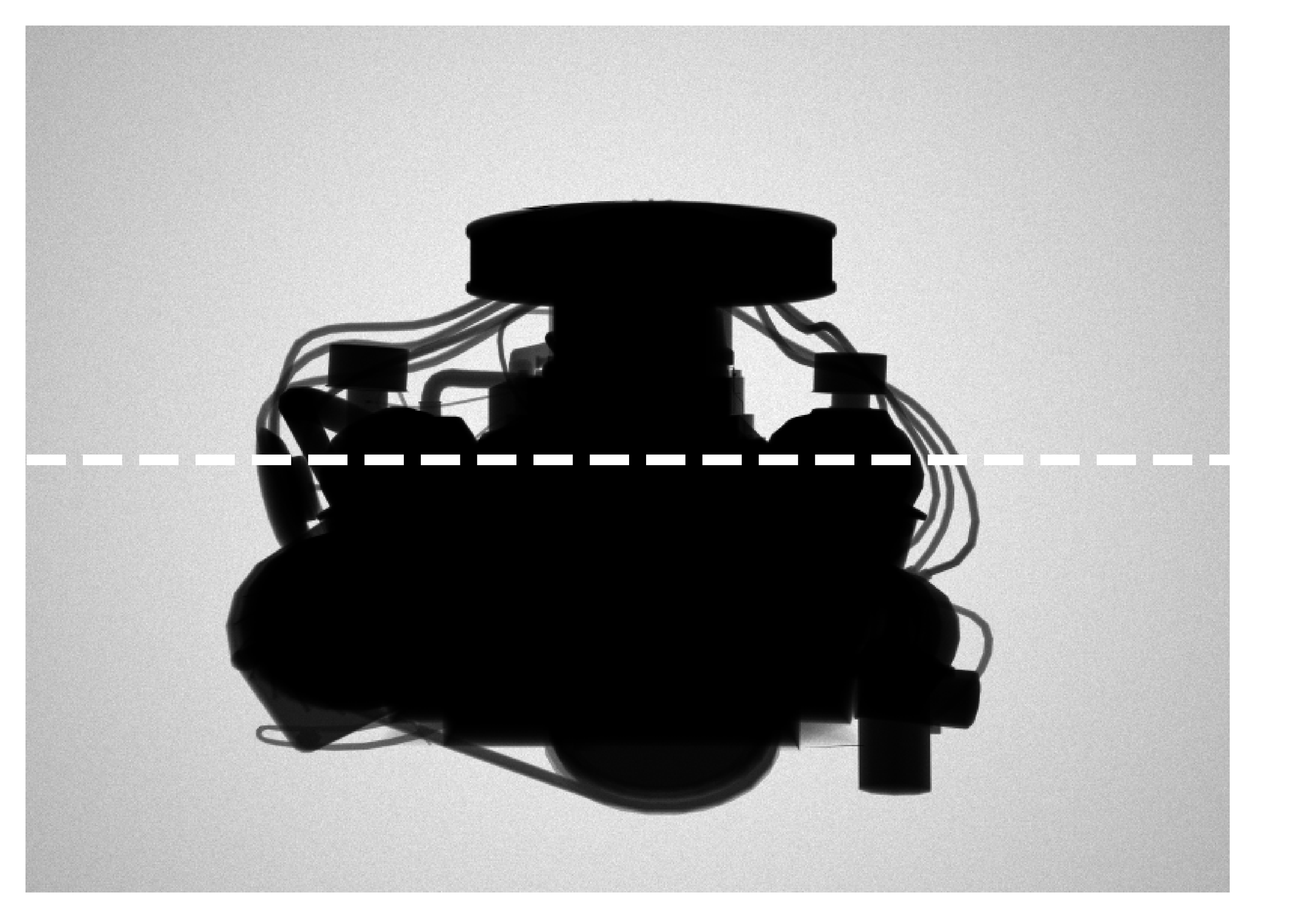}}{0.1in}{.1in}}
\subfloat{\topinset{\bfseries \textcolor{white}{(b)}}{\includegraphics[width=0.5\linewidth]{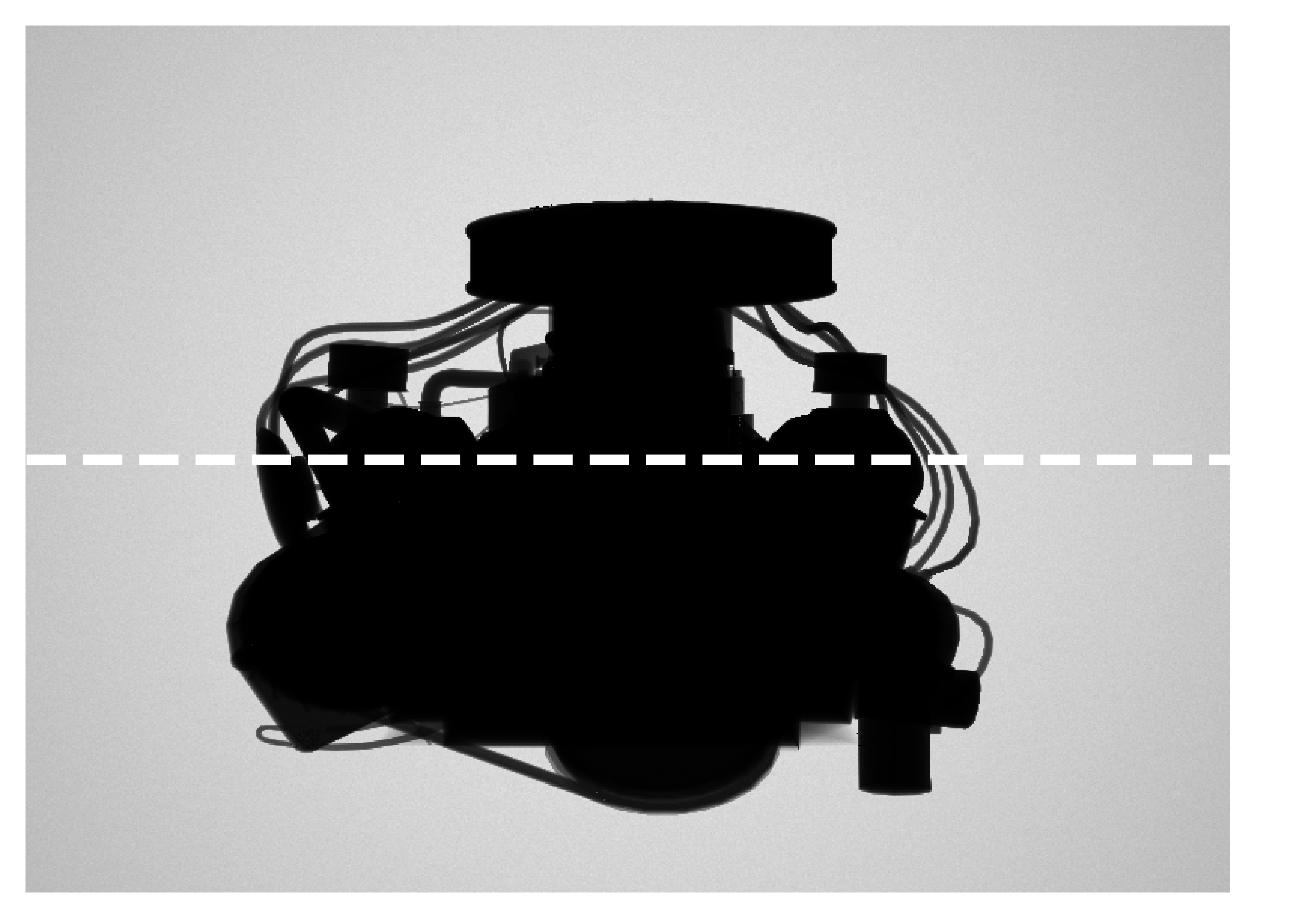}}{0.1in}{.1in}}\\[-0.15ex]
\subfloat{\topinset{\bfseries \textcolor{white}{(c)}}{\includegraphics[width=0.5\linewidth]{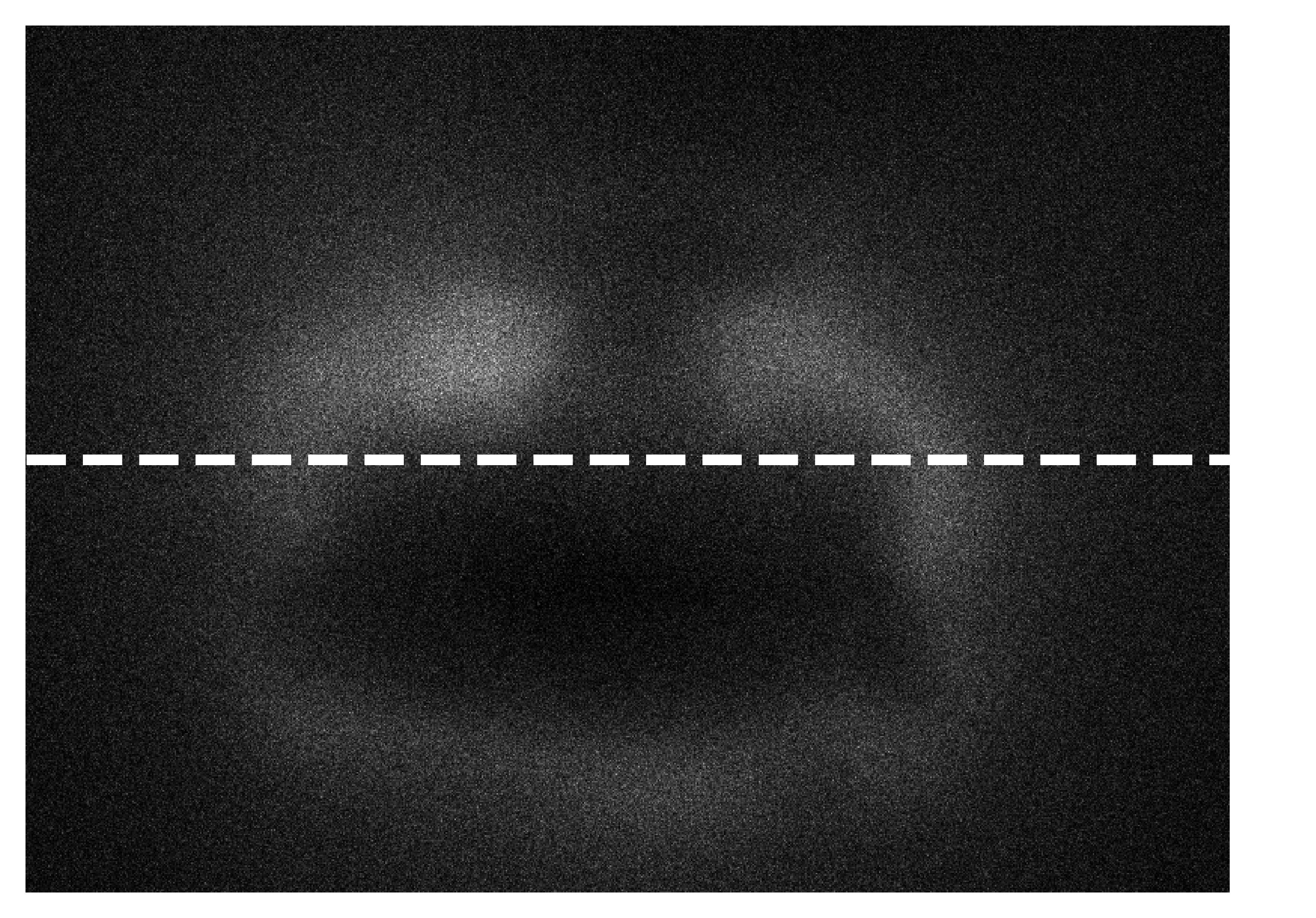}}{0.1in}{.1in}}
\subfloat{\topinset{\bfseries \textcolor{white}{(d)}}{\includegraphics[width=0.5\linewidth]{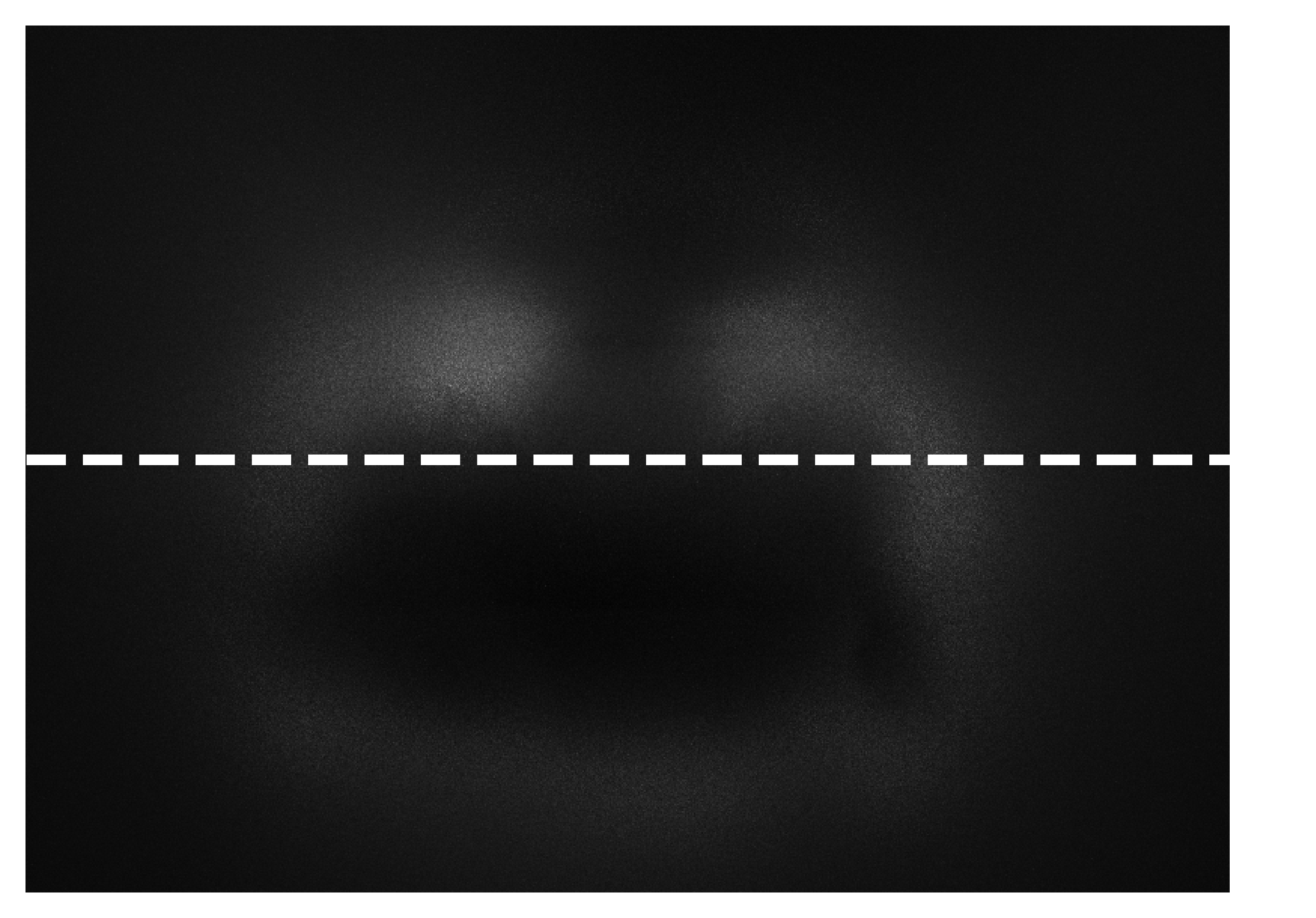}}{0.1in}{.1in}}\\[-0.05ex]
\subfloat{\topinset{\bfseries \textcolor{black}{(e)}}{\includegraphics[width=0.49\linewidth]{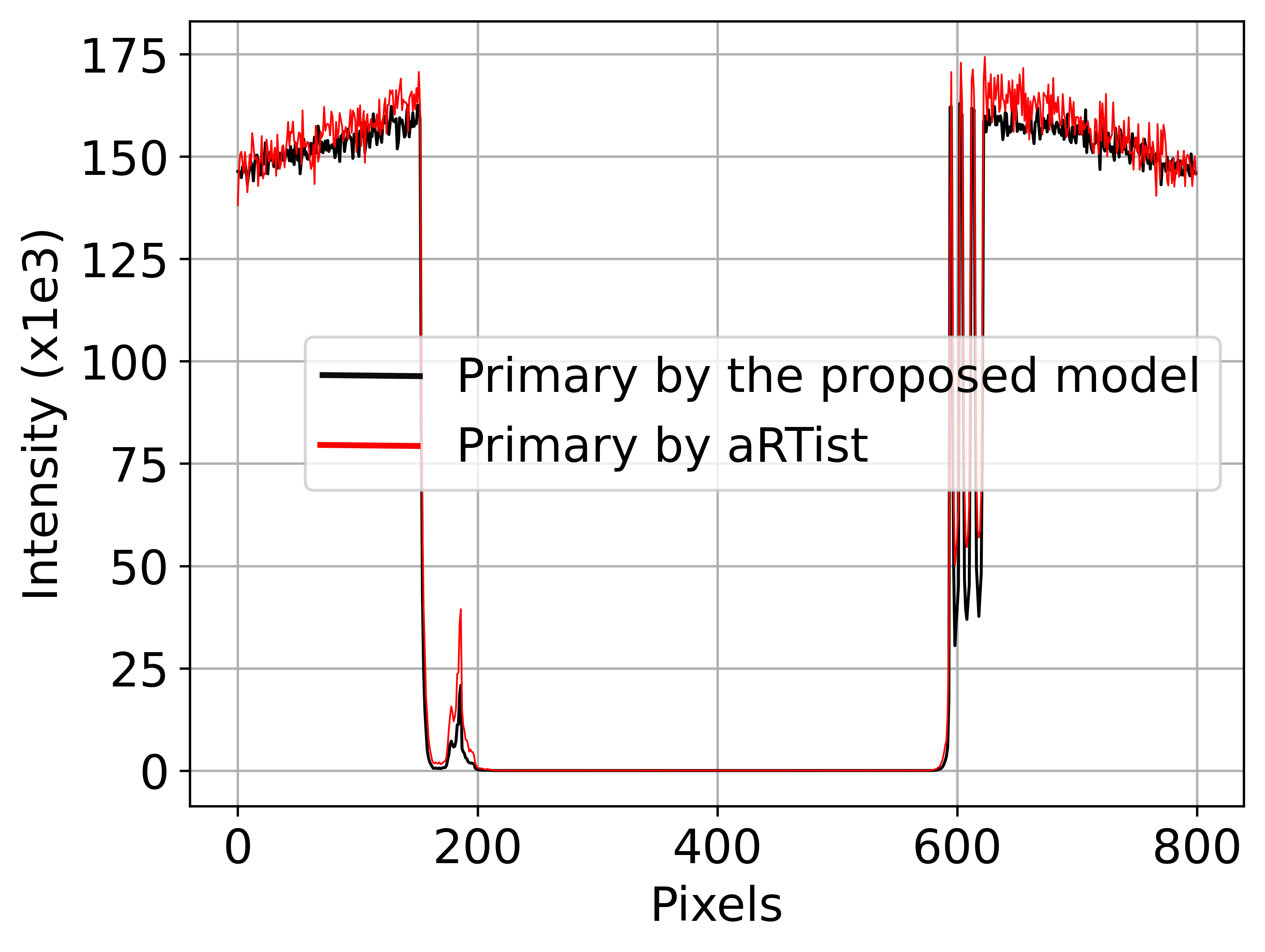}}{0.28in}{.3in}}
\subfloat{\topinset{\bfseries \textcolor{black}{(f)}}{\includegraphics[width=0.49\linewidth]{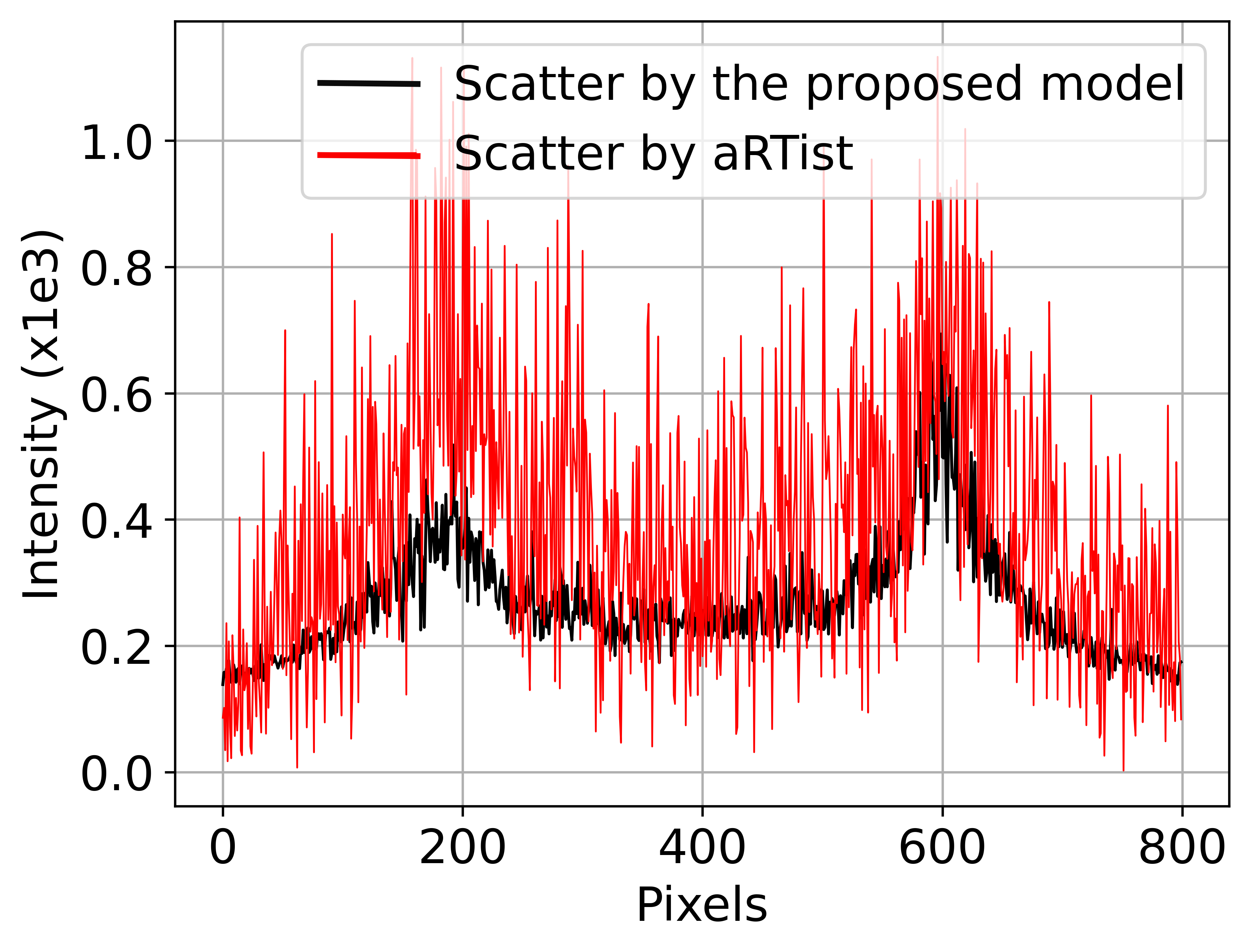}}{0.28in}{.3in}}\\[-0.1ex]

\caption{Comparison of the simulation results between the aRTist simulator and the proposed MC model for an iron engine with 200 keV and $3.9\times10^9$ photons. (a) Primary intensity from the aRTist simulator, (b) primary intensity from the proposed MC model, (c) scatter intensity from the aRTist simulator, (d) scatter intensity from the proposed MC model, (e) central profile lines of the images in (a) and (b), (f) central profile lines of the images in (c) and (d). The profiles are marked by white dashed lines in (a),(b), (c), and (d).}
\label{fig:ComparsionMustang}
\end{figure}

Fig. \ref{fig:ComparsionMustang} shows the scatter and the primary intensity projections of a car's engine simulated using the aRTist simulator and the proposed MC model. In this figure, the comparisons between the scatter and the primary projections from the two simulators show very good agreement.

\subsubsection{Correction of Scatter-Corrupted Projections from EGSnrc Simulator}
\begin{figure}[ht!]
\centering

\def\stackalignment{l}
\subfloat{\topinset{\bfseries \textcolor{white}{(a)}}{\includegraphics[width=0.5\linewidth]{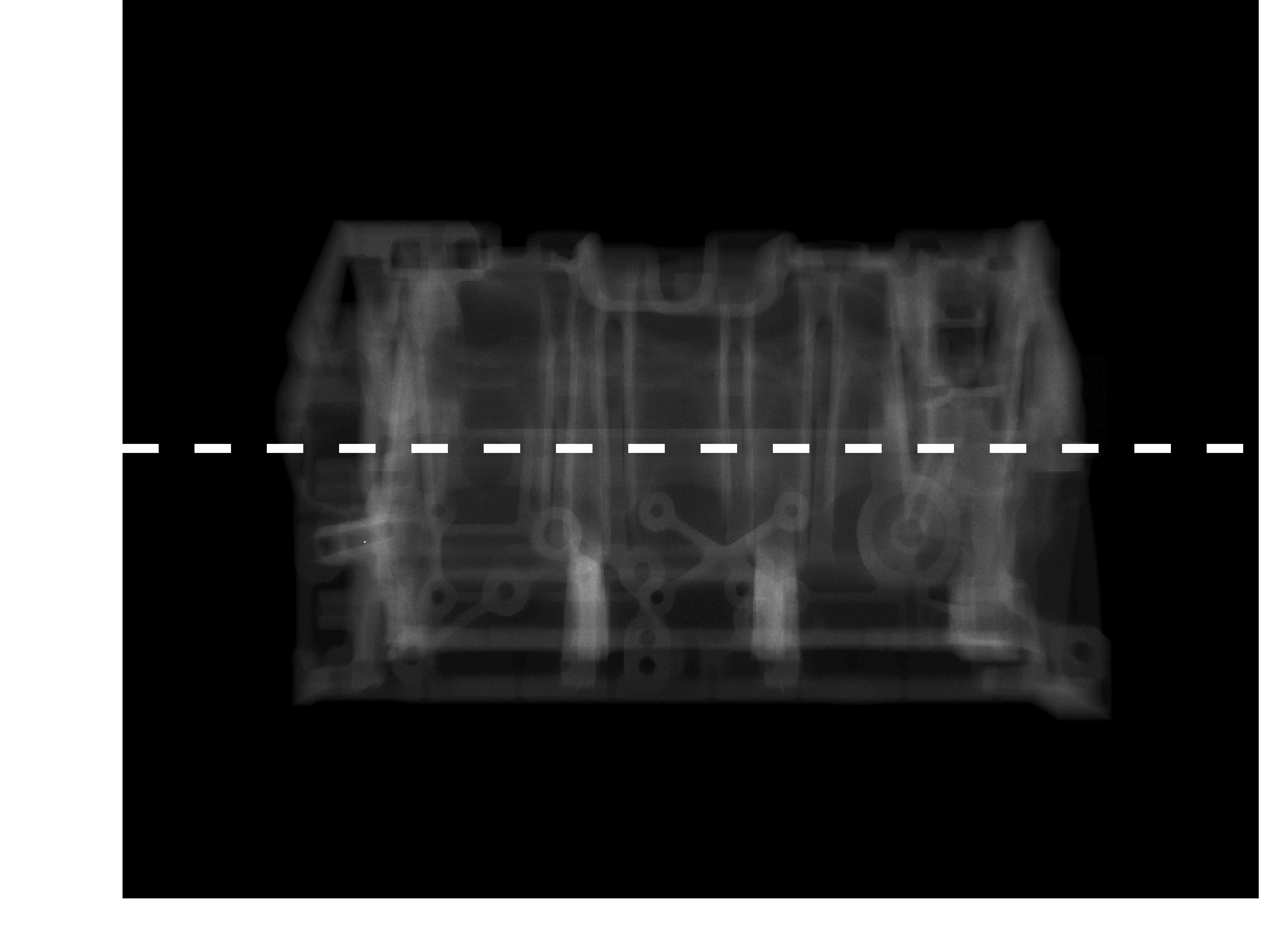}}{0.08in}{.2in}}
\subfloat{\topinset{\bfseries \textcolor{white}{(b)}}{\includegraphics[width=0.5\linewidth]{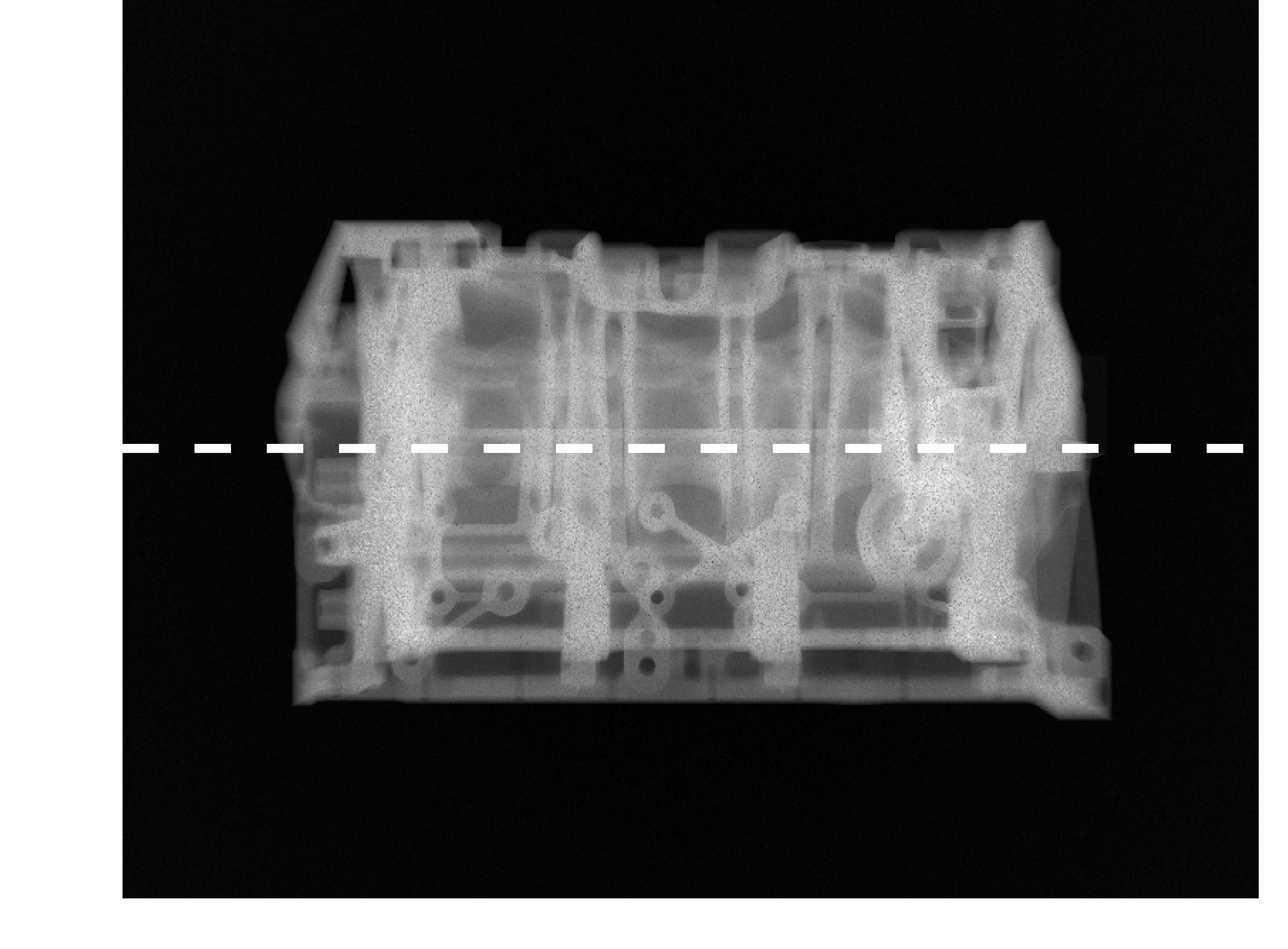}}{0.08in}{.2in}}\\[-0.15ex]
\subfloat{\topinset{\bfseries \textcolor{white}{(c)}}{\includegraphics[width=0.5\linewidth]{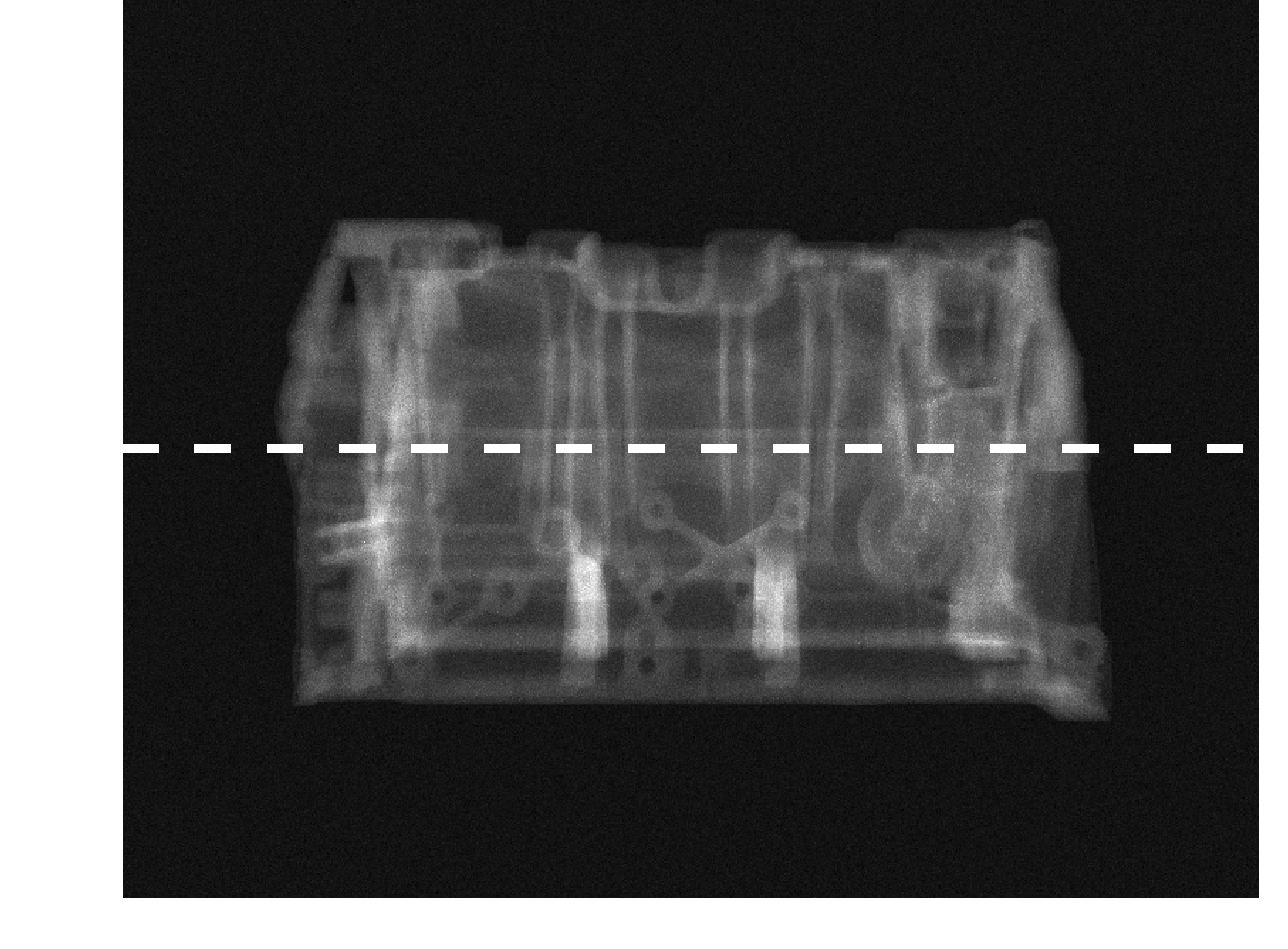}}{0.08in}{.2in}}
\subfloat{\topinset{\bfseries \textcolor{black}{(d)}}{\includegraphics[width=0.5\linewidth]{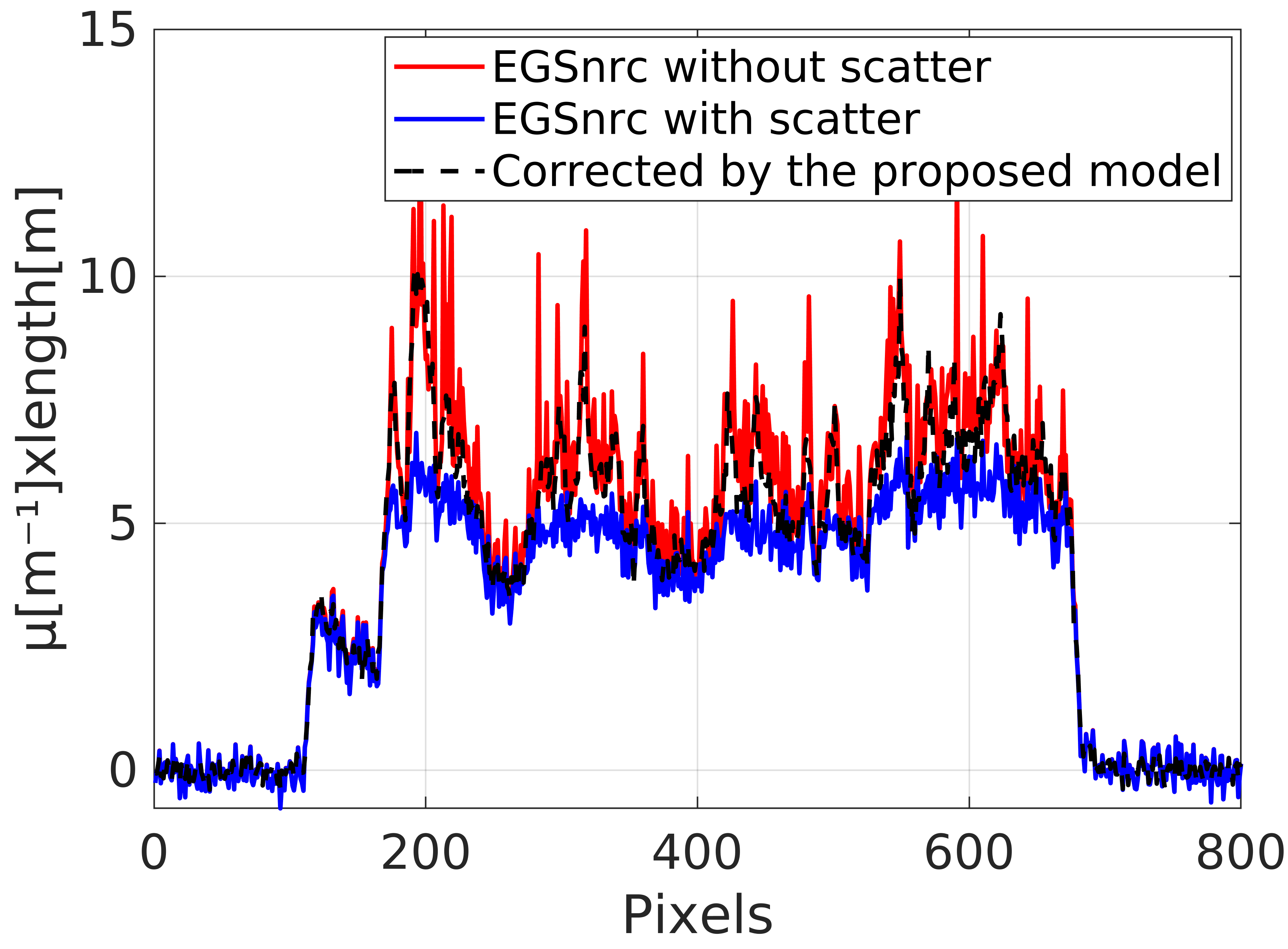}}{0.08in}{.25in}}\\[-0.1ex]

\caption{Correction of EGSnrc scatter-corrupted projection using the proposed MC model for an iron engine with 200 keV and $1\times10^9$ photons. (a) Primary linear attenuation projection from the EGSnrc simulator, (b) scatter-corrupted linear attenuation projection from the EGSnrc simulator, (c) result of the scatter correction using the proposed MC model, (d) central profile lines of the images in (a),(b), and (c). The profiles are marked by white dashed lines in (a), (b), and (c).}
\label{fig:CorrectionEGSnrc}
\end{figure}

A part of the verification process, the proposed MC model was used to correct scatter-corrupted projection from the EGSnrc simulator. The correction was done again using (\ref{eqn:Correction}). The result of the scatter correction in this case and the comparison with the scatter-free projection from the same simulator are shown in Fig. \ref{fig:CorrectionEGSnrc}.

\subsection{Scatter Correction for Motorcycle Cylinder Head} 
\begin{figure}[ht]
\centering

\def\stackalignment{l}
\subfloat{\topinset{\bfseries \textcolor{white}{(a)}}{\includegraphics[width=0.4\linewidth]{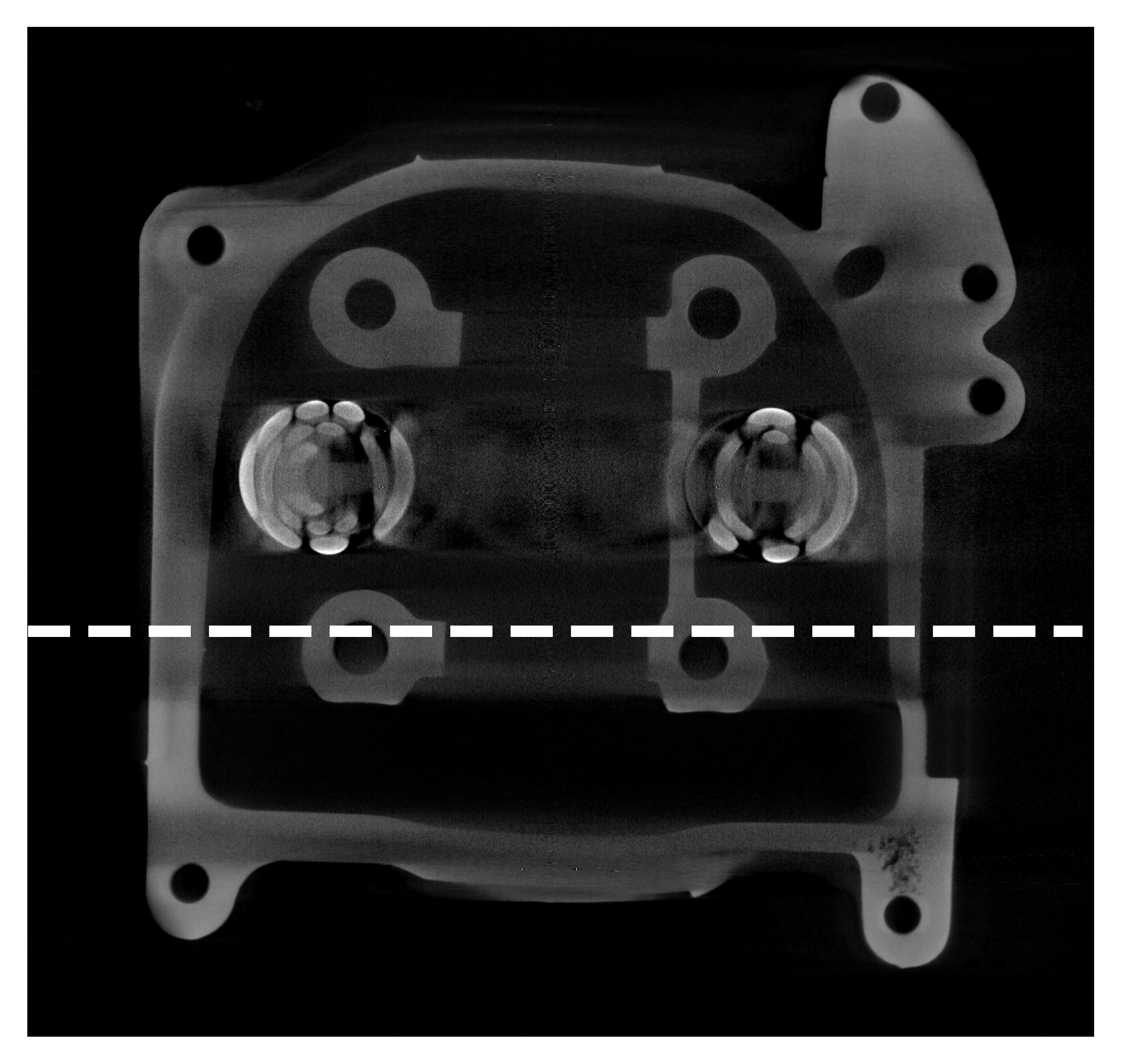}}{0.1in}{.1in}}
\subfloat{\topinset{\bfseries \textcolor{white}{(b)}}{\includegraphics[width=0.4\linewidth]{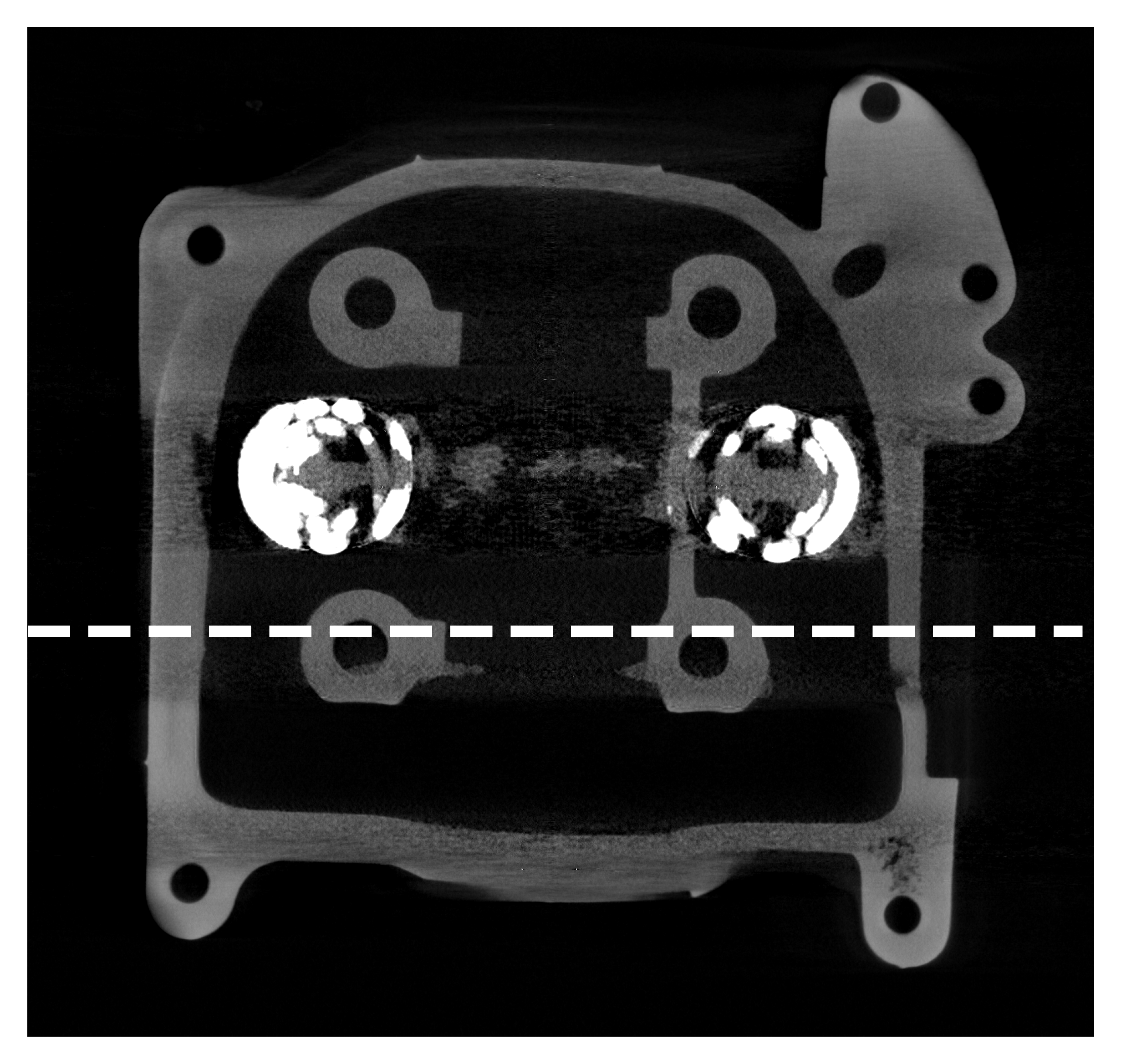}}{0.1in}{.1in}}\\[-0.15ex]
\subfloat{\topinset{\bfseries \textcolor{white}{(c)}}{\includegraphics[width=0.4\linewidth]{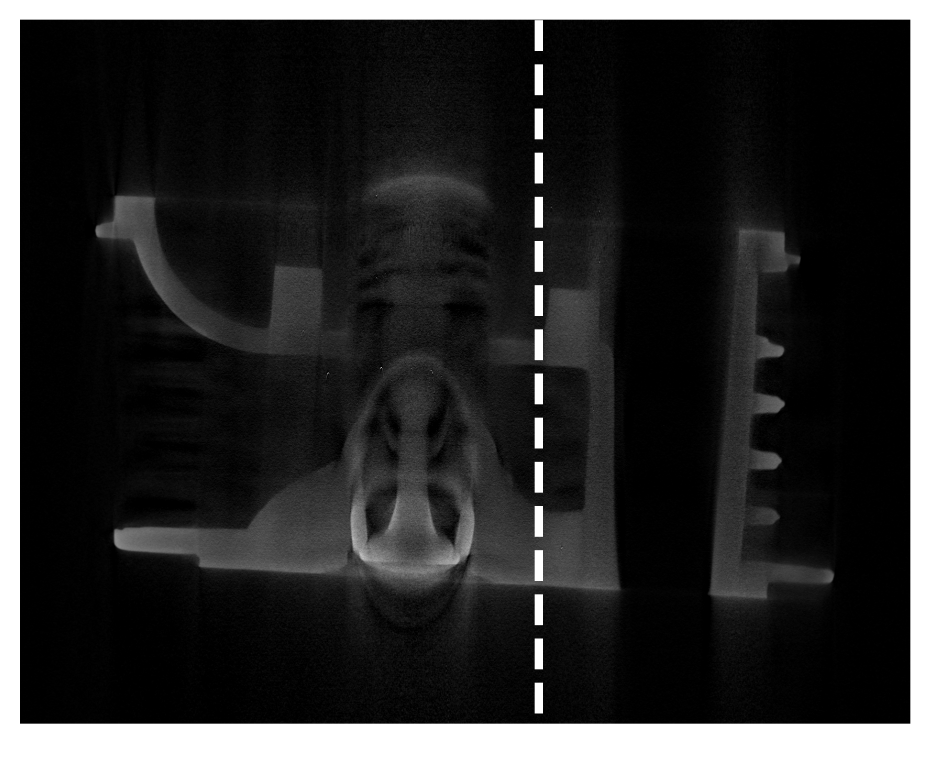}}{0.1in}{.1in}}
\subfloat{\topinset{\bfseries \textcolor{white}{(d)}}{\includegraphics[width=0.4\linewidth]{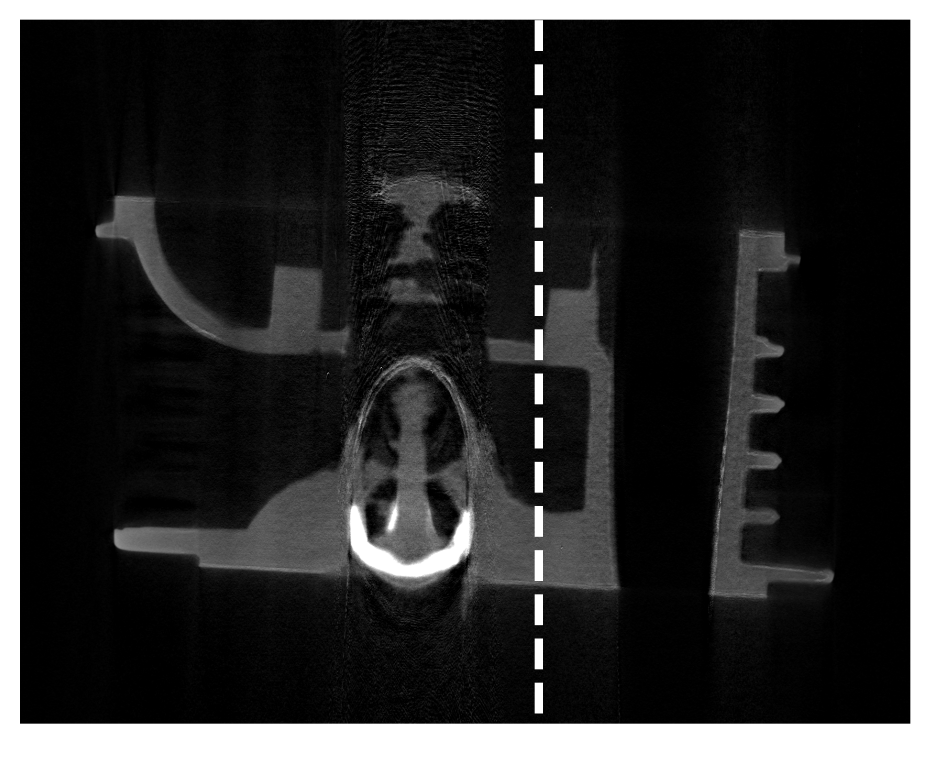}}{0.1in}{.1in}}\\[-0.15ex]
\subfloat{\topinset{\bfseries \textcolor{black}{(e)}}{\includegraphics[width=0.49\linewidth]{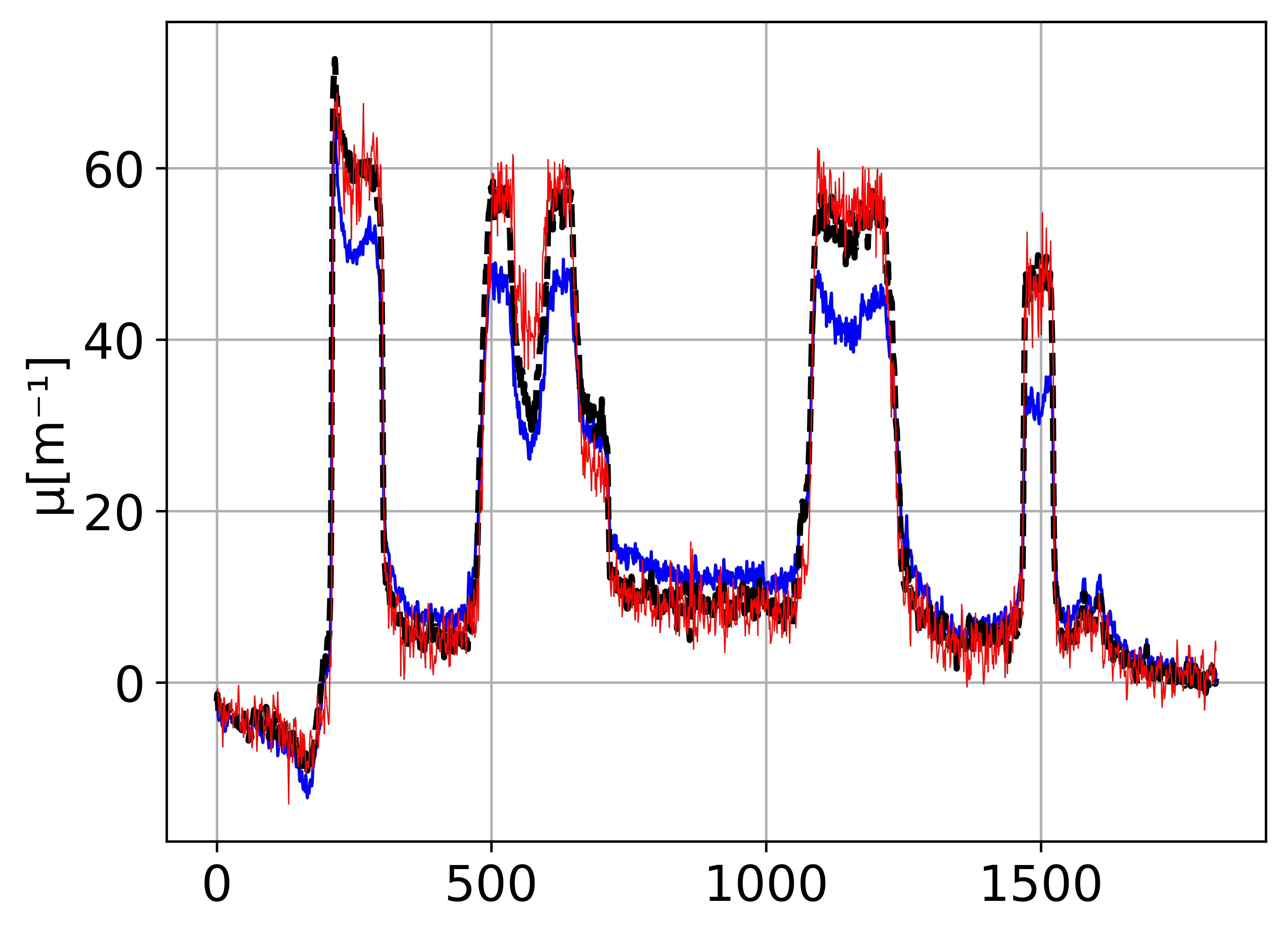}}{0.5in}{.25in}}
\subfloat{\topinset{\bfseries \textcolor{black}{(f)}}{\includegraphics[width=0.49\linewidth]{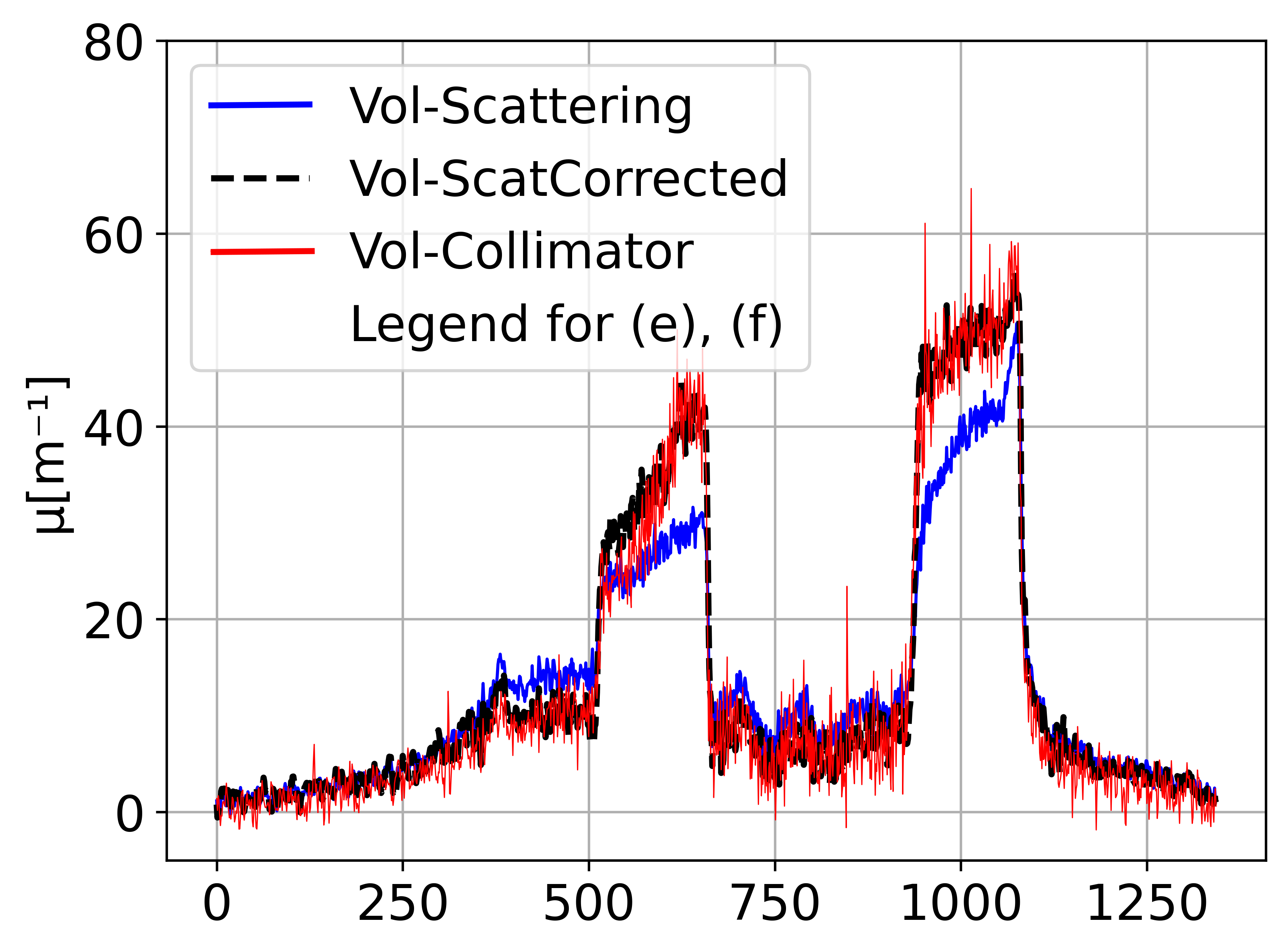}}{0.5in}{.25in}}\\[-1ex]

\caption{Results of the iterative scatter correction algorithm (with the use of the interpolation technique). (a) Slice no. 1613 of the scatter-corrupted volume (front view), (b) same slice from the corrected volume (third iteration), (c) slice no. 1450 of the scatter-corrupted volume (side view), (d) same slice from the corrected volume (third iteration), (e) profile lines of (a), (b), and the near scatter-free volume's slice from the collimator, (f) profile lines of (a), (c), and the near scatter-free volume's slice from the collimator. The profile lines in this example is averaged over multiple rows to suppress the noise and to make the comparison more clear. The profiles are marked by white dashed lines in (a), (b), (c), and (d).}
\label{fig:correction of CylinderHead}

\end{figure}

Using the proposed iterative scatter correction algorithm, three iterations were adequate to correct the scatter for an aluminum motorcycle cylinder head. Fig. \ref{fig:correction of CylinderHead} shows the results of the third iteration of the scatter correction. 1500 scatter and 3000 primary projections were simulated using the proposed MC model with $2.5\times10^8$ photons and a (576,800) resolution.

\begin{figure}[ht]
\centering

\def\stackalignment{l}
\subfloat{\topinset{\bfseries \textcolor{white}{(a)}}{\label{fig:All iterations a}\includegraphics[width=0.25\linewidth]{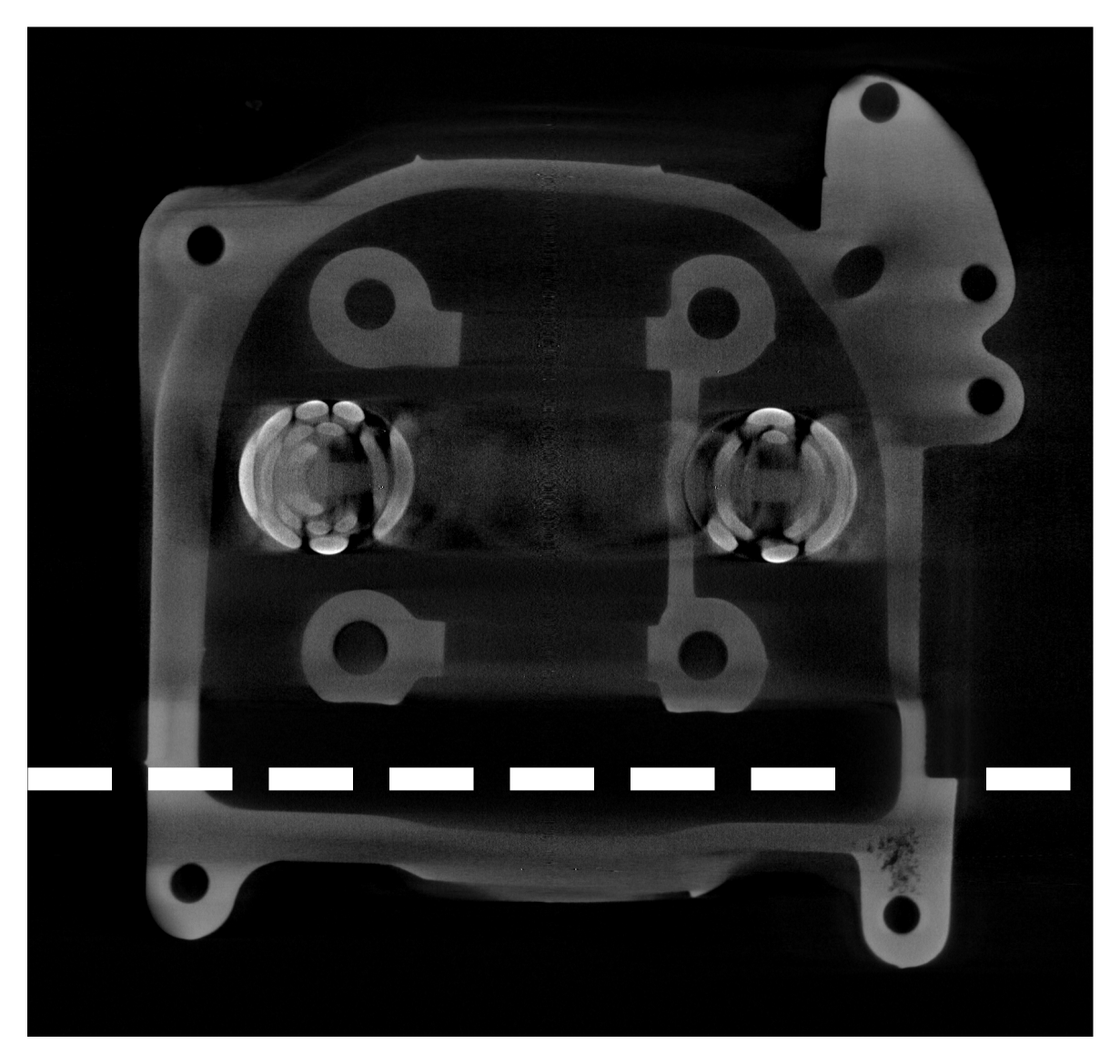}}{0.06in}{.06in}}
\subfloat{\topinset{\bfseries \textcolor{white}{(b)}}{\label{fig:All iterations b}\includegraphics[width=0.25\linewidth]{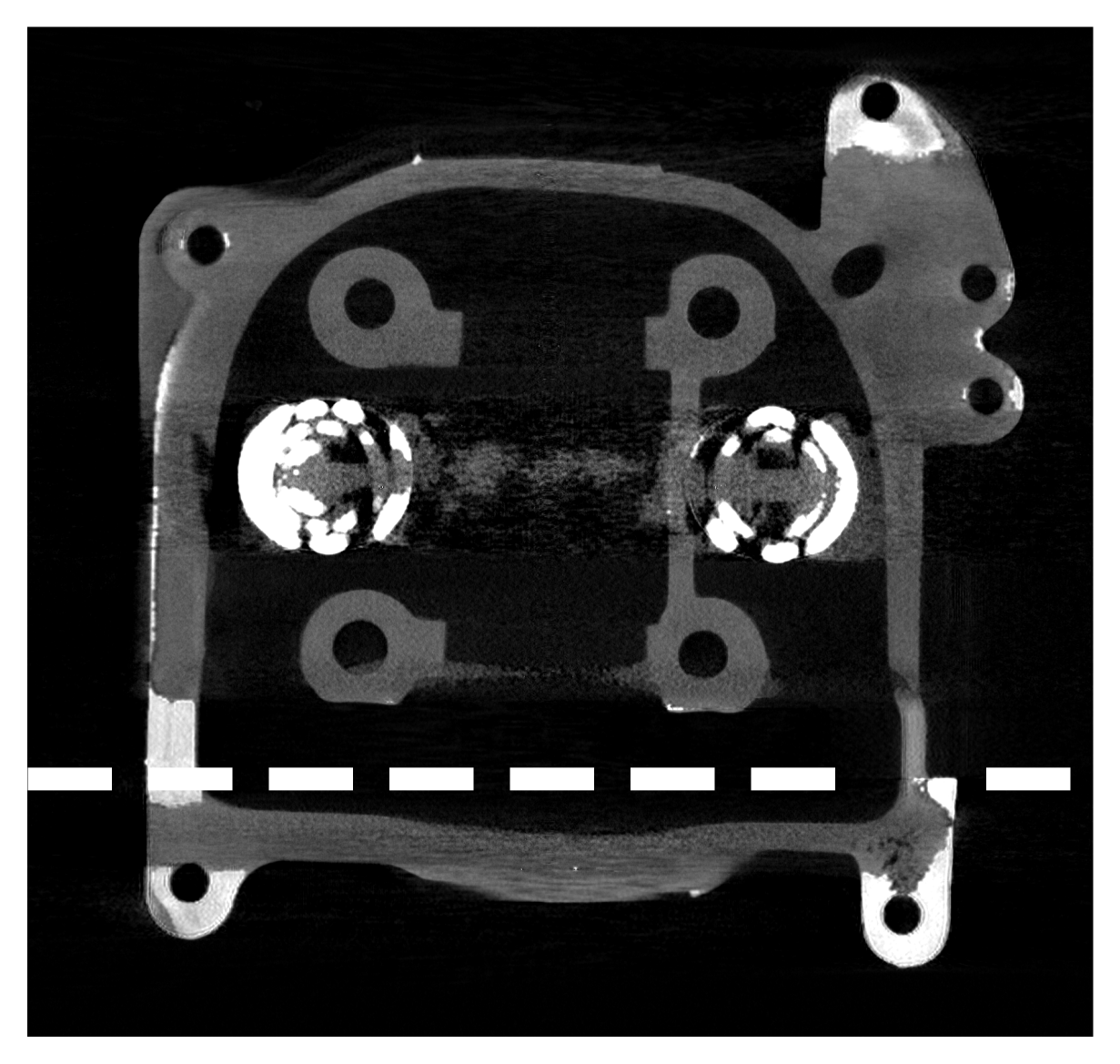}}{0.06in}{.06in}}
\subfloat{\topinset{\bfseries \textcolor{white}{(c)}}{\label{fig:All iterations c}\includegraphics[width=0.25\linewidth]{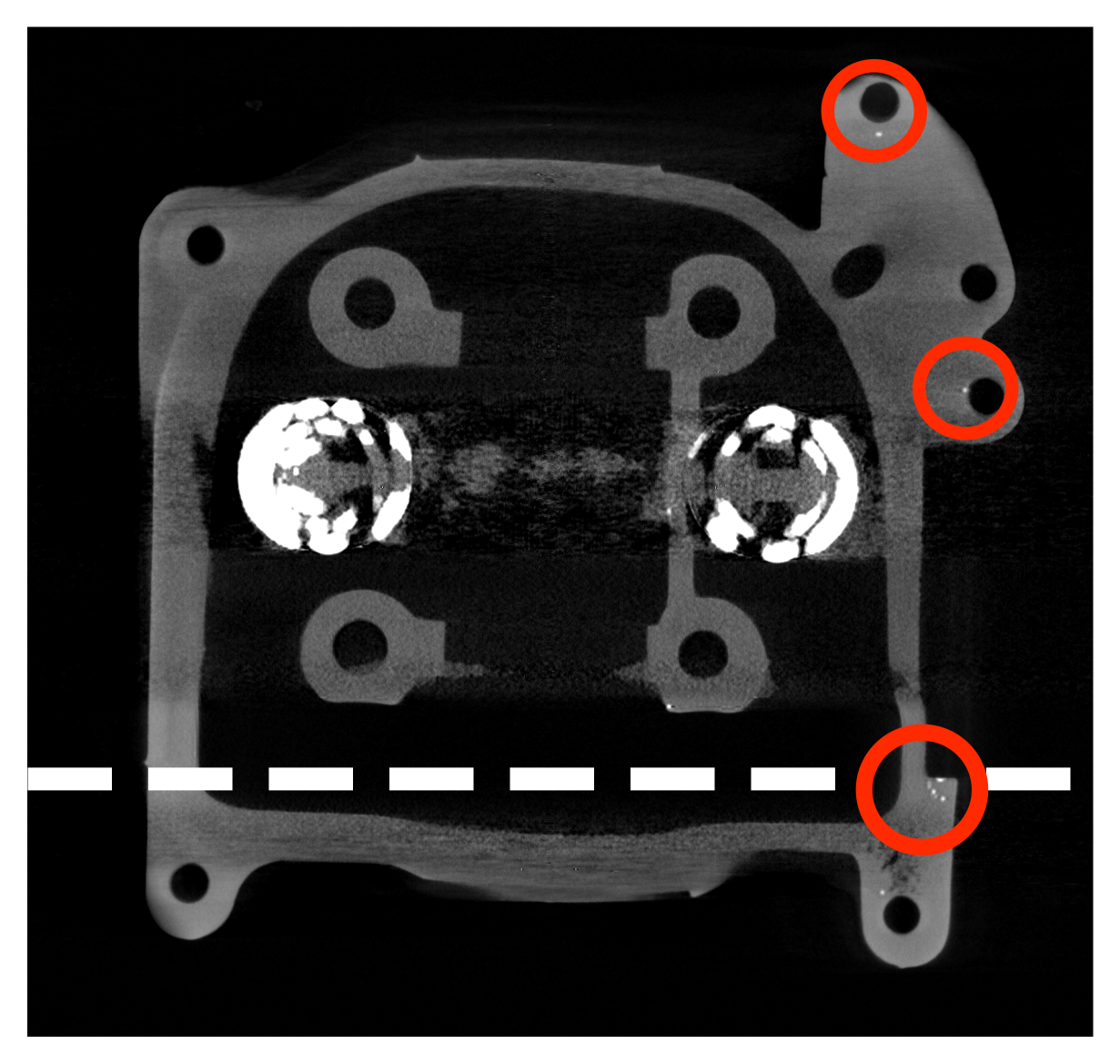}}{0.06in}{.06in}}
\subfloat{\topinset{\bfseries \textcolor{white}{(d)}}{\label{fig:All iterations d}\includegraphics[width=0.25\linewidth]{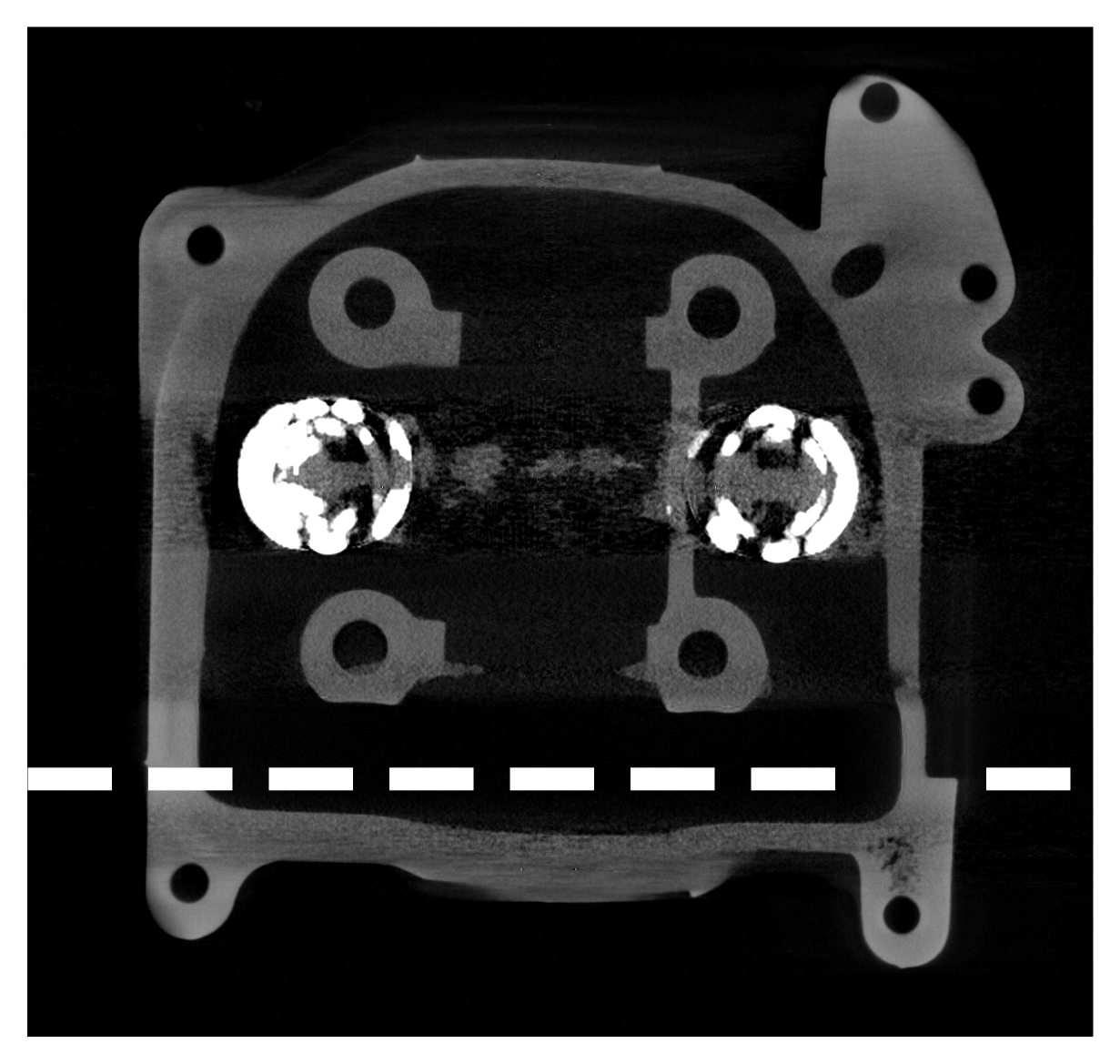}}{0.06in}{.06in}}\\[-0.15ex]
\subfloat{\topinset{\bfseries \textcolor{black}{(e)}}{\includegraphics[width=0.7\linewidth]{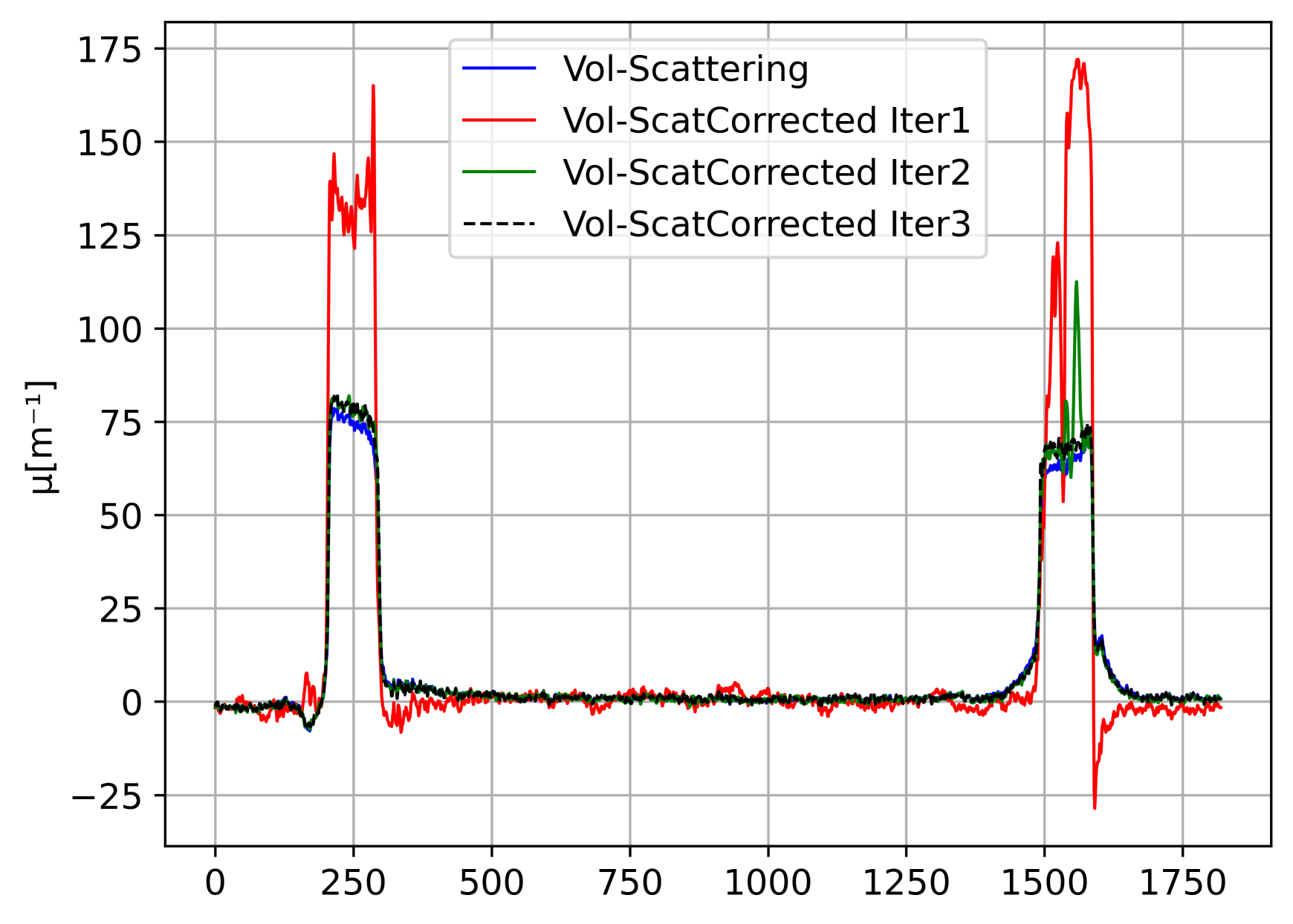}}{0.1in}{.35in}}\\[-1ex]

\caption{Results of the three iterations of the iterative scatter correction algorithm (with the use of the interpolation technique). (a) Slice no. 1613 of the scatter-corrupted volume (front view), (b) slice from the corrected volume in the first iteration, (c) slice from the corrected volume in the second iteration (the red circles mark the regions which are still considered wrongly as steel while they are aluminum), (d) slice from the corrected volume third iteration, (e) profile lines of (a), (b), (c), and (d). The profile lines in this example are averaged over multiple rows to suppress the noise and to make the comparison more clear. The profiles are marked by white dashed lines in (a), (b), (c), and (d). The result from the collimator is not shown here.}

\label{fig:All iterations}

\end{figure}

 The scatter projections were linearly interpolated to 3000 projections and then up-sampled along with the primary projections to the (2304,3200) resolution using cubic interpolation to acquire the same resolution and number of projections from the scanner. As shown in the figure, the results of the scatter correction using the iterative algorithm coincide with the collimator results. The scatter correction results in each one of the three iterations are shown in Fig. \ref{fig:All iterations} (the collimator result is not shown in this figure). Several regions of the aluminum part were considered wrongly as steel in the initial segmented scatter-corrupted volume from the real-scanner. Since the springs, which are formed from steel in this object, were heavily affected by the scatter in a way that the linear attenuation values of these springs were dropped down to be equal to some of the aluminum parts of this object (since the aluminum is a lighter material than the steel, it is less pronouns to the scatter effect i.e., the two materials are not equally attenuated). Fig.~\ref{fig:All iterations}\subref{fig:All iterations b} shows that the result of the first iteration of the scatter correction, which is based on the wrong initial segmentation, in this iteration the scatter correction algorithm produces wrong linear attenuation values of the aluminum material (the correct values should be around 60 $m^{-1}$ for this material according to the collimator result shown in Fig. \ref{fig:correction of CylinderHead}). Better thresholds were derived by the Otsu method in the second iteration, because the segmentation was based on the partially corrected volume of the first iteration. As the steel and the aluminum parts in this iteration retrieve a big portion of their values. This improved segmentation produces an enhanced scatter correction result in the second iteration as shown in Fig.~\ref{fig:All iterations}\subref{fig:All iterations c}. However, some very small parts of the corrected volume in this iteration were still wrongly assumed as steel while they are aluminum (these regions are marked by red circles in Fig.~\ref{fig:All iterations}\subref{fig:All iterations c}). In the last iteration, the aluminum and the steel parts were correctly corrected as shown in Fig.~\ref{fig:All iterations}\subref{fig:All iterations d}. The setting of the scan parameters used in this example is shown in table \ref{tab:scan parameters}. 
 
\begin{table}[ht]
\captionsetup{justification=centering}
\centering
\caption{MEASUREMENTS PARAMETERS FOR THE MOTORCYCLE CYLINDER HEAD SCAN.}
\begin{tabular}{ | C{2cm} | C{2.5cm}| } 
\hline
Parameter & Value\\ 
\hline
SDD & 1.282 m \\ 
\hline
SOD & 0.862 m\\ 
\hline
Resolution  & (2304,3200) pixels\\
\hline
Pixel size & 0.127 $\mu$m\\
\hline
X-ray voltage  & 200 kV\\
\hline
X-ray current & 36 $\mu$A\\
\hline
Exposure time & 1 s\\
\hline
\end{tabular}
\label{tab:scan parameters}
\end{table}

\begin{table}[ht]
\captionsetup{justification=centering}
\centering
\caption{TIME REQUIRED FOR THREE ITERATIONS OF SCATTER CORRECTION USING FOUR GPUs WITH AND WITHOUT THE USE OF THE INTERPOLATION TECHNIQUE.”}
\begin{threeparttable}[t]
\begin{tabular}{ | C{1.5cm}| C{1.5cm} | C{2cm}| C{2cm} | } 
\hline
 Method & No. of photons &Iterative correction & MC simulation (one iteration)\tnote{(1)}  \\ 
\hline
With interpolation& $2.5\times10^8$  & 39240 s &  8110 s  \\ 
\hline
Without interpolation & $7\times10^8$ \tnote{(2)} & 175405 s &  50000 s    \\ 
\hline
\end{tabular}
\begin{tablenotes}\footnotesize
  \item [(1)] With interpolation: 1500 and 3000 scatter and primary projections respectively of size (576,800); Without using interpolation: 3000 scatter and primary projections of size (2304,3200).
  \item [(2)] As in this case, 4$\times$ the resolution of (576,800) is used, the number of photons utilized is $\sim$3$\times$ the number of photons in case of using interpolation.
\end{tablenotes}
\end{threeparttable}%
\label{tab:time4}
\end{table}

The time required for the scatter correction of this object using three iterations in case of with and without the use of the interpolation technique is shown in table \ref{tab:time4}.

\subsection{Scatter Correction of Cement Objects with Steel Rods} 

\begin{figure}[ht]
\centering

\def\stackalignment{l}
\subfloat{\topinset{\bfseries \textcolor{white}{(a)}}{\label{fig:object Alone a}\includegraphics[width=0.32\linewidth]{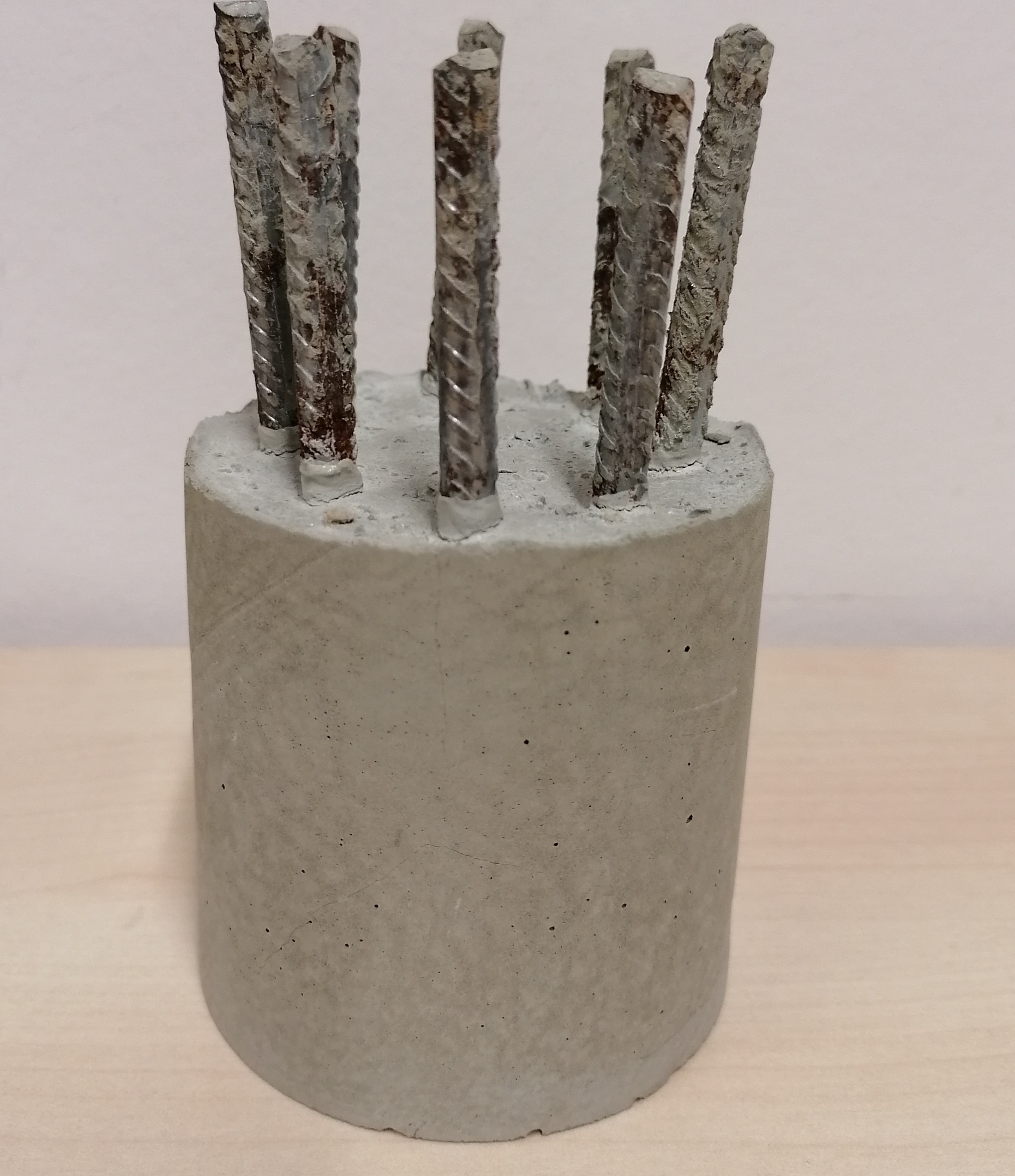}}{0.05in}{.05in}}\hspace*{0.15em}
\subfloat{\topinset{\bfseries \textcolor{white}{(b)}}{\label{fig:object Alone b}\includegraphics[width=0.32\linewidth]{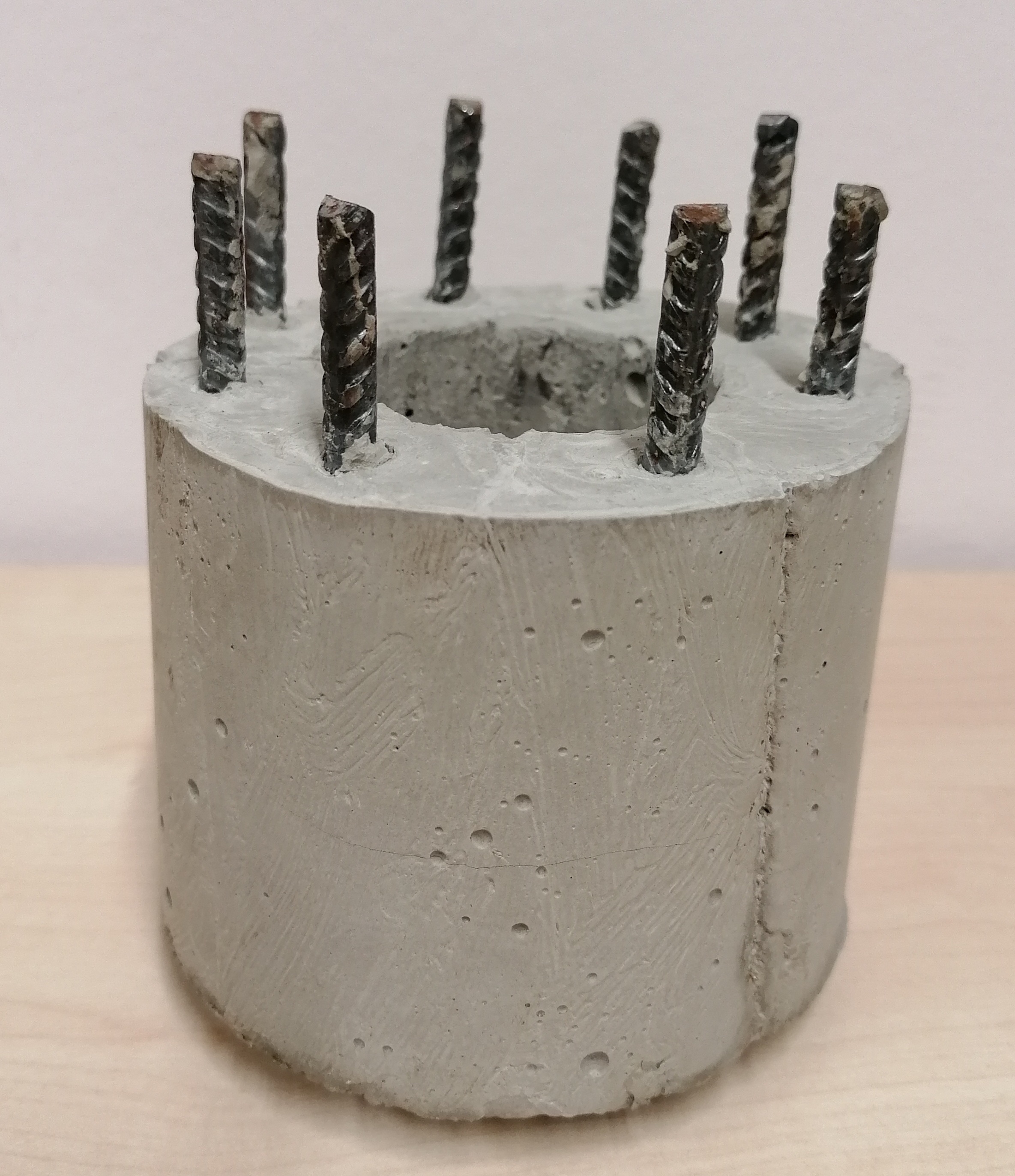}}{0.05in}{.05in}}\hspace*{0.15em}
\subfloat{\topinset{\bfseries \textcolor{white}{(c)}}{\label{fig:object Alone c}\includegraphics[width=0.319\linewidth]{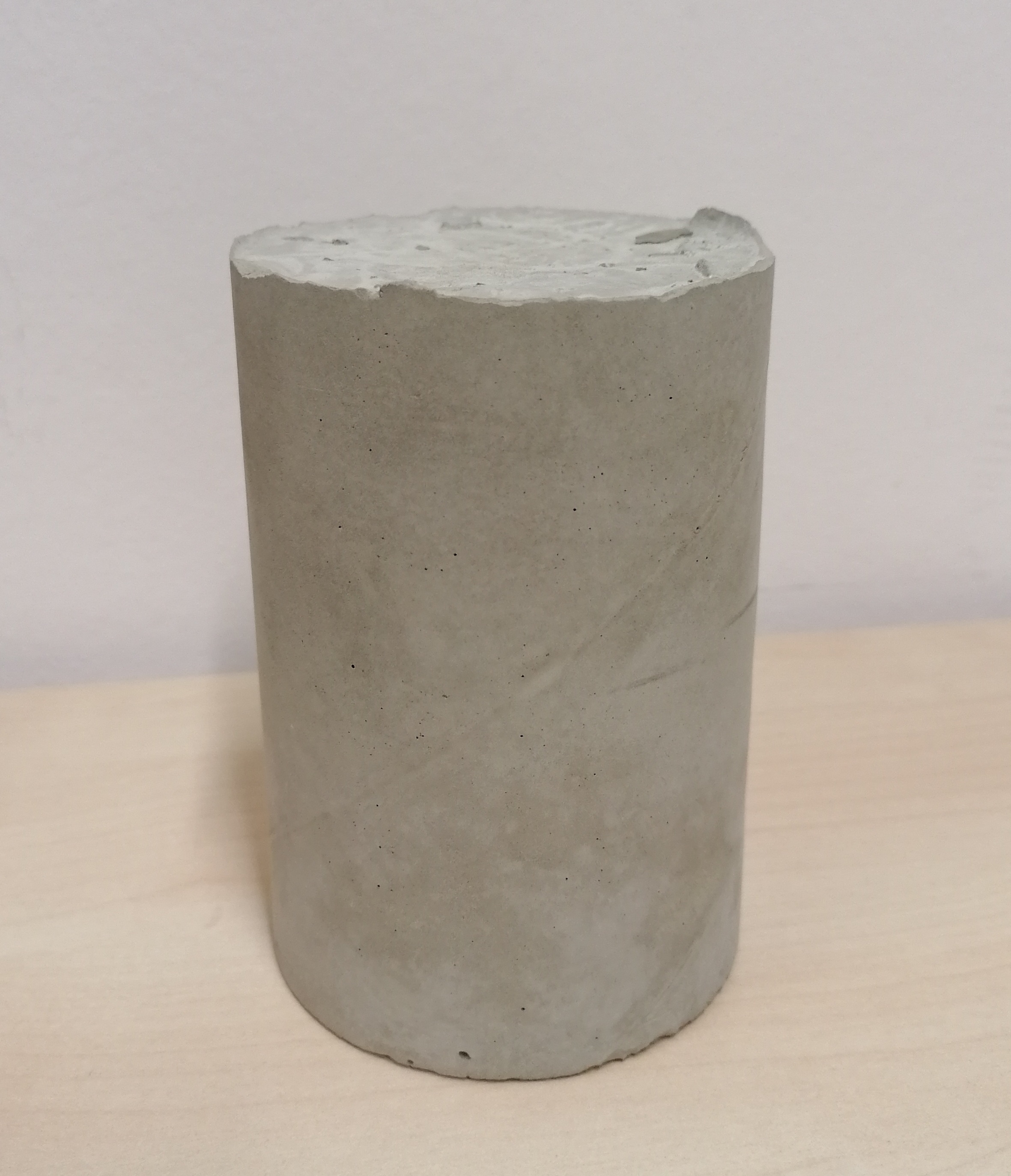}}{0.05in}{.05in}}\\[-0.15ex]

\caption{Cement based objects. (a) Cement cylinder of 7 $cm$ diameter with eight steel rods, (b) cement cylinder of 11 $cm$ outer diameter and 5 $cm$ inner diameter with eight steel rods, (c) cement cylinder of 7 $cm$ diameter without any insertion.}
\label{fig:object Alone}

\end{figure}

\begin{figure}[ht]
\centering

\def\stackalignment{l}
\subfloat{\topinset{\bfseries \textcolor{white}{(a)}}{\includegraphics[width=0.33\linewidth]{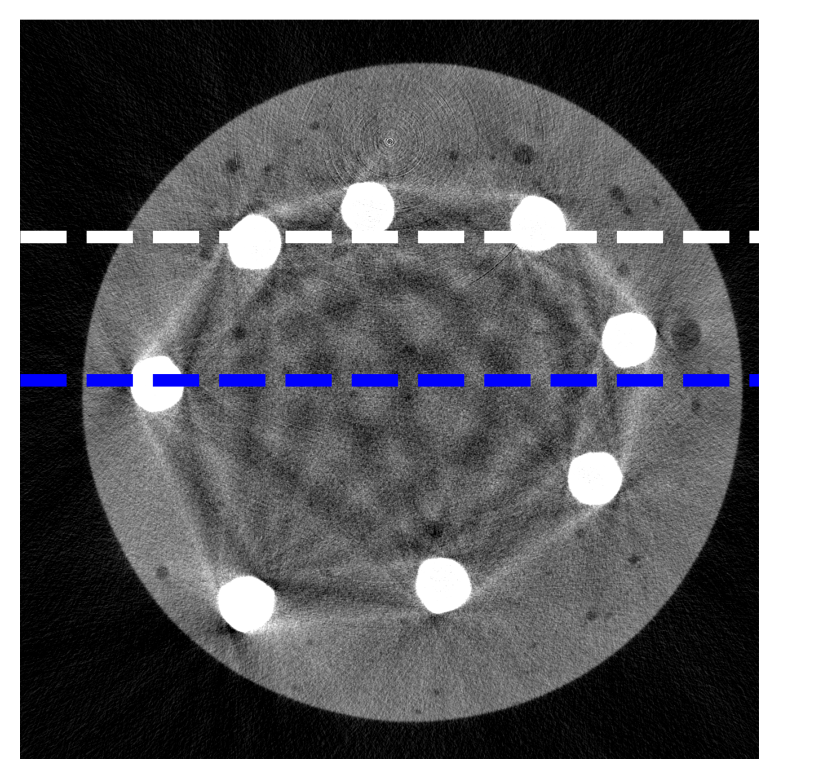}}{0.1in}{.1in}}
\subfloat{\topinset{\bfseries \textcolor{white}{(b)}}{\includegraphics[width=0.33\linewidth]{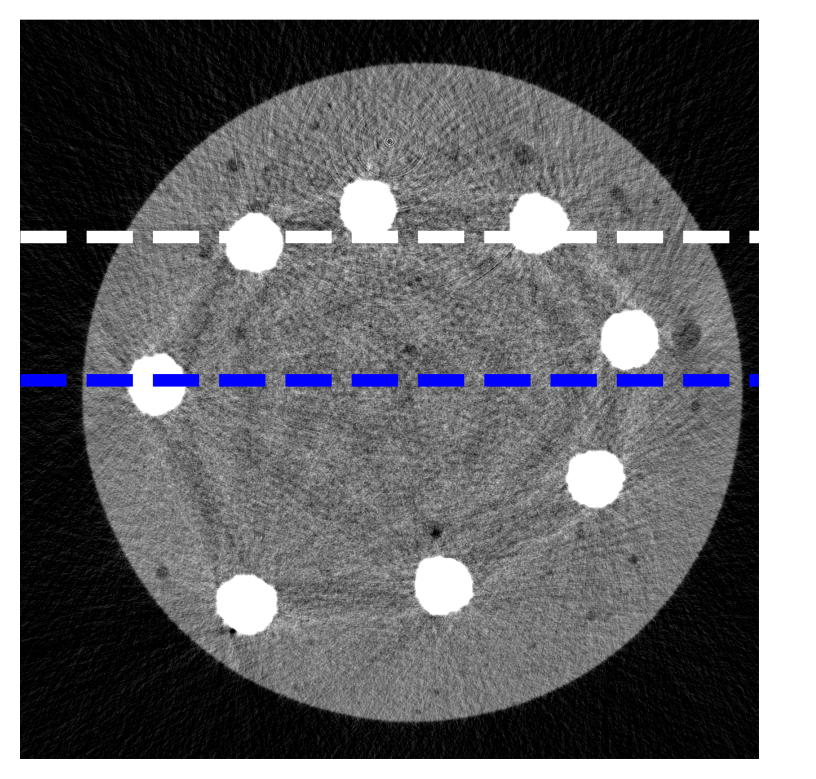}}{0.1in}{.1in}}
\subfloat{\topinset{\bfseries \textcolor{white}{(c)}}{\includegraphics[width=0.33\linewidth]{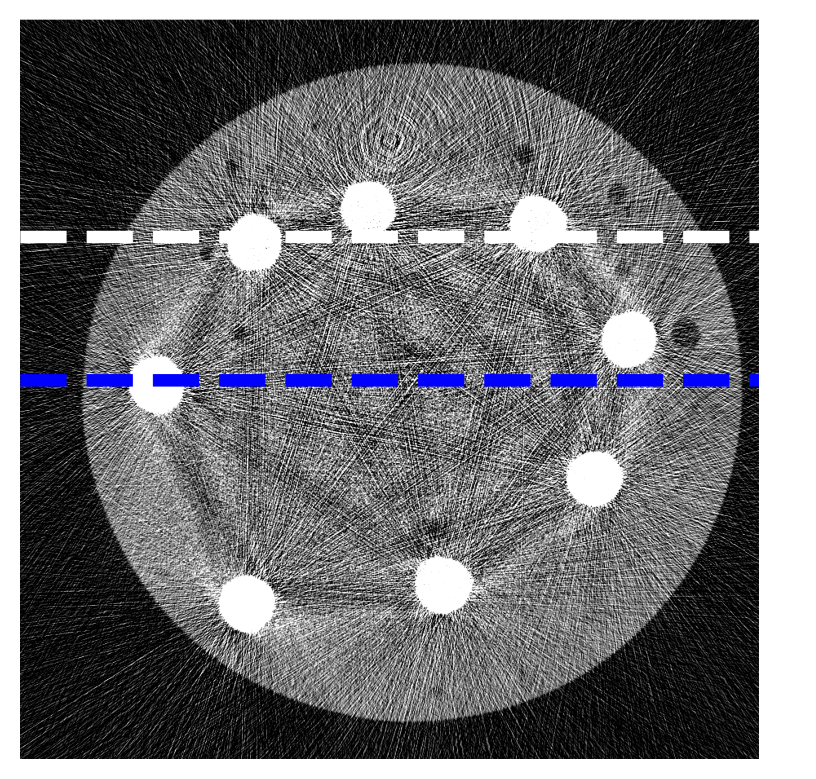}}{0.1in}{.1in}}\\[-0.15ex]
\subfloat{\topinset{\bfseries \textcolor{black}{(d)}}{\includegraphics[width=0.49\linewidth]{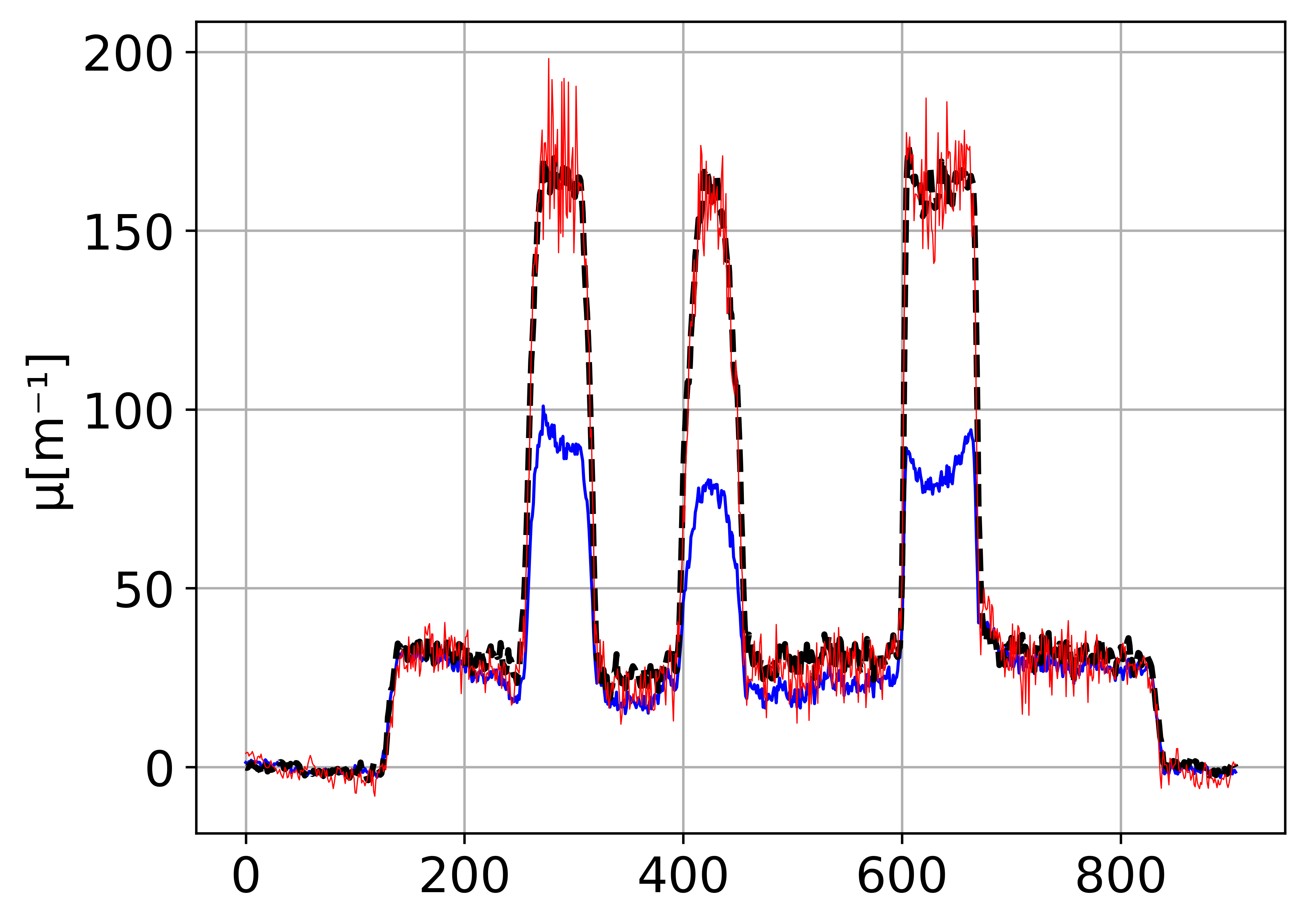}}{0.1in}{.3in}}
\subfloat{\topinset{\bfseries \textcolor{black}{(e)}}{\includegraphics[width=0.49\linewidth]{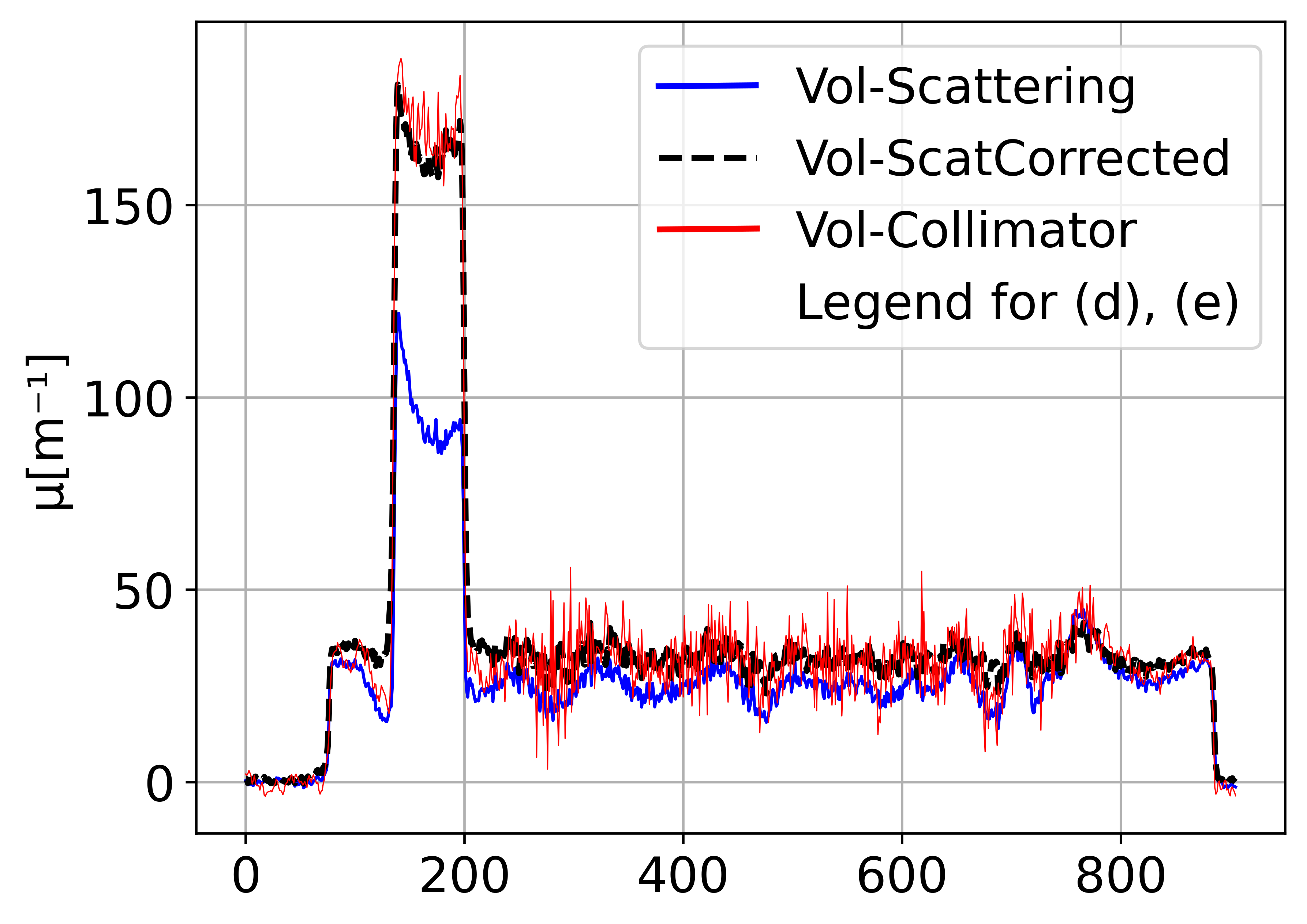}}{0.1in}{.3in}}

\caption{Scatter correction result (with the use of the interpolation technique) of the $7$ $cm$ diameter cement cylinder with eight steel rods. (a) Slice no. 875 of the scatter-corrupted volume, (b) same slice from the corrected volume, (c) same slice from the collimator volume, (d) profile lines (white dotted lines) of (a), (b), and (c), (e) profile lines (blue dotted lines) of (a), (b), and (c). The profile lines in this example are averaged over multiple rows to suppress the noise and to make the comparison more clear. The profiles are marked by white and blue dashed lines in (a), (b), and (c).}
\label{fig:correction of cementSteel}

\end{figure}

\begin{figure}[ht]
\centering

\def\stackalignment{l}
\subfloat{\topinset{\bfseries \textcolor{white}{(a)}}{\includegraphics[width=0.33\linewidth]{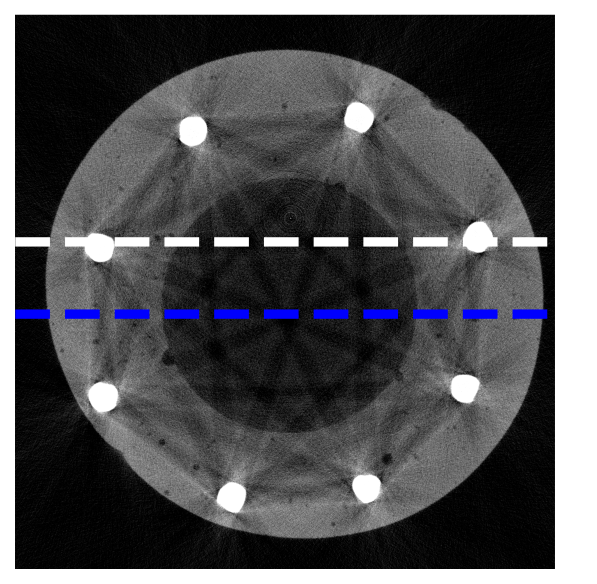}}{0.1in}{.1in}}
\subfloat{\topinset{\bfseries \textcolor{white}{(b)}}{\includegraphics[width=0.33\linewidth]{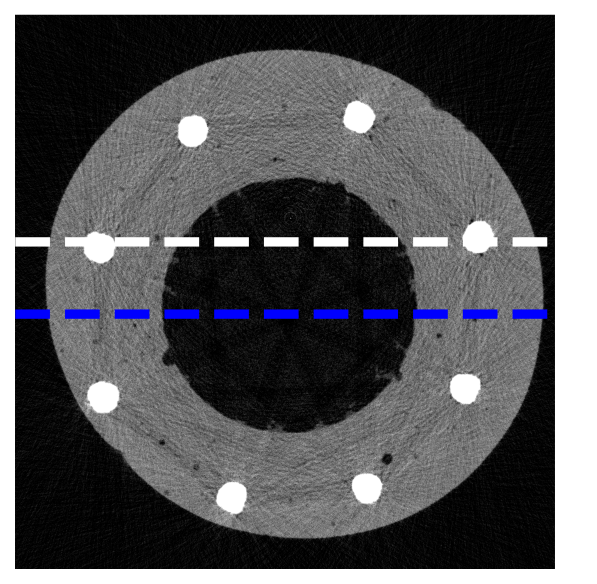}}{0.1in}{.1in}}
\subfloat{\topinset{\bfseries \textcolor{white}{(c)}}{\includegraphics[width=0.33\linewidth]{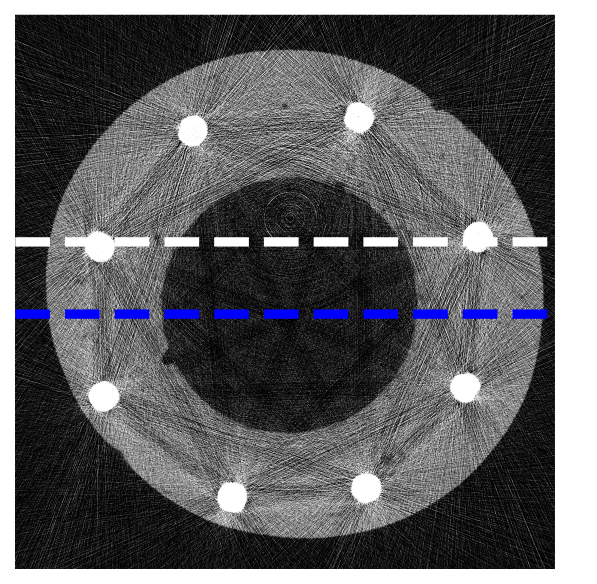}}{0.1in}{.1in}}\\[-0.15ex]
\subfloat{\topinset{\bfseries \textcolor{black}{(d)}}{\includegraphics[width=0.49\linewidth]{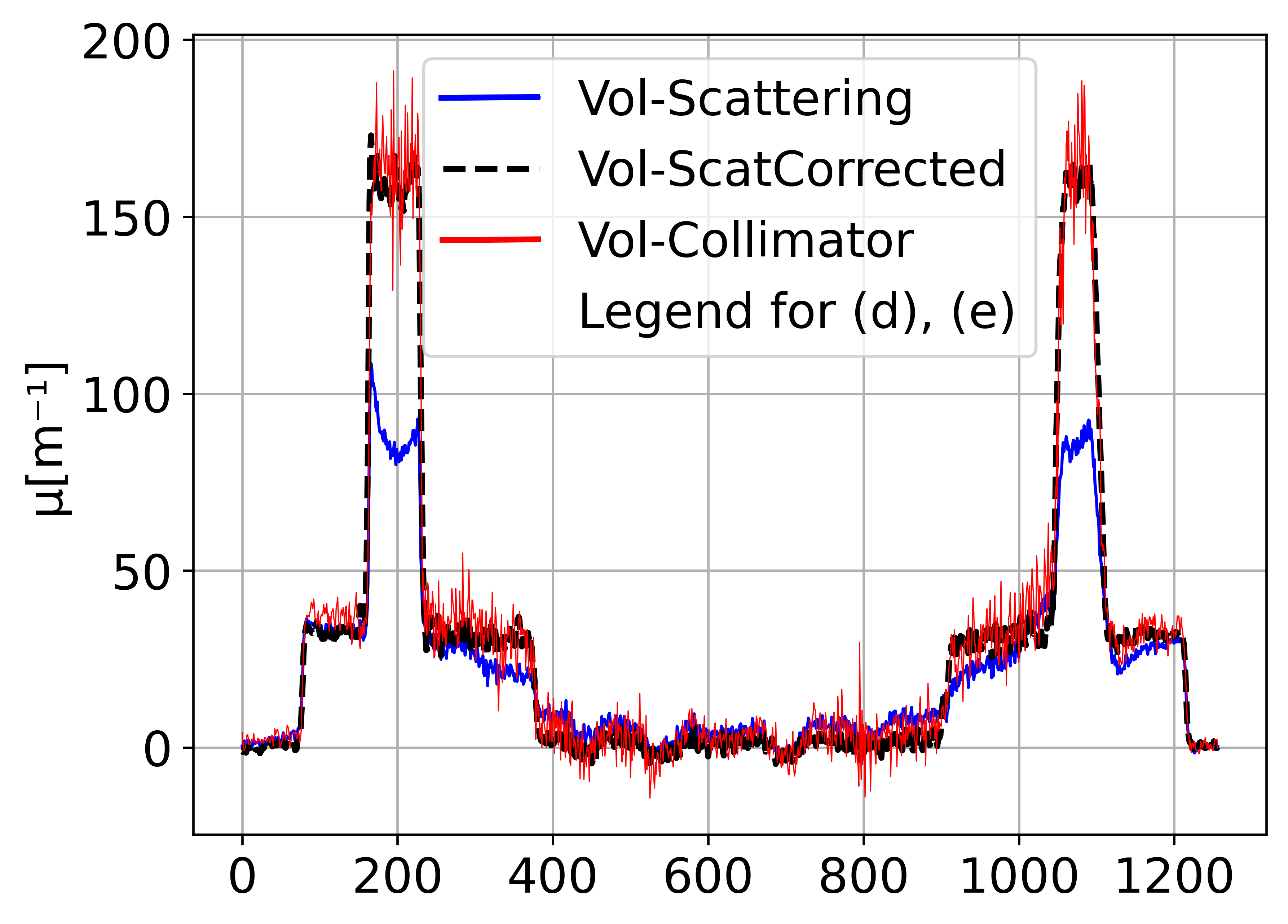}}{0.1in}{.28in}}
\subfloat{\topinset{\bfseries \textcolor{black}{(e)}}{\includegraphics[width=0.48\linewidth]{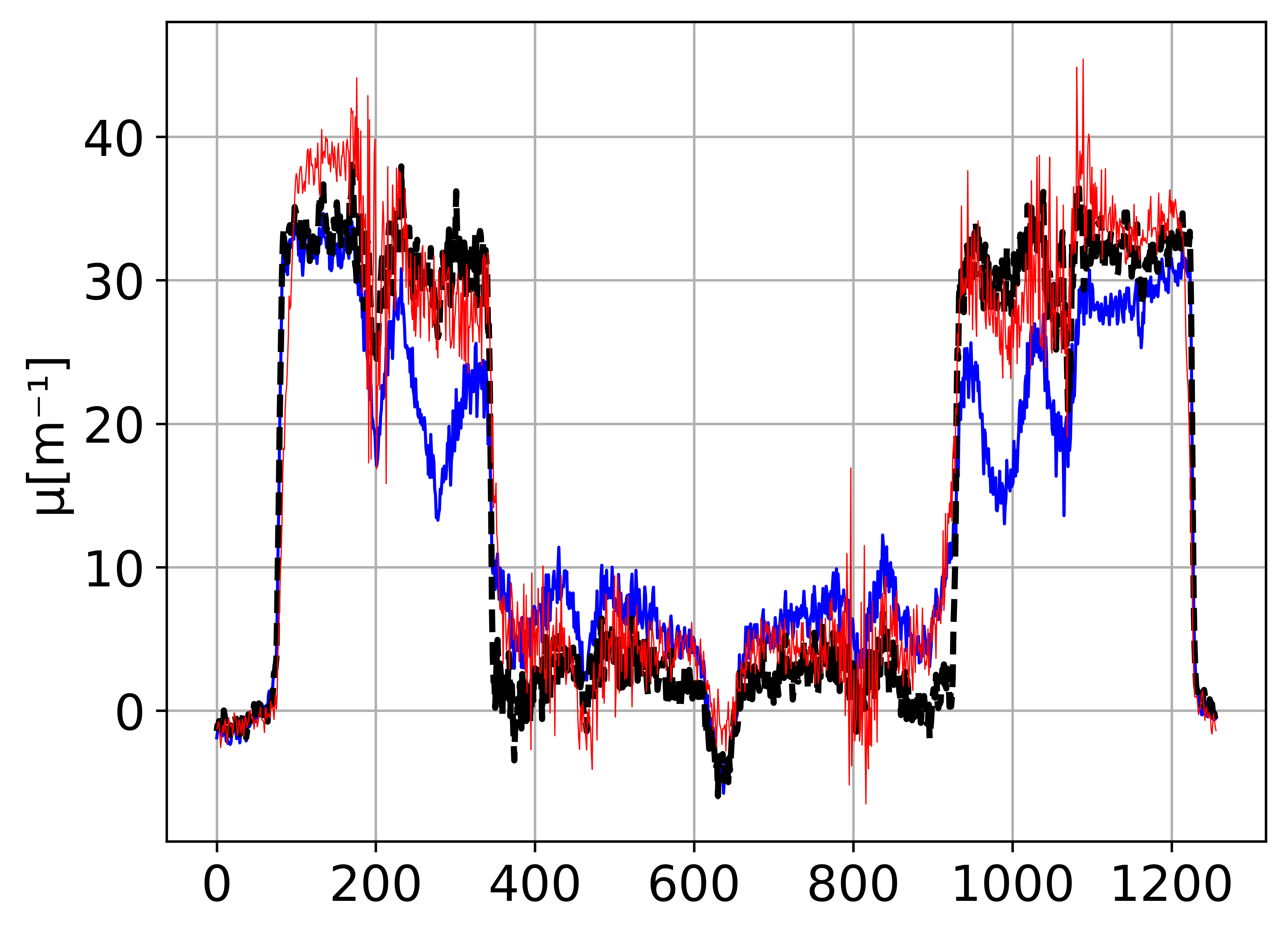}}{0.1in}{.28in}}

\caption{Scatter correction result (with the use of the interpolation technique) of the $11$ $cm$ outer diameter cement cylinder with steel rods. (a) Slice no. 871 of the scatter-corrupted volume, (b) same slice from the corrected volume, (c) same slice from the collimator volume, (d) profile lines (white dotted lines) of (a), (b), and (c), (e) profile lines (blue dotted lines) of (a), (b), and (c). The profile lines in this example are averaged over multiple rows to suppress the noise and to make the comparison more clear. The profiles are marked by white and blue dashed lines in (a), (b), and (c).}
\label{fig:correction of cementRing}

\end{figure}

\begin{figure}[ht]
\centering

\def\stackalignment{l}
\subfloat{\topinset{\bfseries \textcolor{white}{(a)}}{\includegraphics[width=0.33\linewidth]{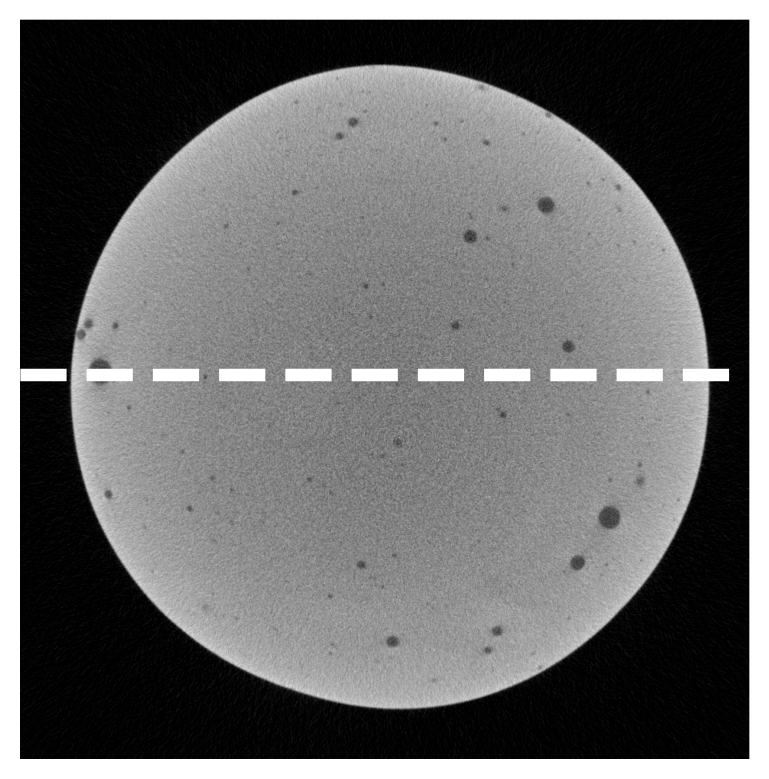}}{0.1in}{.1in}}
\subfloat{\topinset{\bfseries \textcolor{white}{(b)}}{\includegraphics[width=0.33\linewidth]{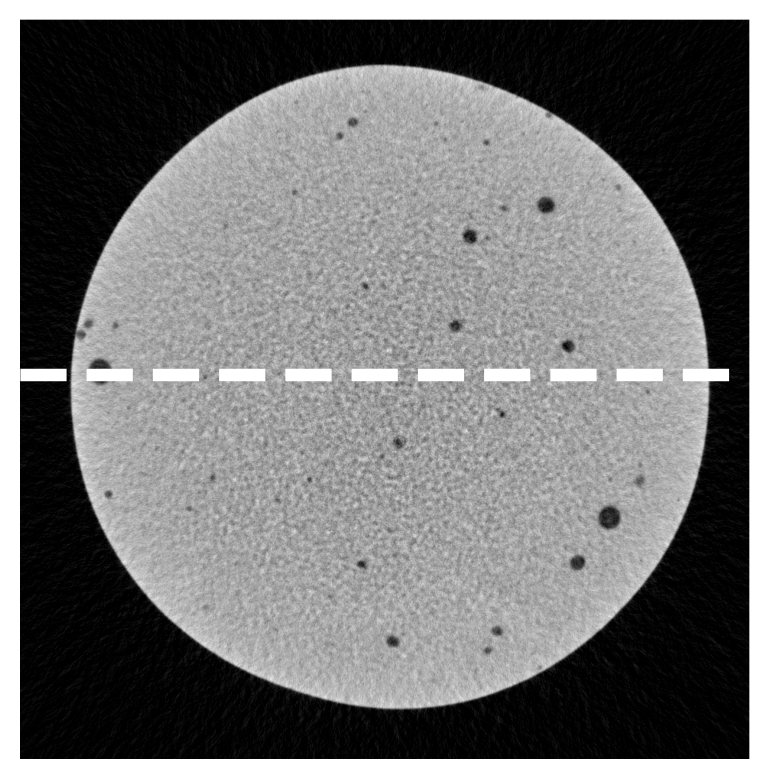}}{0.1in}{.1in}}
\subfloat{\topinset{\bfseries \textcolor{white}{(c)}}{\includegraphics[width=0.33\linewidth]{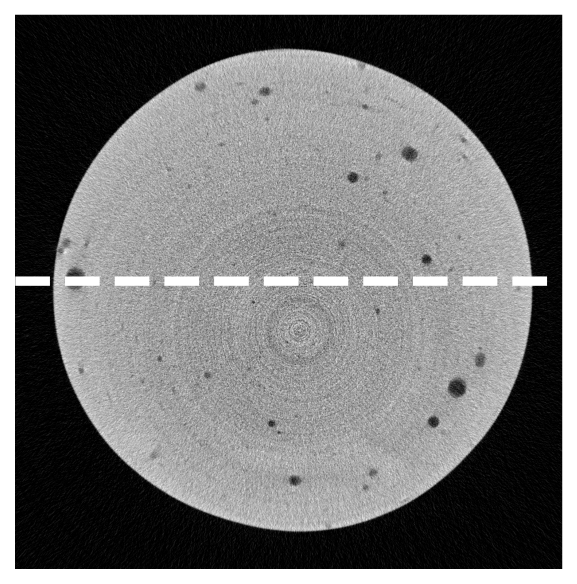}}{0.1in}{.1in}}\\[-0.15ex]
\subfloat{\topinset{\bfseries \textcolor{black}{(d)}}{\includegraphics[width=0.7\linewidth]{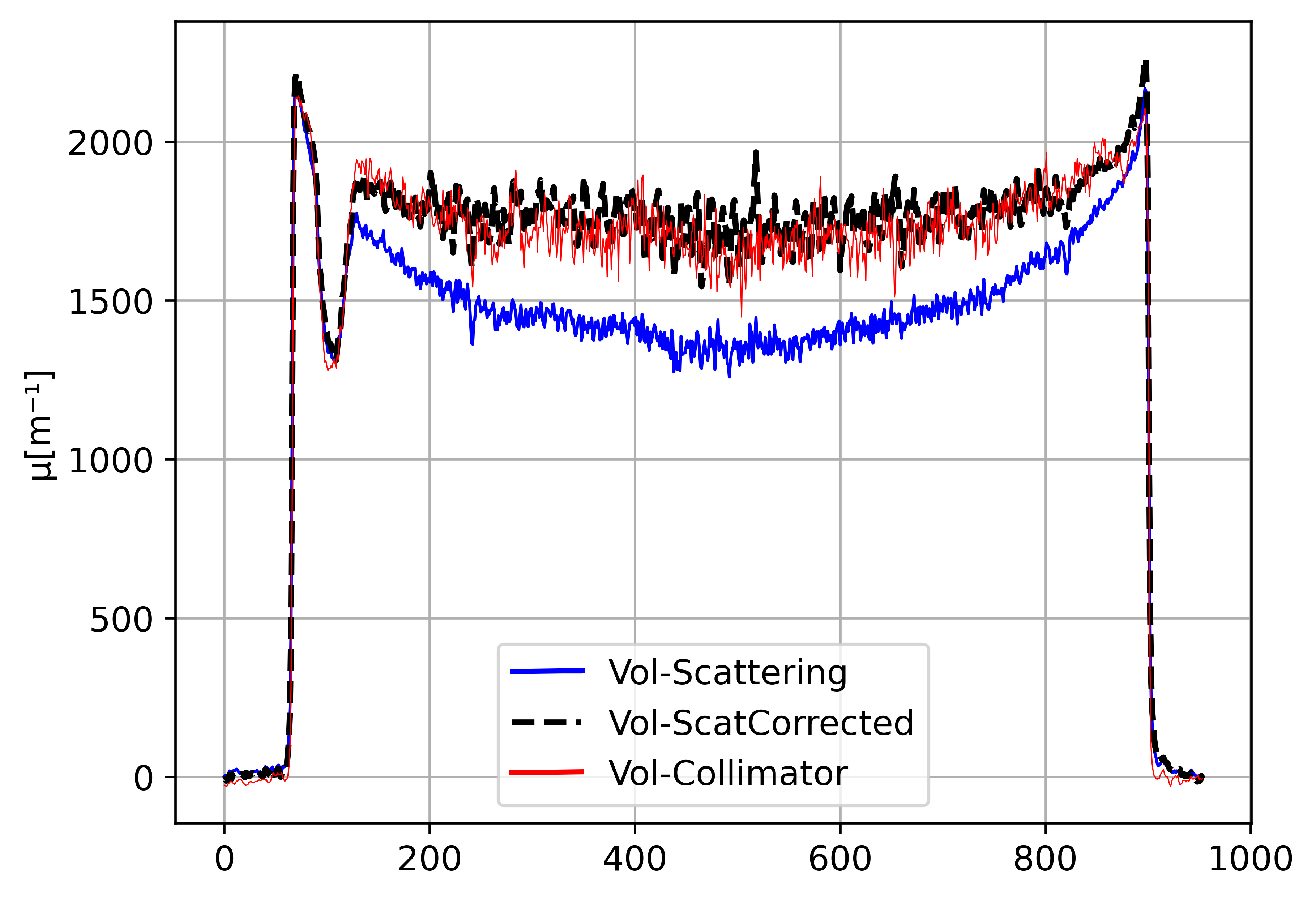}}{0.1in}{.35in}}\\[-1ex]

\caption{Scatter correction result (with the use of the interpolation technique) of the $7$ $cm$ cement cylinder without steel rods. (a) Slice no. 1316 of the scatter-corrupted volume, (b) same slice from the corrected volume, (c) same slice from the collimator volume, (d) profile lines of (a), (b), and (c). The profile lines in this example are averaged over multiple rows to suppress the noise and to make the comparison more clear. The profiles are marked by white dashed lines in (a), (b), and (c).}
\label{fig:correction of cementcylinder}

\end{figure}

Fig. \ref{fig:correction of cementSteel}, Fig. \ref{fig:correction of cementRing}, and Fig. \ref{fig:correction of cementcylinder} show the scatter correction results of different cement based objects. Fig. \ref{fig:correction of cementSteel} shows the scatter correction results of a 7 $cm$ diameter cement cylinder with eight steel rods shown in Fig.~\ref{fig:object Alone}\subref{fig:object Alone a}. Fig. \ref{fig:correction of cementRing} shows the scatter correction results of a cement cylinder with 11 $cm$ outer diameter and 5 $cm$ inner diameter with eight steel rods inserted in this cylinder. This object is shown in Fig.~\ref{fig:object Alone}\subref{fig:object Alone b}. Fig. \ref{fig:correction of cementcylinder} shows a 7 $cm$ diameter cement cylinder without any insertion which is shown in Fig.~\ref{fig:object Alone}\subref{fig:object Alone c}. 3000 projections from the scanner for each one of these objects were corrected using the implemented iterative scatter correction algorithm. Interpolation was also used in this case, in which 1500 scatter projections with (576,800) resolution were simulated using the proposed MC model and then linearly interpolated to 3000 projections. On the other hand, 3000 primary projections were simulated using the proposed MC model. The primary and the scatter projections were then up-sampled to the same resolution of the scanner. The number of photons used in the MC simulation was $4.9\times10^8$ photons. Since the reconstructed volumes from the scanners' projections are well segmented, only a single iteration was used for the correction. The good segmentation result is a consequence of the high diversity between the linear attenuation values of the cement and the steel. Even when the steel is highly affected by the scatter in these examples, the value of the linear attenuation of the steel is still easily distinguishable from the one of the cement. The primary and the scatter projections, simulated using the proposed MC model in this single iteration, were accurate enough for the scatter correction process.

The results of the scatter correction of these objects resemble the near scatter-free results acquired using the collimator. Besides, it is also shown that the scatter correction successfully reduces the cupping and the streak artifacts that are severely affecting the original volumes from the scanner. The parameters of the scan used to acquire the raw projections from the scanner for all the objects are shown in table \ref{tab:scan parameters2}.

Table \ref{tab:time5} shows the time required for a single iteration of the scatter correction and for the MC simulation of the three objects. 

\begin{table}[ht]
\captionsetup{justification=centering}
\centering
\caption{MEASUREMENTS PARAMETERS FOR THE CEMENT BASED OBJECTS.}
\begin{tabular}{ | C{2cm} | C{2.5cm}| } 
\hline
Parameter & Value\\ 
\hline
SDD & 1.282 m \\ 
\hline
SOD & 0.907 m\\ 
\hline
Resolution  & (2304,3200) pixels\\
\hline
Pixel size & 0.127 $\mu$m\\
\hline
X-ray voltage  & 200 kV\\
\hline
X-ray current & 50 $\mu$A\\
\hline
Exposure time & 1 s\\
\hline
\end{tabular}
\label{tab:scan parameters2}
\end{table}

\begin{table}[ht]
\captionsetup{justification=centering}
\centering
\caption{TIME REQUIRED FOR A SINGLE ITERATION OF SCATTER CORRECTION USING FOUR GPUs WITH AND WITHOUT THE USE OF THE INTERPOLATION TECHNIQUE.}
\begin{threeparttable}[t]
\begin{tabular}{ |C{1.4cm}|C{1cm}| C{1.4cm} | C{1.3cm}| C{1.4cm} | } 
\hline
 Method&   Objects & No. of photons  &Iterative correction & MC simulation \tnote{(1)} \\ 
\hline

\multirow{3} {*}   & Cement object1 & $4.9\times10^8$& 19615 s & 16606 s\\ \cline{2-5}

 \centering  {With interpolation}& Cement object2 & $4.9\times10^8$& 22861 s & 19982 s\\ \cline{2-5}

&Cement object3  & $4.9\times10^8$& 18727 s & 15950 s\\\hline 

\multirow{3}*   & Cement object1 & $1.9\times10^9$ \tnote{(2)}& 72473 s & 69464 s \\ \cline{2-5}

\centering  Without interpolation  & Cement object2 & $1.9\times10^9$ \tnote{(2)}& 84323 s & 81444 s\\ \cline{2-5}

&Cement object3  &$1.9\times10^9$ \tnote{(2)}& 65409 s & 62632 s\\\hline

\end{tabular}
\begin{tablenotes}\footnotesize
  \item [(1)] With interpolation: 1500 and 3000 scatter and primary projections respectively of size (576,800); Without using interpolation: 3000 scatter and primary projections of size (2304,3200).
  \item [(2)] As in this case, 4$\times$ the resolution of (576,800) is used, the number of photons utilized is $\sim$4$\times$ the number of photons in case of using interpolation.
\end{tablenotes}
\end{threeparttable}%
\label{tab:time5}
\end{table}

\subsection{Effectiveness of Interpolation on Scatter Correction} 
\label{interpolation effect}
\begin{figure}[ht]
\centering

\def\stackalignment{l}
\subfloat{\topinset{\bfseries \textcolor{white}{(a)}}{\label{fig:a}\includegraphics[width=0.25\linewidth]{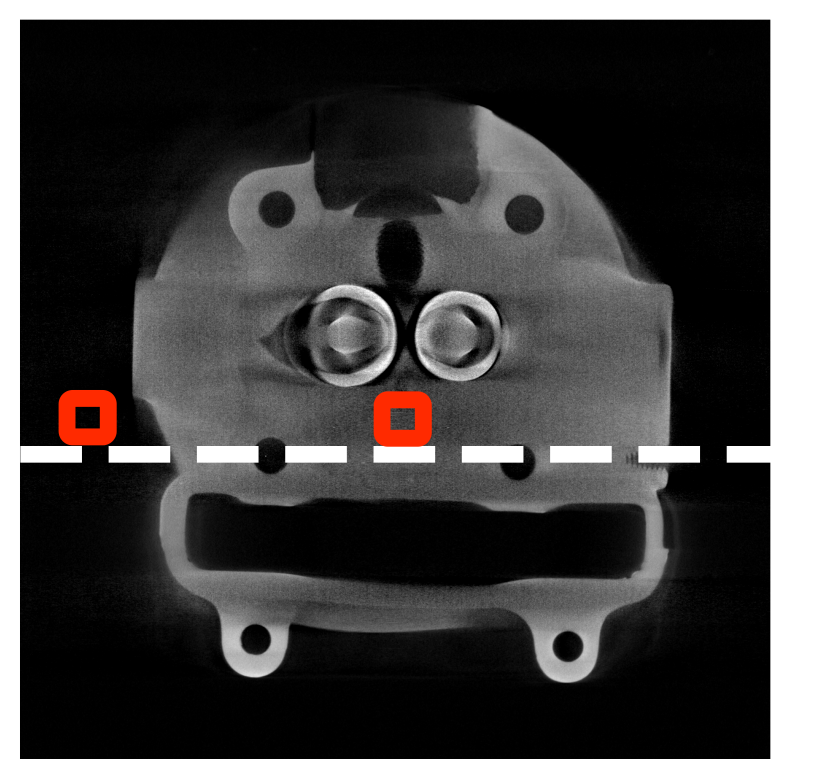}}{0.06in}{.06in}}
\subfloat{\topinset{\bfseries \textcolor{white}{(b)}}{\label{fig:b}\includegraphics[width=0.25\linewidth]{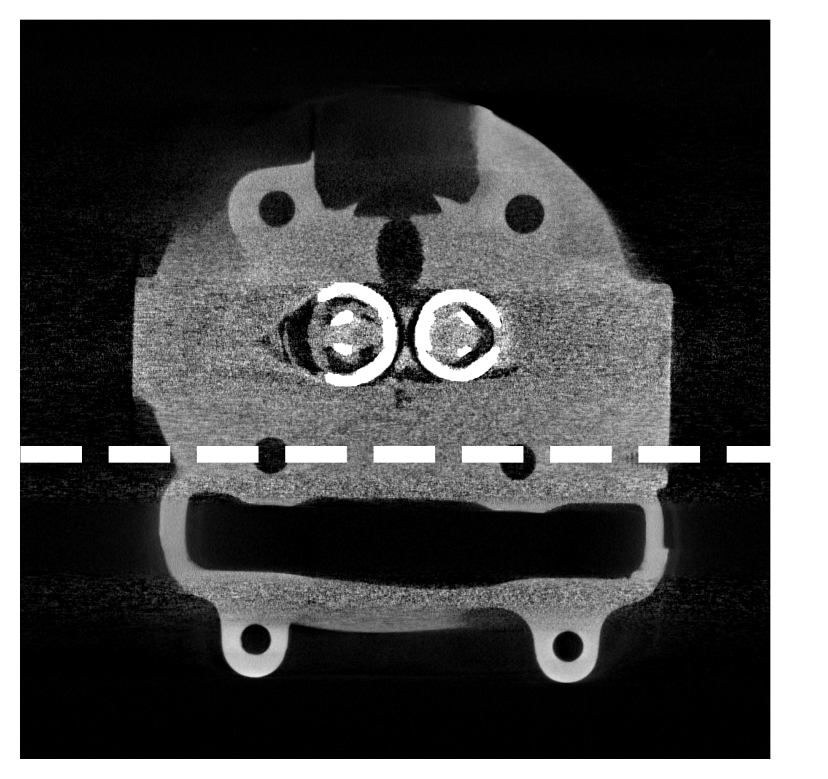}}{0.06in}{.06in}}
\subfloat{\topinset{\bfseries \textcolor{white}{(c)}}{\label{fig:c}\includegraphics[width=0.25\linewidth]{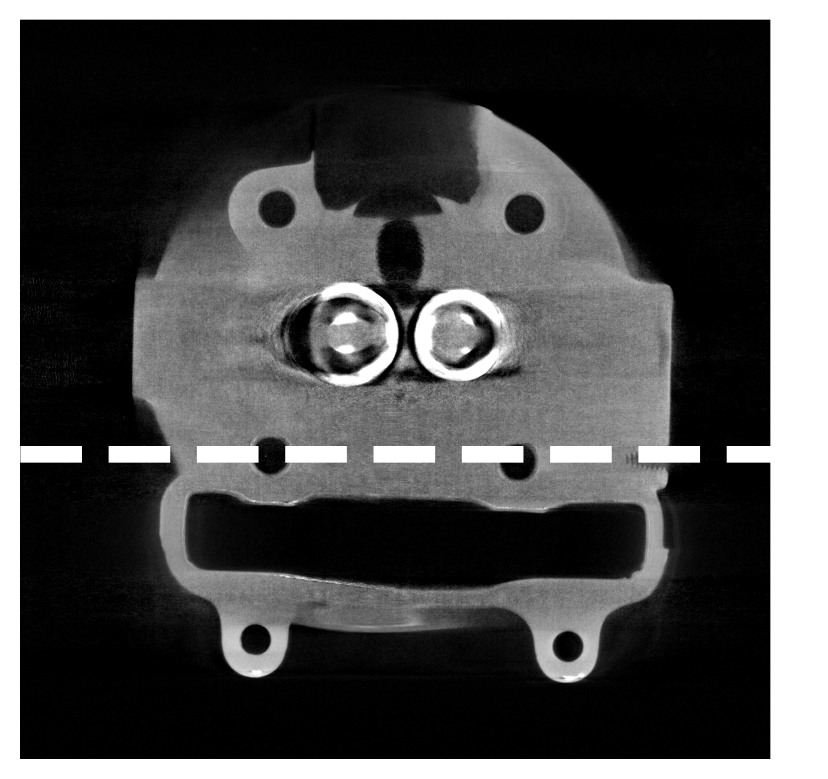}}{0.06in}{.06in}}
\subfloat{\topinset{\bfseries \textcolor{white}{(d)}}{\label{fig:d}\includegraphics[width=0.25\linewidth]{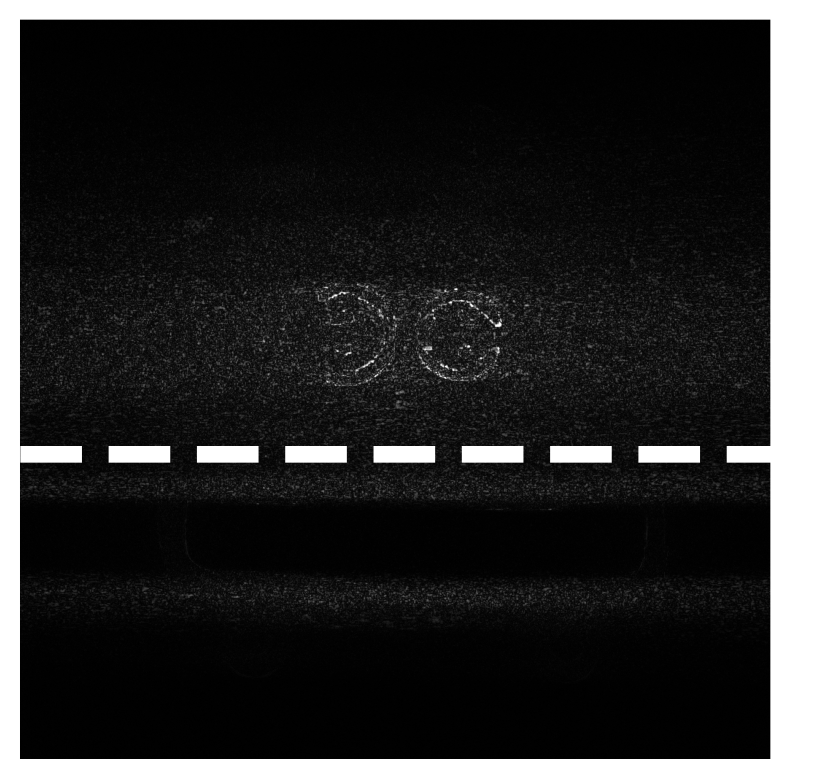}}{0.06in}{.06in}}\\[-0.15ex]
\subfloat{\topinset{\bfseries \textcolor{black}{(e)}}{\label{fig:e}\includegraphics[width=0.49\linewidth]{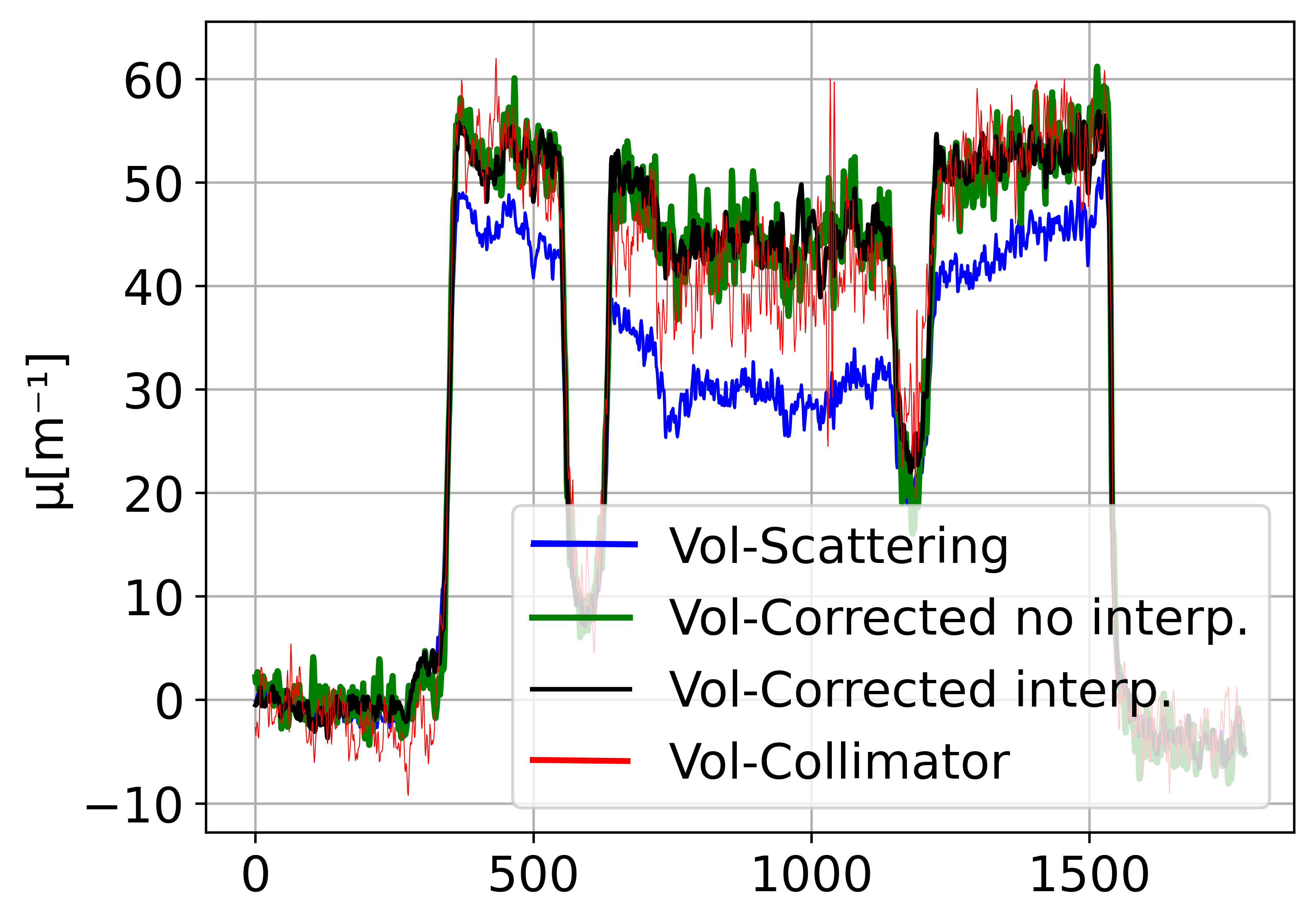}}{0.1in}{.28in}}
\subfloat{\topinset{\bfseries \textcolor{black}{(f)}}{\label{fig:f}\includegraphics[width=0.47\linewidth]{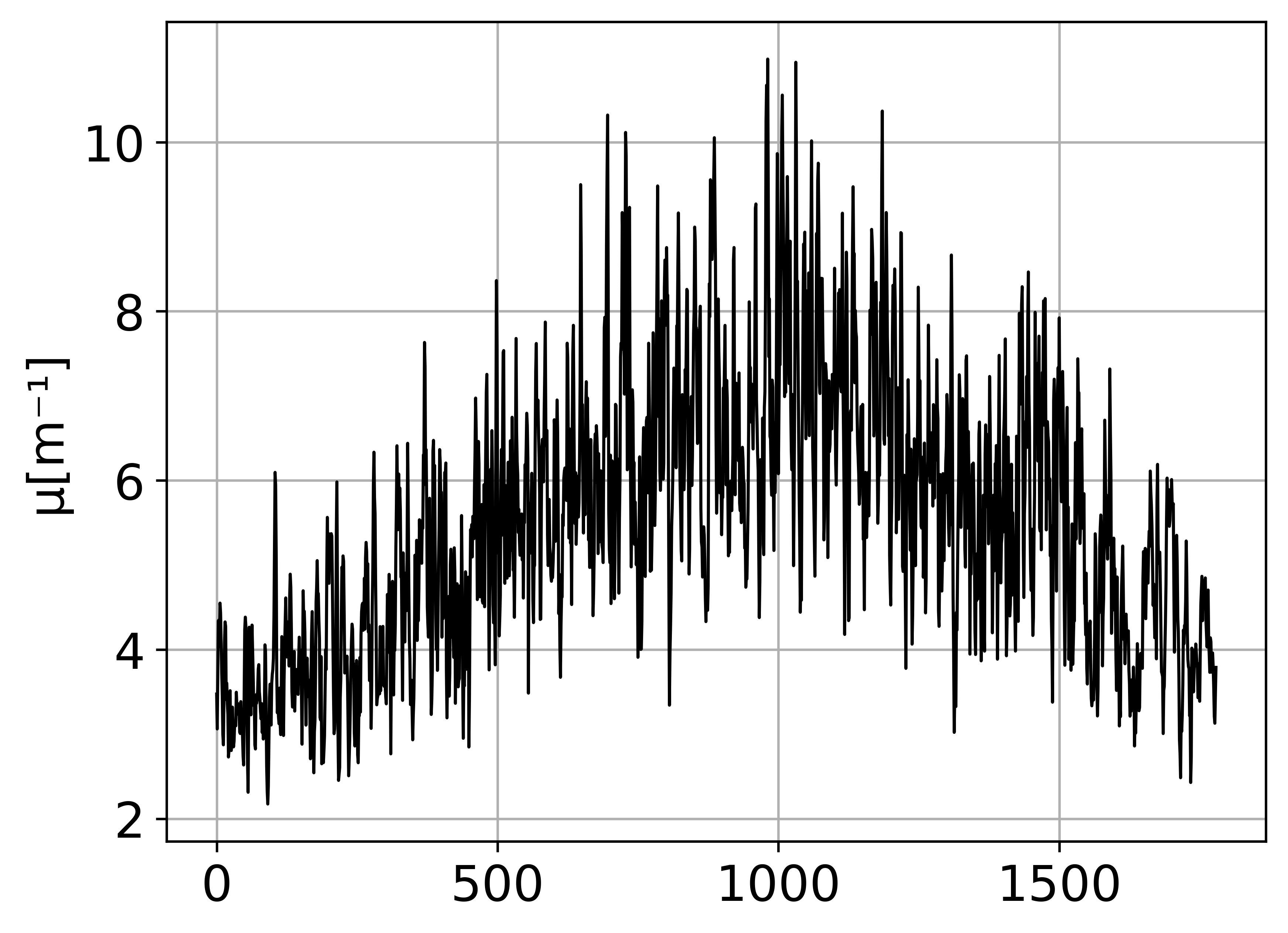}}{0.1in}{.27in}}

\caption{Comparison of the correction results with and without the use of the interpolation technique. (a) Scatter-corrupted volume's slice no. 2110 (front view)(the small red squares represent the ROI and the background which are used in the quantitative evaluation in subsections \ref{optimization effect} and \ref{interpolation effect}), (b) scatter-corrected without using the interpolation, (c) scatter-corrected with the use of the interpolation (d) error image between (b) and (c), (e) profile lines of (a), (b), (c) and with the near scatter-free volume's slice from collimator, (f) profile line of (d).The profile lines in this example are averaged over multiple rows to suppress the noise and to make the comparison more clear. The profiles are marked by white dashed lines in (a), (b), (c), and (d).}

\label{fig:Effect of Interpolation}
\end{figure}

To validate the use of the interpolation in the scatter correction process, the scatter correction was performed with and without the use of the interpolation. As mentioned before, two kinds of interpolation were used here. One between the simulated scatter projections and one to up-sample the simulated scatter and primary projections from low-resolution to a higher resolution. In case the interpolation is applied, the number of photons used in the MC simulation to simulate the primary and the scatter projections is $2.5\times10^8$ with (576,800) resolution. Besides, only half the number of the scatter projections is simulated in this case. While for primary projections, the full set of the projections is simulated. For the case of the interpolation is not used, the original resolution and the same number of projections from the scanner have been used in the proposed MC model to derive the scatter and the primary projections using $7\times10^8$ photons. Fig. \ref{fig:Effect of Interpolation} shows the results of the two cases. The error image in Fig.~\ref{fig:Effect of Interpolation}\subref{fig:d} shows that the use of the interpolation in the correction process only has a minor effect on the results. Quantitative evaluation between the two cases has been done using the mean square error (MSE), and the normalized cross-correlation (NCC). The contrast-to-noise ratio (CNR) for the regions of interest shown in Fig.~\ref{fig:Effect of Interpolation}\subref{fig:a} (red squares) has been used to evaluate the quality of the two methods individually. Table \ref{tab:statistics1} shows the result of this evaluation. From this table, the results of the MSE and the NCC show that the images from both cases are close to each other. On the other hand, the CNR shows superior of the case of using the interpolation techniques. This is mainly because using $2.5\times10^8$ photons in the low-resolution case produces a low noise image, by interpolating it to the high-resolution case it maintains its quality. For the case of without interpolation, the use of $7\times10^8$ photons to simulate high-resolution projection produces noisy scatter estimates. This impacts the quality of the scatter correction result. The object of the study in this case is the aluminum motorcycle cylinder head.

\begin{table}[ht]
\captionsetup{justification=centering}
\caption{QUANTITATIVE EVALUATION OF THE SCATTER CORRECTION RESULTS WITH AND WITHOUT THE USE OF THE INTERPOLATION TECHNIQUE IN THE SIMULATION OF THE SCATTER.}
\centering
\begin{threeparttable}[t]
\begin{tabular}{| C{0.8cm} | C{1.8cm}| C{0.7cm}|}
\hline
MSE& \multicolumn{1}{c} {26.4}&\\  
\hline
NCC& \multicolumn{1}{c} {0.97}&\\ 
\hline

\multirow{2}{*}{CNR} & With Interp.& 16\\
\cline{2-3}

\centering  & Without Interp.&9.4 \\
\cline{2-3}

\hline
\end{tabular}
\end{threeparttable}%
\label{tab:statistics1}
\end{table}

\subsection{Evaluation on Execution Time of the Proposed MC Model}
\label{Evaluation on Execution Time}
Three different objects were used to evaluate the required execution time of the proposed MC model against other simulators. Object (2) is shown in Fig. \ref{fig:ComparsionMustang}, and object (3) is shown in Fig. \ref{fig:CorrectionEGSnrc}, (object (1) is not shown here however, the occupation of this object on the screen is the same as the object in Fig. \ref{fig:ComparsionMustang}). The dimensions of the objects are $18.4\,cm\times20\,cm\times18.4\,cm$, $18.1\,cm\times22.4\,cm\times13.5\,cm$, and $16.7\,cm\times13.7\,cm\times9.8\,cm$ for object (1), (2) and (3), respectively. Voxelized objects were used in the simulations of the proposed  GPU MC model and the EGSnrc simulator, while CAD objects of closed triangulated surfaces were used in the aRTist and the CAD model (the CAD model is the CPU version of the proposed MC GPU model). The platform of the execution of the CPU simulators is Intel(R) Core(TM) i7-5820k CPU 3.30 GHz, while a GeForce GTX 1080 Nvidia GPU has been used for the proposed MC model. The number of photons used in the simulation was $4.2\times10^8$ photons from 190 energy bins. The detector resolution in these simulations was of (576,800) pixels. The evaluation of the execution time was done with three different cases. 

\subsubsection{Evaluation of the Execution Time Using a Single-CPU Thread and a Single-GPU}

In the first case, a single thread was used for the simulation of the CPU simulators. While for the proposed MC model, a single-GPU was used in the simulation. The time required for a single projection is shown in table \ref{tab:time1}. Considering the comparison case of the car engine (2), 347$\times$ acceleration is achieved in comparison to the EGSnrc simulator, while more than 18$\times$ acceleration is achieved in comparison to the aRTist simulator. It is worth mentioning that the aRTist simulator is one of the fastest simulators available on the CPU platform \cite{aRTist}.

\begin{table}[ht]
\captionsetup{justification=centering}
\centering
\caption{TIME FOR THE EXECUTION ON DIFFERENT PLATFORMS USING SINGLE-GPU FOR THE PROPOSED MC MODEL AND SINGLE THREAD FOR THE CPU MODELS.}
\begin{tabular}{ | C{6.0em} | C{1.2cm}| C{1.2cm} | C{1.2cm} | C{1.2cm} | } 
\hline
& GPU model& CAD model (CPU) & EGSnrc (CPU) & aRTist (CPU)\\ 
\hline
Car engine(1) & 50 s & 1860 s & 8500 s & 1008 s\\ 
\hline
Car engine(2)& 44 s & 2016 s & 15300 s & 832 s\\ 
\hline
Car engine(3) & 44.2 s & 5860 s & 4400 s & 1218 s\\
\hline
\end{tabular}
\label{tab:time1}
\end{table}

\subsubsection{Evaluation of the Execution Time Using 12 CPU Threads and a Single-GPU}

In the second case, 12 threads were used for the CPU simulators. While only a single-GPU was used for the proposed MC model for the same test objects of the first case. Considering the same comparison case of the second car engine, 49$\times$ acceleration is achieved in comparison to the EGSnrce simulator, and 2$\times$ in the case of the aRTist simulator. The results of this case are shown in table \ref{tab:time2}.

\begin{table}[ht]
\captionsetup{justification=centering}
\centering
\caption{TIME OF THE EXECUTION ON DIFFERENT PLATFORMS USING SINGLE-GPU FOR THE PROPOSED MC MODEL AND 12 THREADS FOR THE CPU MODELS.}
\begin{tabular}{ | C{6.0em} | C{1.2cm}| C{1.2cm} | C{1.2cm} | C{1.2cm} | } 
\hline
& GPU model& CAD model (CPU) & EGSnrc (CPU) & aRTist (CPU)\\ 
\hline
Car engine(1) & 50 s & 203 s & 1214 s & 130 s\\ 
\hline
Car engine(2)& 44 s & 189 s & 2185 s & 106 s\\ 
\hline
Car engine(3) & 44.2 s & 444 s & 628 s & 145 s\\
\hline
\end{tabular}
\label{tab:time2}
\end{table}

\subsubsection{Evaluation of the Execution Time Using 12 CPU Threads and Four GPUs}

As mentioned before, the proposed MC model performs the acceleration on a multi-GPU system by distributing the number of projections on the available GPUs. These projections are then simultaneously simulated. The evaluation is done by comparing the total simulation time required for the full set of projections. Thus, 3000 projections were simulated on four GPUs using the proposed MC model. Whereas 12 threads were used in the simulation of the CPU simulators. Table \ref{tab:time3} shows the time required for the simulation. Multi-GPU implementation achieves a 202$\times$ and 9$\times$ acceleration in comparison to the EGSnrce simulator and the aRTist simulator respectively.

\begin{table}[ht]
\captionsetup{justification=centering}
\centering
\caption{TIME FOR SIMULATING 3000 PROJECTIONS ON DIFFERENT PLATFORMS USING FOUR GPUs FOR THE PROPOSED MC MODEL AND 12 THREADS FOR THE CPU MODELS (NO INTERPOLATION APPLIED)}
\begin{tabular}{ | C{6.0em} | C{1.2cm}| C{1.2cm} | C{1.2cm} | C{1.2cm} | } 
\hline
& GPU model& CAD model (CPU) & EGSnrc (CPU) & aRTist (CPU)\\ 
\hline
Car engine(1) & 10.4 h & 169 h & 1011 h & 108 h\\ 
\hline
Car engine(2)& 9 h & 157.5 h & 1820 h & 88.3 h\\ 
\hline
Car engine(3) & 9.2 h & 370 h & 523 h & 120.8 h\\
\hline
\end{tabular}
\label{tab:time3}
\end{table}

\subsection{Optimization of the Runtime of MC Simulation}
\label{optimization effect}

In this work, key parameters are made controllable to optimize the runtime for the MC simulation as it is the most expensive part of the correction process. Table (\ref{tab:timeOptimized}) shows five cases of simulations. The second column shows the setting of the standard MC simulation with no optimization of the key parameters and no smoothing and interpolation techniques were used.

\begin{figure}[ht]
\centering

\def\stackalignment{l}
\subfloat{\topinset{\bfseries \textcolor{white}{(a)}}{\label{fig:Effect of optimization a}\includegraphics[width=0.25\linewidth]{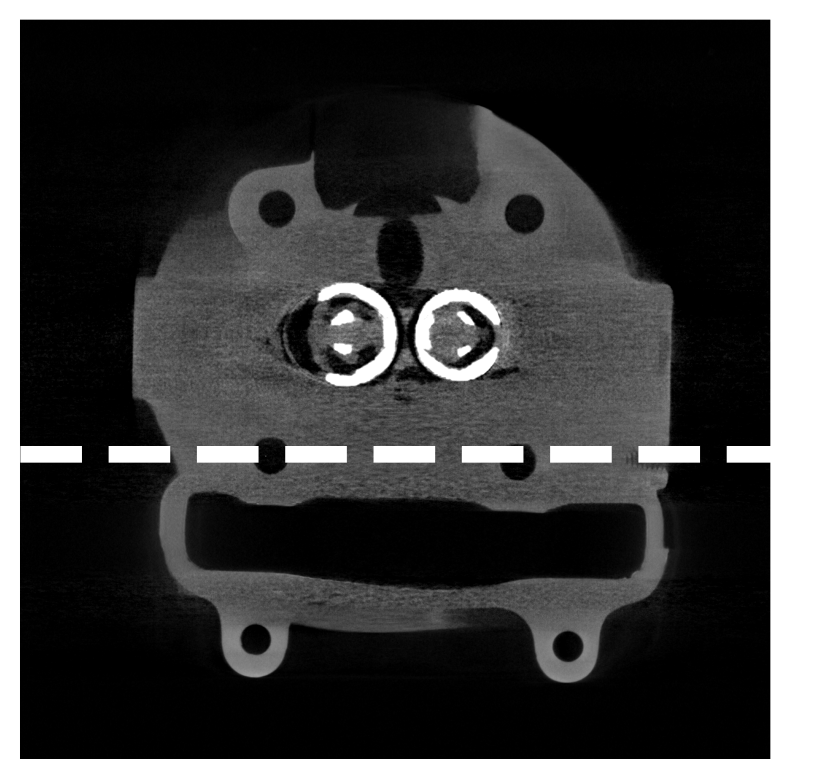}}{0.06in}{.06in}}
\subfloat{\topinset{\bfseries \textcolor{white}{(b)}}{\label{fig:Effect of optimization b}\includegraphics[width=0.25\linewidth]{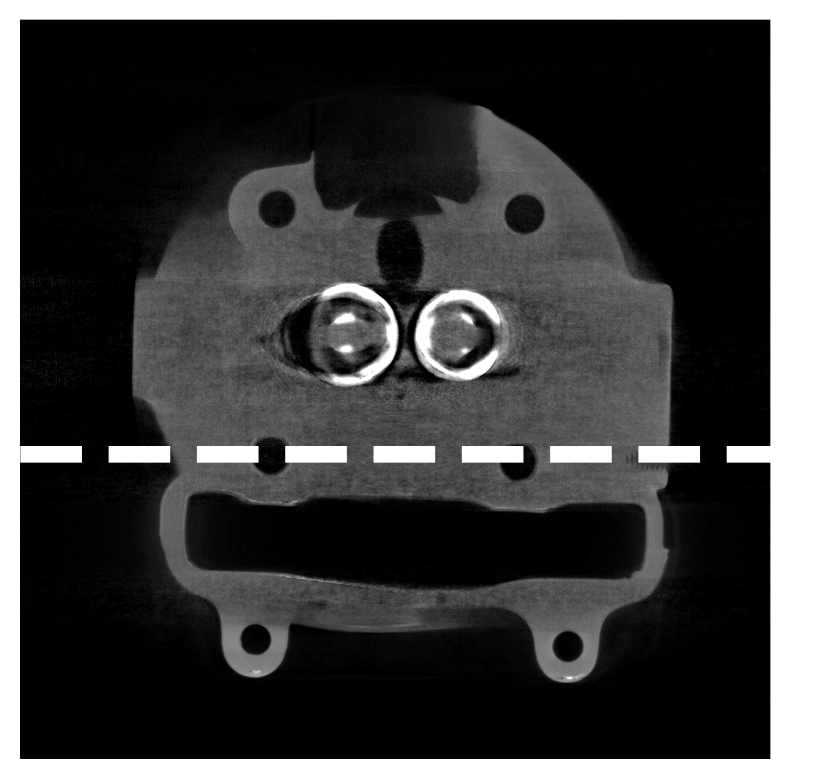}}{0.06in}{.06in}}
\subfloat{\topinset{\bfseries \textcolor{white}{(c)}}{\label{fig:Effect of optimization c}\includegraphics[width=0.25\linewidth]{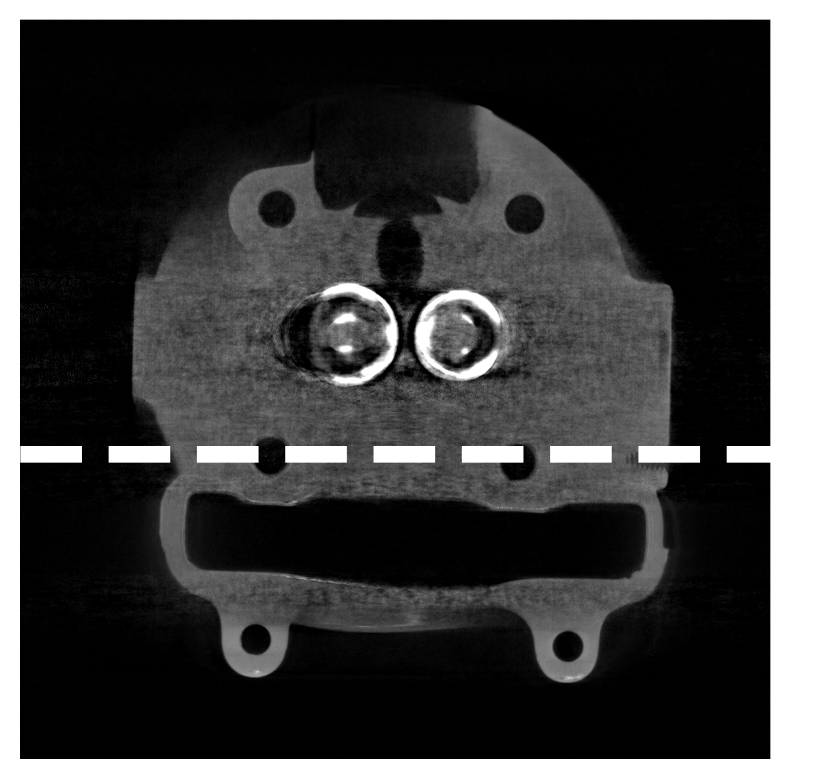}}{0.06in}{.06in}}
\subfloat{\topinset{\bfseries \textcolor{white}{(d)}}{\label{fig:Effect of optimization d}\includegraphics[width=0.25\linewidth]{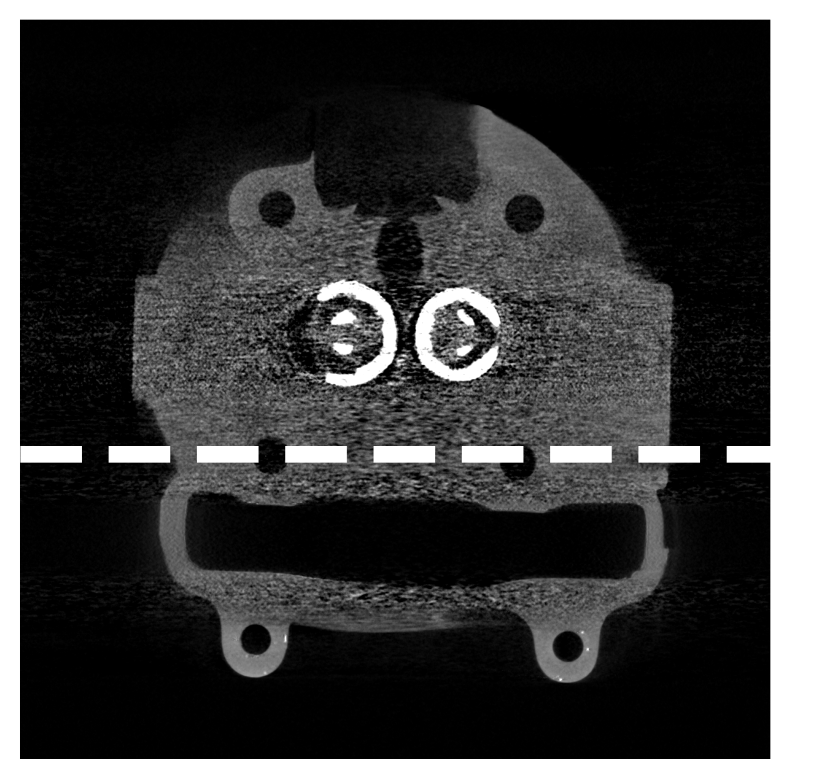}}{0.06in}{.06in}}\\[-0.15ex]
\subfloat{\topinset{\bfseries \textcolor{black}{(e)}}{\includegraphics[width=0.7\linewidth]{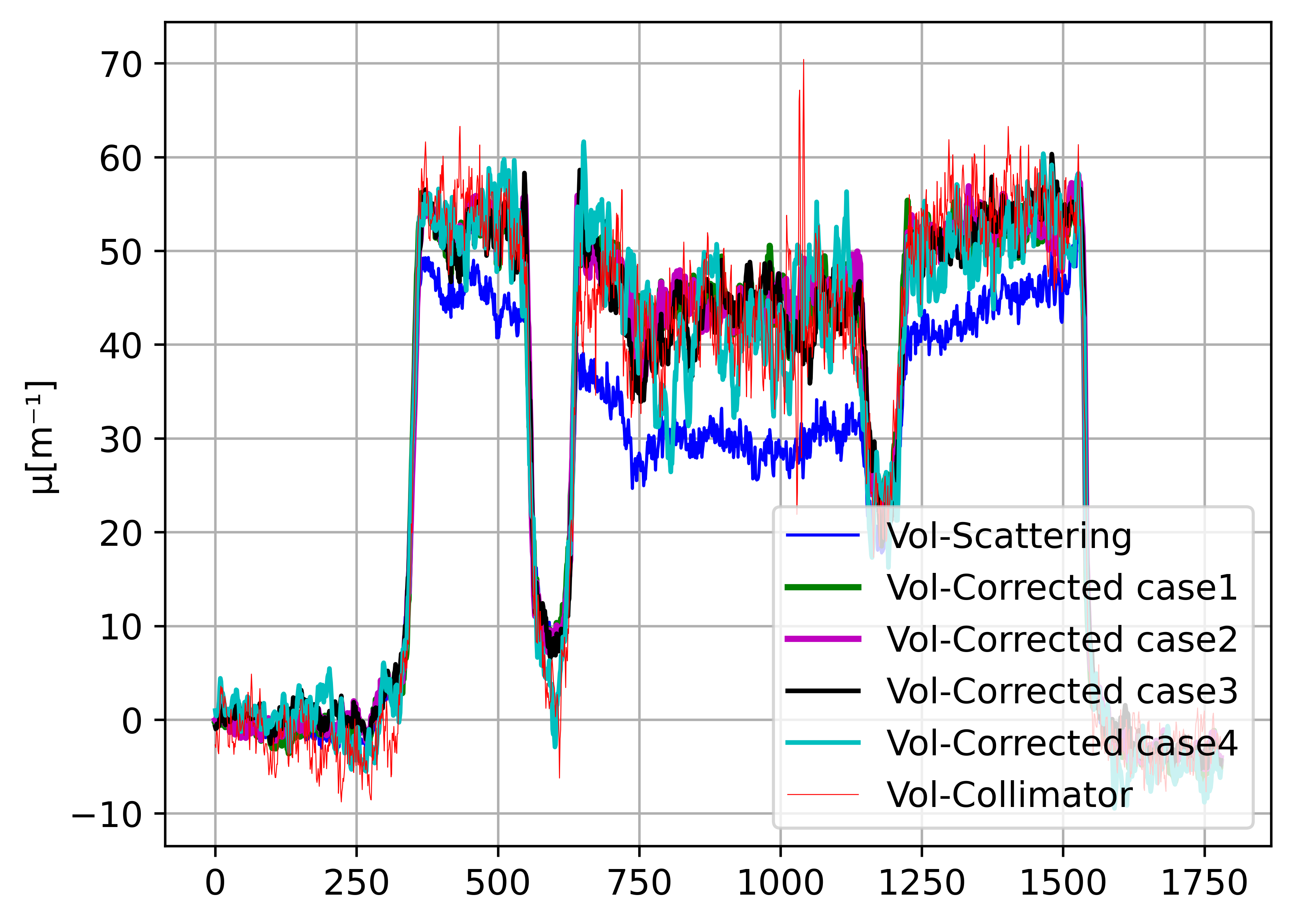}}{0.1in}{.4in}}

\caption{Comparison of the results of the correction with different parameters' optimization. (a) Scatter-corrected case 1 (standard parameters), (b) scatter-corrected case 2, (c) scatter-corrected case 3, (d) scatter-corrected case 4 (e) profile lines of the images in (a), (b), (c), (d), and from the near scatter-free volume's slice from the collimator. The profile lines in this example are averaged over multiple rows to suppress the noise and to make the comparison more clear. The profiles are marked by white dashed lines in (a), (b), (c), and (d).}
\label{fig:Effect of optimization}
\end{figure}


 It is shown that following this approach results in a very long computation time for the MC simulation and the scatter correction algorithm. In the rest of the simulation cases, both the interpolation and the smoothing techniques were used to accelerate the simulation in addition to the optimization of the MC key parameters. Columns 3-6 show different settings of these key parameters. These settings gradually decrease the simulation time of both the MC simulation and the scatter correction algorithm. The quality of the result of the scatter correction has been assessed using the quantitative evaluation methods mentioned in subsection \ref{Quantitative Evaluation}. The results of this evaluation are shown in table (\ref{tab:casesquanitative}). According to this evaluation, the results of both cases 2 and 3 which are shown in Fig.~\ref{fig:Effect of optimization}\subref{fig:Effect of optimization b} and Fig.~\ref{fig:Effect of optimization}\subref{fig:Effect of optimization c} respectively, produce good scatter correction quality with a great enhancement of the computation time in comparison to case 1 (shown in Fig.~\ref{fig:Effect of optimization}\subref{fig:Effect of optimization a}) and the standard case (not shown). Case 3 for example is able to simulate a single scatter projection within 2.7 seconds only, whereas the full set of the projections is calculated within 1.12 h using a single-GPU. This gives a 162$\times$ speed-up in comparison to the standard case (column 2) which requires 181.6 h to acquire the full set of the projections. Although, further optimization of the key parameters in case 4 results in less required computation time in comparison to the other cases, the result of the scatter correction, in this case, is noisy as shown in Fig.~\ref{fig:Effect of optimization}\subref{fig:Effect of optimization d}.

\begin{table}[ht]
\captionsetup{justification=centering}
\caption{SUMMARY OF THE SIMULATION PARAMETERS AND THE TIME REQUIRED FOR THE MC SIMULATION AND THE SCATTER CORRECTION FOR THE TEST CASES (1-4).}
\begin{threeparttable}[t]
\centering
\begin{tabular}{| C{1.6cm} | C{0.9cm} | C{0.9cm} | C{0.9cm} | C{0.9cm} | C{0.9cm} |}
\hline
& Standard & Case1 & Case2  & Case3 & Case4\\ 
\hline
Projections for FBP & 3000 & 3000 &  3000 & 3000 &3000\\ 
\hline
FBP time & 1670 s & 1670 s  & 1670 s & 1670 s &1670 s\\
\hline
Photons& $2.4e9$ & $2.4e8$ &  $1.2e8$ & $8.5e7$ & $5e7$\\
\hline
Splitting & 20 & 20 & 10 & 5 & 10\\
\hline
Step size & 1 & 1 &  2 &3&2\\
\hline
Projections simulated& 3000 & 1500 & 1500 & 1500 & 1500\\
\hline
MC time \tnote{(1)} /projection& 218 s & 21.6 s &  4.6 s & 2.7 s & 2 s\\
\hline
MC time /iteration & 45.4 h & 2.25 h &  0.48 h & 0.28 h & 0.2 h\\
\hline
Correction /iteration& 46.3 h & 3.6 h & 1.9 h  & 1.7 h & 1.2 h\\
\hline
Correction / 3 iterations& 139 h & 11 h & 5.8 h &  5.2 h & 4.9 h\\
\hline
\end{tabular}
\begin{tablenotes}\footnotesize
  \item [(1)] This represents the simulation time of a single projection over single-GPU.
\end{tablenotes}
\end{threeparttable}
\label{tab:timeOptimized}
\end{table}

\begin{table}[ht]
\captionsetup{justification=centering}
\caption{QUANTITATIVE EVALUATION OF THE SCATTER CORRECTION RESULTS FOR THE TEST CASES (1-4).}
\centering
\begin{tabular}{| C{1cm} | C{1cm} | C{1cm} | C{1cm} | C{1cm} | C{1cm} |}
\hline
& Corrupted& Case1 & Case2  & Case3 & Case4\\ 
\hline
CNR & 13.1 & 16 &  18.5 & 13.5 &7\\ 
\hline
MSE &- & - & 74.8 & 84.7  & 117 \\
\hline
NCC& - &-& 0.956 & 0.951 & 0.937\\
\hline

\end{tabular}
\label{tab:casesquanitative}
\end{table}

\section*{Acknowledgment}
This work was supported by the German Academic Exchange Service (DAAD, No. 57381412) and the German Research Foundation (DFG, Germany) under the DFG-project SI 587/18-1 in the priority program SPP 2187.

\section{Conclusion}

In this work, a multi-GPU accelerated MC photon transport model has been implemented and embedded into the iterative scatter correction algorithm for high-resolution flat-panel CT. Specially, fundamental physics including Compton scattering, Rayleigh scattering, and photoelectric absorption are implemented in the proposed MC model. The MC model is accelerated by splitting the projections equally over multi-GPU to enable a full parallelization. In the experiment, we have validated the proposed MC model by comparing it with the state-of-the-art MC simulators, i.e., EGSnrc and aRTist, and with the real-world scanner. In comparison with the multi-threaded MC simulator EGSnrc, a 202$\times$ speed-up is achieved for a (2304,3200) pixels detector using four GPUs. In addition, an iterative scatter correction algorithm is introduced by combining the proposed MC model with FBP. Compared to the reference images acquired with the collimator, the scattering artifacts are effectively suppressed within one to three iterations of the scatter correction algorithm. The work in this paper has demonstrated that the proposed MC photon transport model is both sufficiently fast compared to FBP and sufficiently accurate for scattering artifact correction in high-resolution flat-panel CT reconstruction.





%

\appendices




\ifCLASSOPTIONcaptionsoff
  \newpage
\fi

\bibliographystyle{IEEEtran}
{\footnotesize
\bibliography{IEEEabrv,bare_jrnl}}

\end{document}